\documentclass[10pt,english]{IEEEtran}
\usepackage[T1]{fontenc}
\usepackage[latin9]{inputenc}
\usepackage{array}
\usepackage{verbatim}
\usepackage{amsmath}
\usepackage{amssymb}
\usepackage{graphicx}

\makeatletter

\providecommand{\tabularnewline}{\\}

\makeatother

\usepackage{amsthm}
\usepackage{dsfont}
\usepackage{array}
\usepackage{mathrsfs}
\usepackage{cite}
\usepackage{comment}
\usepackage{mathrsfs}
\usepackage{hyperref}

\makeatother

\usepackage{babel}

\makeatother

\usepackage{babel}

\makeatother

\usepackage{babel}
\begin{document}
\bibliographystyle{IEEEtran}

\title{Shannon Meets Nyquist: \\Capacity of Sampled Gaussian Channels}

\author{Yuxin Chen, Yonina C. Eldar, and Andrea J. Goldsmith%
\thanks{Y. Chen is with the Department of Electrical Engineering and the Department
of Statistics, Stanford University, Stanford, CA 94305, USA (email:
yxchen@stanford.edu). Y. C. Eldar is with the Department of Electrical
Engineering, Technion, Israel Institute of Technology Haifa, Israel
32000, and has been a visiting professor at Stanford University (email:
yonina@ee.technion.ac.il). A. J. Goldsmith is with the Department
of Electrical Engineering, Stanford University, Stanford, CA 94305,
USA (email: andrea@wsl.stanford.edu). The contact author is Y. Chen.
This work was supported in part by the NSF Center for Science of Information,
the Interconnect Focus Center of the Semiconductor Research Corporation,
and BSF Transformative Science Grant 2010505. It was presented in
part at the IEEE International Conference on Acoustics, Speech and
Signal Processing (ICASSP) 2011, the 49 Annual Allerton Conference
on Communication, Control, and Computing, and IEEE Information Theory
Workshop 2011. 

Copyright (c) 2012 IEEE. Personal use of this material is permitted.
However, permission to use this material for any other purposes must
be obtained from the IEEE by sending a request to pubs-permissions@ieee.org.%
}}
\maketitle
\begin{abstract}
We explore two fundamental questions at the intersection of sampling
theory and information theory: how channel capacity is affected by
sampling below the channel's Nyquist rate, and what sub-Nyquist sampling
strategy should be employed to maximize capacity. In particular, we
derive the capacity of sampled analog channels for three prevalent
sampling strategies: sampling with filtering, sampling with filter
banks, and sampling with modulation and filter banks. These sampling
mechanisms subsume most nonuniform sampling techniques applied in
practice. Our analyses illuminate interesting connections between
under-sampled channels and multiple-input multiple-output channels.
The optimal sampling structures are shown to extract out the frequencies
with the highest SNR from each aliased frequency set, while suppressing
aliasing and out-of-band noise. We also highlight connections between
undersampled channel capacity and minimum mean-squared error (MSE)
estimation from sampled data. In particular, we show that the filters
maximizing capacity and the ones minimizing MSE are equivalent under
both filtering and filter-bank sampling strategies. These results
demonstrate the effect upon channel capacity of sub-Nyquist sampling
techniques, and characterize the tradeoff between information rate
and sampling rate.\end{abstract}
\theoremstyle{plain}\newtheorem{lem}{\textbf{Lemma}}\newtheorem{theorem}{\textbf{Theorem}}\newtheorem{corollary}{\textbf{Corollary}}\newtheorem{prop}{\textbf{Proposition}}\newtheorem{fct}{Fact}\newtheorem{remark}{\textbf{Remark}}

\theoremstyle{definition}\newtheorem{definition}{\textbf{Definition}}\newtheorem{example}{\textbf{Example}}

\section{Introduction}

The capacity of continuous-time Gaussian channels and the corresponding
capacity-achieving water-filling power allocation strategy over frequency
are well-known \cite{Gallager68}, and provide much insight and performance
targets for practical communication system design. These results implicitly
assume sampling above the Nyquist rate at the receiver end. However,
channels that are not bandlimited have an infinite Nyquist rate and
hence cannot be sampled at that rate. Moreover, hardware and power
limitations often preclude sampling at the Nyquist rate associated
with the channel bandwidth, especially for wideband communication
systems. This gives rise to several natural questions at the intersection
of sampling theory and information theory, which we will explore in
this paper: (1) how much information, in the Shannon sense, can be
conveyed through undersampled analog channels; (2) under a sub-Nyquist
sampling-rate constraint, which sampling structures should be chosen
in order to maximize information rate.

\subsection{Related Work}

The derivation of the capacity of linear time-invariant (LTI) channels
was pioneered by Shannon \cite{Sha48}. Making use of the asymptotic
spectral properties of Toeplitz operators \cite{GreSze1984}, this
capacity result established the optimality of a water-filling power
allocation based on signal-to-noise ratio (SNR) across the frequency
domain \cite{Gallager68}. Similar results for discrete-time Gaussian
channels have also been derived using Fourier analysis \cite{HirtMassey1988}.
On the other hand, the Shannon-Nyquist sampling theorem, which dictates
that channel capacity is preserved when the received signal is sampled
at or above the Nyquist rate, has frequently been used to transform
analog channels into their discrete counterparts (e.g.\cite{Bello1963,ForUng1998}).
For instance, this paradigm of discretization was employed by Medard
to bound the maximum mutual information in time-varying channels \cite{Med2000}.
However, all of these works focus on analog channel capacity sampled
at or above the Nyquist rate, and do not account for the effect upon
capacity of reduced-rate sampling. 

The Nyquist rate is the sampling rate required for perfect reconstruction
of bandlimited analog signals or, more generally, the class of signals
lying in shift-invariant subspaces. Various sampling methods at this
sampling rate for bandlimited functions have been proposed. One example
is recurrent non-uniform sampling proposed by Yen \cite{Yen1956},
which samples the signal in such a way that all sample points are
divided into blocks where each block contains $N$ points and has
a recurrent period. Another example is generalized multi-branch sampling
first analyzed by Papoulis \cite{Papoulis1977}, in which the input
is sampled through $M$ linear systems. For perfect recovery, these
methods require sampling at an aggregate rate above the Nyquist rate. 

In practice, however, the Nyquist rate may be excessive for perfect
reconstruction of signals that possess certain structures. For example,
consider multiband signals, whose spectral content resides continuously
within several subbands over a wide spectrum, as might occur in a
cognitive radio system. If the spectral support is known \textit{a
priori}, then the sampling rate requirement for perfect recovery is
the sum of the subband bandwidths, termed the \textit{Landau rate}
\cite{Landau1967}. One type of sampling mechanism that can reconstruct
multiband signals sampled at the Landau rate is a filter bank followed
by sampling, studied in \cite{LinVai1998,UnsZer1998}. The basic sampling
paradigm of these works is to apply a bank of prefilters to the input,
each followed by a uniform sampler. 

When the channel or signal structure is unknown, sub-Nyquist sampling
approaches have been recently developed to exploit the structure of
various classes of input signals, such as multiband signals \cite{MisEld2010Theory2Practice}.
In particular, sampling with modulation and filter banks is very effective
for signal reconstruction, where the key step is to scramble spectral
contents from different subbands through the modulation operation.
Examples includes the modulated wideband converter proposed by Mishali
\textit{et al.} \cite{MisEld2010Theory2Practice,MisEldDouSho2011}.
In fact, modulation and filter-bank sampling represents a very general
class of realizable nonuniform sampling techniques applied in practice.

Most of the above sampling theoretic work aims at finding optimal
sampling methods that admit perfect recovery of a class of analog
signals from \textit{noiseless} samples. There has also been work
on minimum reconstruction error from \textit{noisy} samples based
on certain statistical measures (e.g. mean squared error (MSE)). Another
line of work pioneered by Berger \emph{et. al.} \cite{BerTuf1967,BergerThesis,ChaDon1971,Eri1973}
investigated joint optimization of the transmitted pulse shape and
receiver prefiltering in pulse amplitude modulation over a sub-sampled
analog channel. In this work the optimal receiver prefilter that minimizes
the MSE between the original signal and the reconstructed signal is
shown to prevent aliasing. However, this work does not consider optimal
sampling techniques based on capacity as a metric. The optimal filters
derived in \cite{BerTuf1967,BergerThesis}\emph{ }are used to determine
an SNR metric which in turn is used to approximate sampled channel
capacity based on the formula for capacity of bandlimited AWGN channels.
However, this approximation does not correspond to the precise channel
capacity we derive herein, nor is the capacity of more general undersampled
analog channels considered. 

The tradeoff between capacity and hardware complexity has been studied
in another line of work on sampling precision \cite{Sha1994,KochLap2010}.
These works demonstrate that, due to quantization, oversampling can
be beneficial in increasing achievable data rates. The focus of these
works is on the effect of oversampling upon capacity loss due to quantization
error, rather than the effect of quantization-free subsampling upon
channel capacity.

\subsection{Contribution}

In this paper, we explore sampled Gaussian channels with the following
three classes of sampling mechanisms: (1) a filter followed by sampling:
the analog channel output is prefiltered by an LTI filter followed
by an ideal uniform sampler (see Fig. \ref{fig:SingleFilter}); (2)
filter banks followed by sampling: the analog channel output is passed
through a bank of LTI filters, each followed by an ideal uniform sampler
(see Fig. \ref{fig:SingleAntenna}); (3) modulation and filter banks
followed by sampling: the channel output is passed through $M$ branches,
where each branch is prefiltered by an LTI filter, modulated by different
modulation sequences, passed through another LTI filter and then sampled
uniformly. Our main contributions are summarized as follows.
\begin{itemize}
\item \textbf{Filtering followed by sampling.} We derive the sampled channel
capacity in the presence of both white and colored noise. Due to aliasing,
the sampled channel can be represented as a multiple-input single
output (MISO) Gaussian channel in the spectral domain, while the optimal
input effectively performs maximum ratio combining. The optimal prefilter
is derived and shown to extract out the frequency with the highest
SNR while suppressing signals from all other frequencies and hence
preventing aliasing. This prefilter also minimizes the MSE between
the original signal and the reconstructed signal, illuminating a connection
between capacity and MMSE estimation.
\item \textbf{Filter banks followed by sampling.} A closed-form expression
for sampled channel capacity is derived, along with analysis that
relates it to a multiple-input multiple-output (MIMO) Gaussian channel.
We also derive optimal filter banks that maximize capacity. The $M$
filters select the $M$ frequencies with highest SNRs and zero out
signals from all other frequencies. This alias-suppressing strategy
is also shown to minimize the MSE between the original and reconstructed
signals. This mechanism often achieves larger sampled channel capacity
than a single filter followed by sampling if the channel is non-monotonic,
and it achieves the analog capacity of multiband channels at the Landau
rate if the number of branches is appropriately chosen. 
\item \textbf{Modulation and filter banks followed by sampling.} For modulation
sequences that are periodic with period $T_{q}$, we derive the sampled
channel capacity and show its connection to a general MIMO Gaussian
channel in the frequency domain. For sampling following a single branch
of modulation and filtering, we provide an algorithm to identify the
optimal modulation sequence for piece-wise flat channels when $T_{q}$
is an integer multiple of the sampling period. We also show that the
optimal single-branch mechanism is equivalent to an optimal filter
bank with each branch sampled at a period $T_{q}$.
\end{itemize}
One interesting fact we discover for all these techniques is the non-monotonicity
of capacity with sampling rate, which indicates that at certain sampling
rates, channel degrees of freedom are lost. Thus, more sophisticated
sampling techniques are needed to maximize achievable data rates at
sub-Nyquist sampling rates in order to preserve all channel degrees
of freedom.

\begin{table}
\caption{\label{tab:Summary-of-Notation}Summary of Notation and Parameters}

\centering{}%
\begin{tabular}{>{\raggedright}p{0.7in}>{\raggedright}p{2.55in}}
$\mathcal{L}_{1}$ & set of measurable functions $f$ such that $\int\left|f\right|\mathrm{d}\mu<\infty$\tabularnewline
$\mathbb{S}_{+}$ & set of positive semidefinite matrices\tabularnewline
$h(t)$,$H(f)$  & impulse response, and frequency response of the analog channel\tabularnewline
$s_{i}(t)$, $S_{i}(f)$  & impulse response, and frequency response of the $i$th post-modulation
filter\tabularnewline
$p_{i}(t)$, $P_{i}(f)$ & impulse response, and frequency response of the $i$th pre-modulation
filter\tabularnewline
$\mathcal{S}_{\eta}(f)$, $\mathcal{S}_{x}(f)$ & power spectral density of the noise $\eta(t)$ and the stationary
input signal $x(t)$\tabularnewline
$M$  & number of prefilters\tabularnewline
$f_{s}$, $T_{s}$ & aggregate sampling rate, and the corresponding sampling interval ($T_{s}=1/f_{s}$)\tabularnewline
$q_{i}(t)$ & modulating sequence in the $i$th channel\tabularnewline
$T_{q}$ & period of the modulating sequence $q_{i}(t)$ \tabularnewline
$\left\Vert \cdot\right\Vert _{\text{F}}$, $\left\Vert \cdot\right\Vert _{2}$ & Frobenius norm, $\ell_{2}$ norm\tabularnewline
$\left[x\right]^{+}$, $\log^{+}x$ & $\max\left\{ x,0\right\} $, $\max\left\{ \log x,0\right\} $\tabularnewline
\end{tabular}
\end{table}

\subsection{Organization}

The remainder of this paper is organized as follows. In Section \ref{sec:Problem-Formulation},
we describe the problem formulation of sampled analog channels. The
capacity results for three classes of sampling strategies are presented
in Sections \ref{sec:General-Uniform-Sampling}--\ref{sec:Multi-channel-Pre-modulated-Pre-filtered}.
In each section, we analyze and interpret the main theorems based
on Fourier analysis and MIMO channel capacity, and identify sampling
structures that maximize capacity. The connection between the capacity-maximizing
samplers and the MMSE samplers is provided in Section \ref{sec:ConnectionCapacityMMSE}.
Proofs of the main theorems are provided in the appendices, and the
notation is summarized in Table \ref{tab:Summary-of-Notation}.$ $

\section{Preliminaries: Capacity of undersampled channels\label{sec:Problem-Formulation}}

\subsection{Capacity Definition}

We consider the continuous-time additive Gaussian channel (see \cite[Chapter 8]{Gallager68}),
where the channel is modeled as an LTI filter with impulse response
$h(t)$ and frequency response $H(f)=\int_{-\infty}^{\infty}h(t)\exp(-j2\pi ft)\text{d}t$.
The transmit signal $x(t)$ is time-constrained to the interval $(0,T]$.
The analog channel output is given as
\begin{equation}
r(t)=h(t)*x(t)+\eta(t),\label{eq:ChannelModel}
\end{equation}
 and is observed over%
\footnote{We impose the assumption that both the transmit signal and the observed
signal are constrained to finite time intervals to allow for a rigorous
definition of channel capacity. In particular, as per Gallager's analysis
\cite[Chapter 8]{Gallager68}, we first calculate the capacity for
finite time intervals and then take the limit of the interval to infinity. %
} $\left(0,T\right]$, where $\eta(t)$ is stationary zero-mean Gaussian
noise. We assume throughout the paper that \textit{perfect channel
state information, }\textit{\emph{i.e. perfect knowledge of $h(t)$,}}
is known at both the transmitter and the receiver. The analog channel
capacity is defined as \cite[Section 8.1]{Gallager68} 
\[
C=\lim_{T\rightarrow\infty}\frac{1}{T}\sup I\left(\left\{ x(t)\right\} _{t=0}^{T};\left\{ r(t)\right\} _{t=0}^{T}\right),
\]
where the supremum is over all input distributions subject to an average
power constraint $\mathbb{E}(\frac{1}{T}\int_{0}^{T}\left|x(\tau)\right|^{2}\mathrm{d}\tau)\leq P$.
Since any given analog channel can be converted to a countable number
of independent parallel discrete channels by a Karhunen-Loeve decomposition,
the capacity metric quantifies the maximum mutual information between
the input and output of these discrete channels. If we denote $[x]^{+}=\max\{x,0\}$
and $\log^{+}x=\max\left\{ 0,\log x\right\} $, then the analog channel
capacity is given as follows.

\begin{theorem}\label{thmGallagerChannelCapacity}\cite[Theorem 8.5.1]{Gallager68}
Consider an analog channel with power constraint $P$ and noise power
spectral density $\mathcal{S}_{\eta}(f)$. Assume that $\left|H(f)\right|^{2}/\mathcal{S}_{\eta}(f)$
is bounded and integrable, and that either $\int_{-\infty}^{\infty}\mathcal{S}_{\eta}(f)\mathrm{d}f<\infty$
or that $\mathcal{S}_{\eta}(f)$ is white. Then the analog channel
capacity is given by 
\begin{align}
C & =\frac{1}{2}{\displaystyle \int}_{-\infty}^{\infty}\log^{+}\left(\nu\frac{\left|H\left(f\right)\right|^{2}}{\mathcal{S}_{\eta}(f)}\right)\mathrm{d}f,\label{eq:GallagerCapacity}
\end{align}
 where $\nu$ satisfies 
\begin{equation}
{\displaystyle \int}_{-\infty}^{\infty}\left[\nu-\frac{\mathcal{S}_{\eta}(f)}{\left|H\left(f\right)\right|^{2}}\right]^{+}\mathrm{d}f=P.\label{eq:WaterFillingConstraint}
\end{equation}
\end{theorem} 

$ $For a channel whose bandwidth lies in $[-B,B]$, if we remove
the noise outside the channel bandwidth via prefiltering and sample
the output at a rate $f\geq2B$, then we can perfectly recover all
information conveyed within the channel bandwidth, which allows (\ref{eq:GallagerCapacity})
to be achieved without sampling loss. For this reason, we will use
the terminology \textbf{\textit{Nyquist-rate channel capacity}} for
the analog channel capacity (\ref{eq:GallagerCapacity}), which is
commensurate with sampling at or above the Nyquist rate of the received
signal after optimized prefiltering. 

\begin{figure}[htbp]
\begin{centering}
\textsf{\includegraphics[scale=0.45]{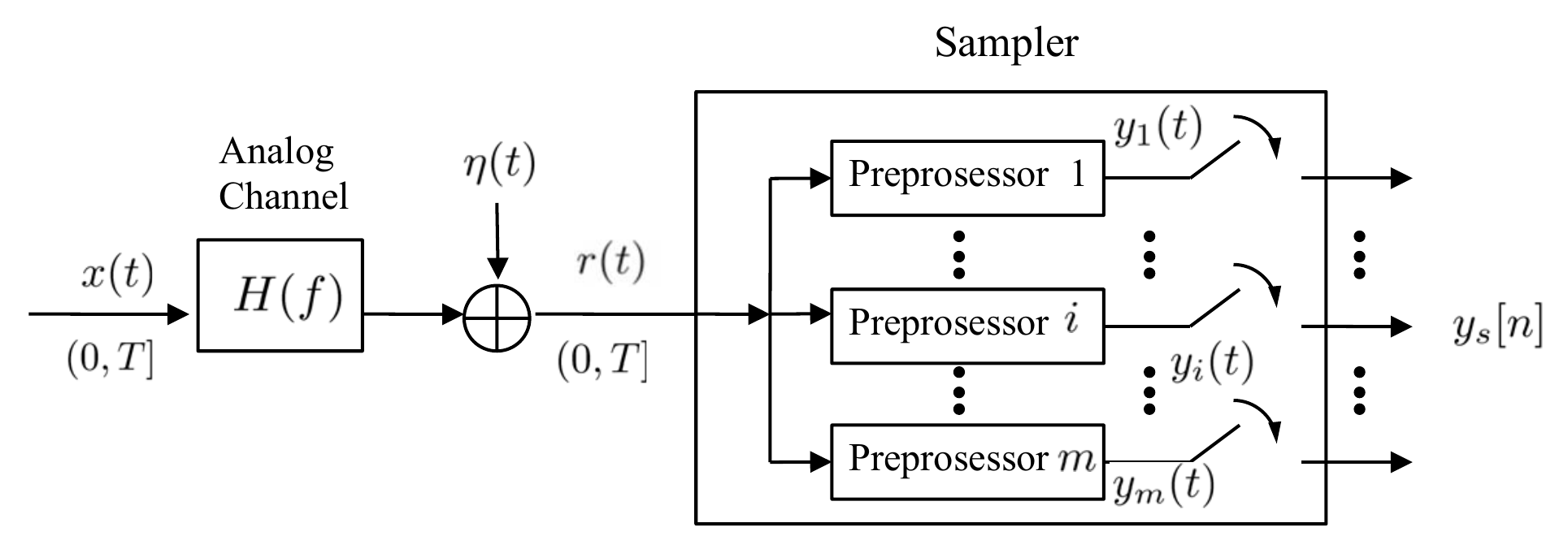}} 
\par\end{centering}

\caption{\label{fig:ProblemFormulation}Sampled Gaussian channel. The input
$x(t)$, constrained to $(0,T]$, is passed through $M$ branches
of the receiver analog front end to yield analog outputs $\left\{ y_{i}(t):\text{ }1\leq i\leq M\right\} $;
each $y_{i}(t)$ is observed over $\left(0,T\right]$ and uniformly
sampled at a rate $f_{s}/M$ to yield the sampled sequence $y_{i}[n]$.
The preprocessor can be a filter, or combination of a filter and a
modulator.}
\end{figure}

Under sub-Nyquist sampling, the capacity depends on the sampling mechanism
and its sampling rate. Specifically, the channel output $r(t)$ is
now passed through the receiver's analog front end, which may include
a filter, a bank of $M$ filters, or a bank of preprocessors consisting
of filters and modulation modules, yielding a collection of analog
outputs $\left\{ y_{i}(t):\text{ }1\leq i\leq M\right\} $. We assume
that the analog outputs are observed over the time interval $\left(0,T\right]$
and then passed through ideal uniform samplers, yielding a set of
digital sequences $\left\{ y_{i}[n]:\text{ }n\in\mathbb{Z},\text{ }1\leq i\leq M\right\} $,
as illustrated in Fig. \ref{fig:ProblemFormulation}. Here, each branch
is uniformly sampled at a sampling rate of $f_{s}/M$ samples per
second.

Define ${\bf y}[n]=\left[y_{1}[n],\cdots,y_{M}[n]\right]$, and denote
by $I(\left\{ x(t)\right\} _{t=0}^{T};\left\{ {\bf y}[n]\right\} _{t=0}^{T})$
the mutual information between the input $x(t)$ on the interval $0<t\leq T$
and the samples $\left\{ y[n]\right\} $ observed on the interval
$0<t\leq T$. We pose the problem of finding the capacity $C(f_{s})$
of sampled channels as quantifying the maximum mutual information
in the limit as $T\rightarrow\infty$. The sampled channel capacity
can then be expressed as
\[
C(f_{s})=\lim_{T\rightarrow\infty}\frac{1}{T}\sup I\left(\left\{ x(t)\right\} _{t=0}^{T};\left\{ {\bf y}[n]\right\} _{t=0}^{T}\right),
\]
where the supremum is over all possible input distributions subject
to an average power constraint $\mathbb{E}(\frac{1}{T}\int_{0}^{T}\left|x(\tau)\right|^{2}\mathrm{d}\tau)\leq P$.
$ $We restrict the transmit signal $x(t)$ to be continuous with
bounded variance (i.e. $\sup_{t}\mathbb{E}\left|x(t)\right|^{2}<\infty$),
and restrict the probability measure of $x(t)$ to be uniformly continuous.
This restriction simplifies some mathematical analysis, while still
encompassing most practical signals of interests %
\footnote{Note that this condition is not necessary for our main theorems. An
alternative proof based on correlation functions is provided in \cite{ChenEldarGoldsmith2012},
which does not require this condition.%
}.

\subsection{Sampling Mechanisms\label{sub:Sampling-Mechanisms}}

In this subsection, we describe three classes of sampling strategies
with increasing complexity. In particular, we start from sampling
following a single filter, and extend our results to incorporate filter
banks and modulation banks.

\subsubsection{Filtering followed by sampling}

\begin{figure}[htbp]
\begin{centering}
\textsf{\includegraphics[scale=0.35]{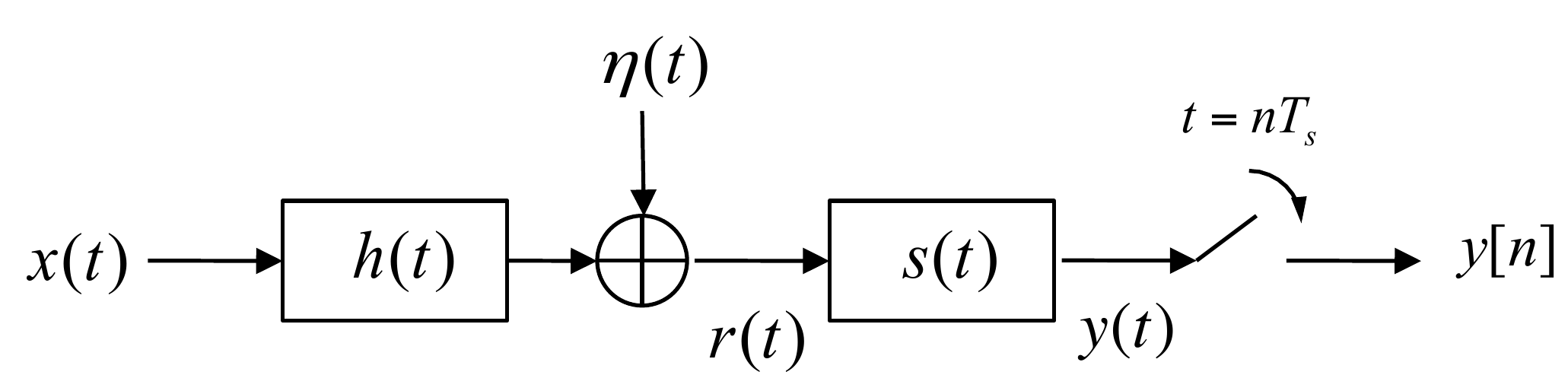}} 
\par\end{centering}

\caption{\label{fig:SingleFilter}Filtering followed by sampling: the analog
channel output $r(t)$ is linearly filtered prior to ideal uniform
sampling.}
\end{figure}

Ideal uniform sampling is performed by sampling the analog signal
uniformly at a rate $f_{s}=T_{s}^{-1}$, where $T_{s}$ denotes the
sampling interval. In order to avoid aliasing, suppress out-of-band
noise, and compensate for linear distortion of practical sampling
devices, a prefilter is often added prior to the ideal uniform sampler
\cite{EldMic2009}. Our sampling process thus includes a general analog
prefilter, as illustrated in Fig. \ref{fig:SingleFilter}. Specifically,
before sampling, we prefilter the received signal with an LTI filter
that has impulse response $s(t)$ and frequency response $S\left(f\right)$,
where we assume that $h(t)$ and $s(t)$ are both bounded and continuous.
The filtered output is observed over $(0,T]$ and can be written as
\begin{equation}
y(t)=s(t)*\left(h(t)*x(t)+\eta(t)\right),\quad t\in\left(0,T\right].\label{eq:PrefilteredReceiveSignals-1}
\end{equation}
 We then sample $y(t)$ using an ideal uniform sampler, leading to
the sampled sequence
\[
y[n]=y(nT_{s}).
\]

\begin{figure}[htbp]
\begin{centering}
\textsf{\includegraphics[scale=0.33]{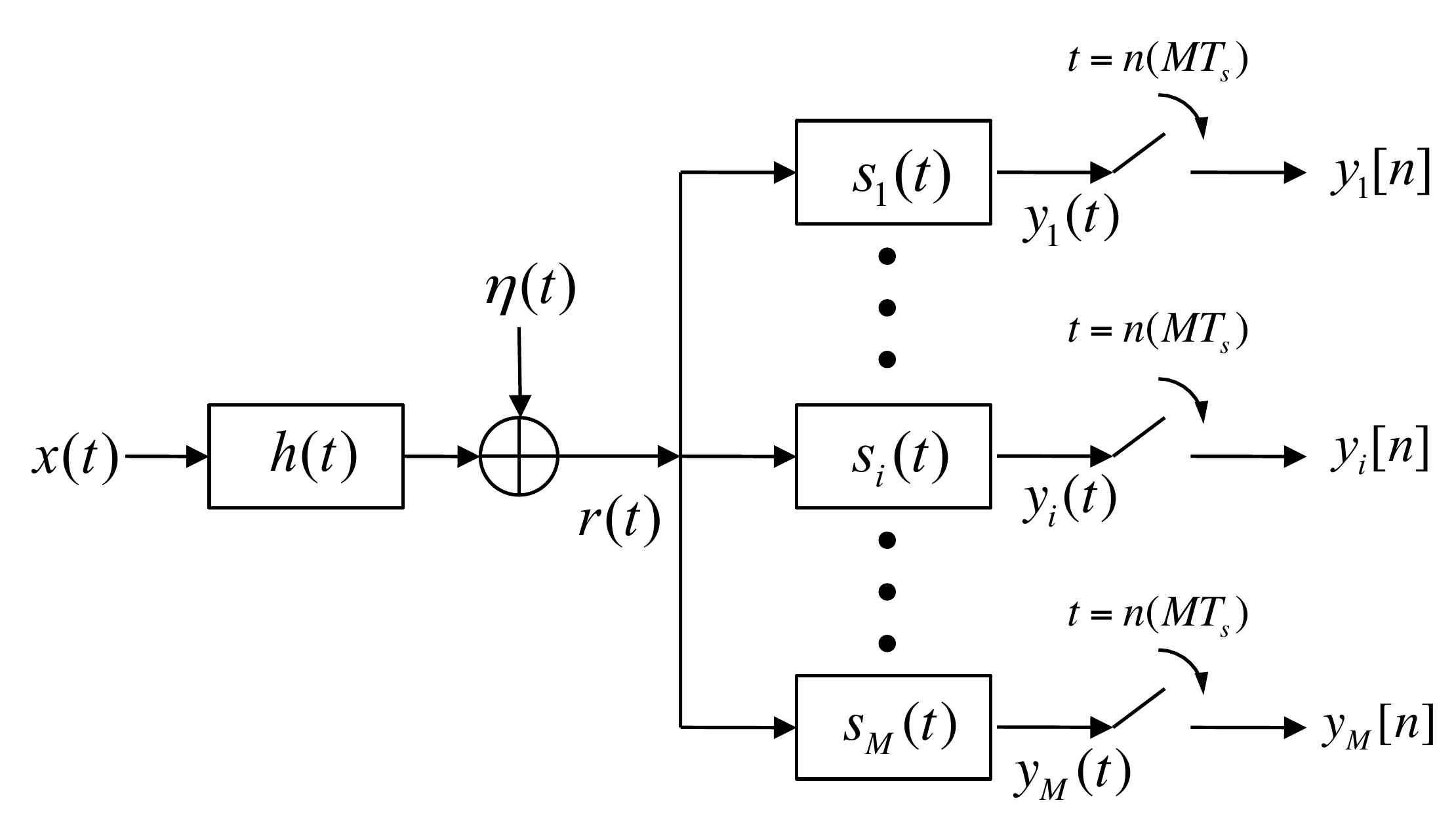}} 
\par\end{centering}

\caption{\label{fig:SingleAntenna}A filter bank followed by sampling: the
received analog signal $r(t)$ is passed through $M$ branches. In
the $i$th branch, the signal $r(t)$ is passed through an LTI prefilter
with impulse response $s_{i}(f)$, and then sampled uniformly at a
rate $f_{s}/M$.}
\end{figure}

\subsubsection{Sampling following Filter Banks}

Sampling following a single filter often falls short of exploiting
channel structure. In particular, although Nyquist-rate uniform sampling
preserves information for bandlimited signals, for multiband signals
it does not ensure perfect reconstruction at the Landau rate (i.e.
the total widths of spectral support). That is because uniform sampling
at sub-Nyquist rates may suppress information by collapsing subbands,
resulting in fewer degrees of freedom. This motivates us to investigate
certain nonuniform sampling mechanisms. We begin by considering a
popular class of non-uniform sampling mechanisms, where the received
signal is preprocessed by a bank of filters. Most practical nonuniform
sampling techniques \cite{Papoulis1977,LinVai1998,UnsZer1998} fall
under filter-bank sampling and modulation-bank sampling (as described
in Section \ref{sub:Sampling-via-Modulation}). Note that the filters
may introduce delays, so that this approach subsumes that of a filter
bank with different sampling times at each branch.

In this sampling strategy, we replace the single prefilter in Fig.
\ref{fig:SingleFilter} by a bank of $M$ analog filters each followed
by ideal sampling at rate $f_{s}/M$, as illustrated in Fig. \ref{fig:SingleAntenna}.
We denote by $s_{i}(t)$ and $S_{i}\left(f\right)$ the impulse response
and frequency response of the $i$th LTI filter, respectively. The
filtered analog output in the $i$th branch prior to sampling is then
given as
\begin{equation}
y_{i}(t)=\left(h(t)*s_{i}(t)\right)*x(t)+s_{i}(t)*\eta(t),\quad t\in\left(0,T\right].\label{eq:ContinuousSignalFilterBank}
\end{equation}
These filtered signals are then sampled uniformly to yield
\[
y_{i}[n]\overset{\Delta}{=}y_{i}(nMT_{s})\quad\text{and}\quad{\bf y}[n]\overset{\Delta}{=}\left[y_{1}[n],y_{2}[n],\cdots,y_{M}[n]\right],
\]
 where $T_{s}=f_{s}^{-1}$.

\subsubsection{Modulation and Filter Banks Followed by Sampling\label{sub:Sampling-via-Modulation}}

\begin{figure}[htbp]
\begin{centering}
\textsf{\includegraphics[scale=0.35]{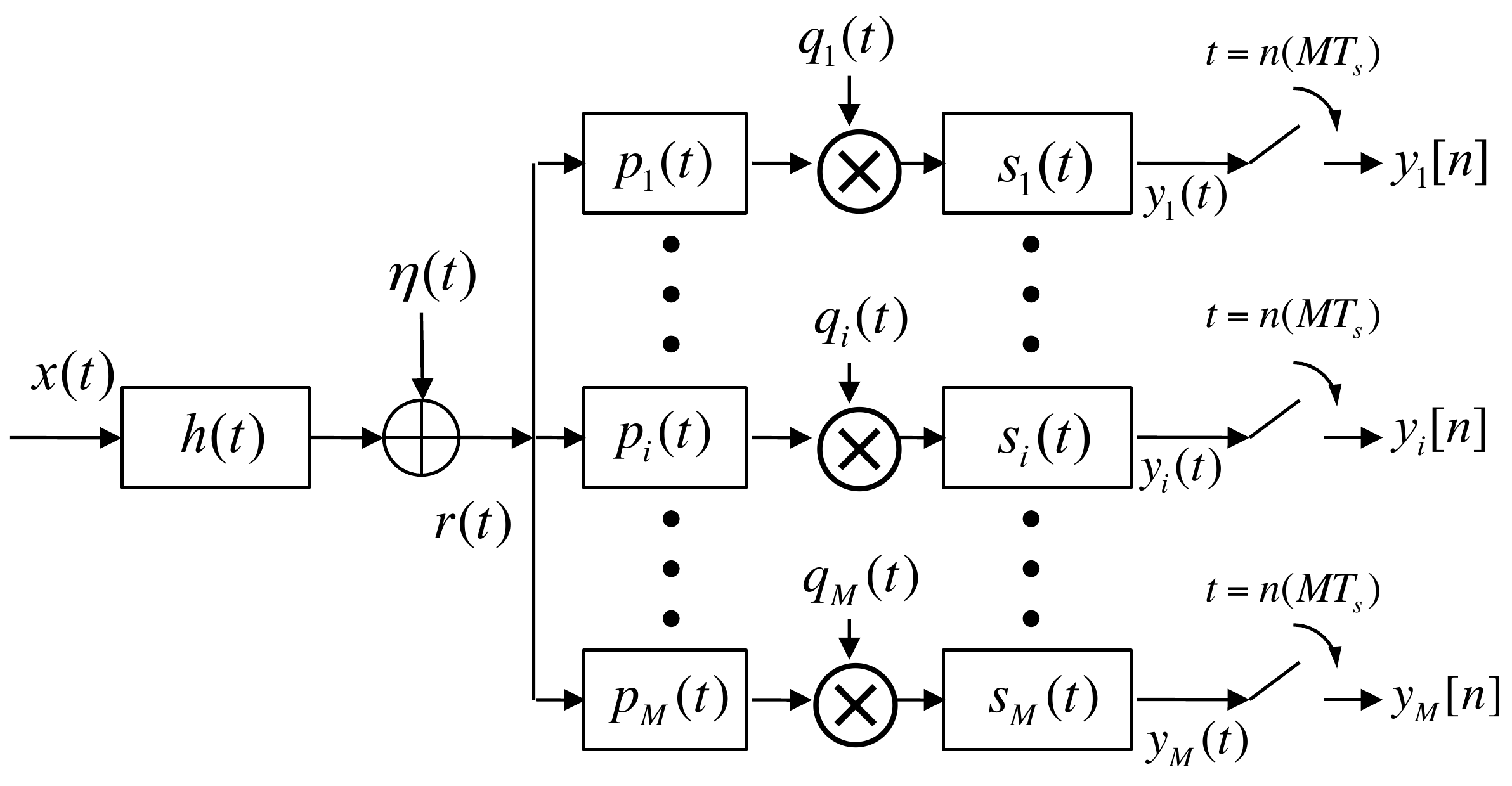}} 
\par\end{centering}

\caption{\label{fig:PremodulatedPrefilteredSampler}Modulation and filter banks
followed by sampling: in each branch, the received signal is prefiltered
by an LTI filter with impulse response $p_{i}(t)$, modulated by a
periodic waveform $q_{i}(t)$, filtered by another LTI filter with
impulse response $s_{i}(t)$, and then sampled at a rate $f_{s}/M$.}
\end{figure}

We generalize the filter-bank sampling strategy by adding an additional
filter bank and a modulation bank, which includes as special cases
a broad class of nonuniform sampling methods that are applied in both
theory and practice. Specifically, the sampling system with sampling
rate $f_{s}$ comprises $M$ branches. In the $i$th branch, the received
signal $r(t)$ is prefiltered by an LTI filter with impulse response
$p_{i}(t)$ and frequency response $P_{i}(f)$, modulated by a periodic
waveform $q_{i}(t)$ of period $T_{q}$, filtered by another LTI filter
with impulse response $s_{i}(t)$ and frequency response $S_{i}(f)$,
and then sampled uniformly at a rate $f_{s}/M=\left(MT_{s}\right)^{-1}$,
as illustrated in Fig. \ref{fig:PremodulatedPrefilteredSampler}.
The first prefilter $P_{i}(f)$ will be useful in removing out-of-band
noise, while the periodic waveforms scramble spectral contents from
different aliased sets, thus bringing in more design flexibility that
may potentially lead to better exploitation of channel structure.
By taking advantage of random modulation sequences to achieve incoherence
among different branches, this sampling mechanism has proven useful
for sub-sampling multiband signals \cite{MisEld2010Theory2Practice}. 

In the $i$th branch, the analog signal after post-modulation filtering
prior to sampling can be written as
\begin{equation}
y_{i}(t)=s_{i}(t)*\left(q_{i}(t)\cdot\left[p_{i}(t)*\left(h(t)*x(t)+\eta(t)\right)\right]\right),\label{eq:ContinuousSignalPremodulatedPrefiltered}
\end{equation}
resulting in the digital sequence of samples 
\[
y_{i}[n]=y_{i}(nMT_{s})\quad\text{and}\quad{\bf y}\left[n\right]=\left[y_{1}\left[n\right],\cdots,y_{M}\left[n\right]\right]^{T}.
\]

\section{A Filter Followed by Sampling\label{sec:General-Uniform-Sampling}}

\subsection{Main Results}

The sampled channel capacity under sampling with filtering is stated
in the following theorem.

\begin{theorem}\label{thmPerfectCSIPrefilteredSamplerRigorous}Consider
the system shown Fig. \ref{fig:SingleFilter}, where $\eta(t)$ is
Gaussian noise with power spectral density $\mathcal{S}_{\eta}(f)$.
Assume that $h(t)$, $s(t)$, $S\left(f\right)\sqrt{\mathcal{S}_{\eta}\left(f\right)}$
are all continuous, bounded and absolutely Riemann integrable. Additionally,
suppose that $h_{\eta}(t):=\mathcal{F}^{-1}\left(\frac{H\left(f\right)}{\sqrt{\mathcal{S}_{\eta}\left(f\right)}}\right)$
satisfies $h_{\eta}(t)=o\left(t^{-\epsilon}\right)$ for some constant
\footnote{This condition is used in Appendix \ref{sec:Proof-of-Theorem-PerfectCSIPrefilteredSampler}
as a sufficient condition to guarantee asymptotic properties of Toeplitz
matrices. A similar condition will be used in Theorems \ref{thmPerfectCSIFilterBankSingleAntenna}
and \ref{thmPremodulatedFilterBank}.%
} $\epsilon>1$. The capacity $C(f_{s})$ of the sampled channel with
a power constraint $P$ is then given parametrically as
\begin{align}
C\left(f_{s}\right) & ={\displaystyle \int}_{-\frac{f_{s}}{2}}^{\frac{f_{s}}{2}}\frac{1}{2}\log^{+}\left(\nu\gamma^{\text{s}}(f)\right)\mathrm{d}f,\label{eq:CapacityGeneralSamplingColorNoise}
\end{align}
 where $\nu$ satisfies 
\begin{equation}
{\displaystyle \int}_{-\frac{f_{s}}{2}}^{\frac{f_{s}}{2}}\left[\nu-1/\gamma^{\text{s}}(f)\right]^{+}\mathrm{d}f=P.\label{eq:WaterFillingConstraintColorNoise}
\end{equation}
Here, we denote 
\[
\gamma^{\text{s}}(f):=\frac{\underset{l\in\mathbb{Z}}{\sum}\left|H(f-lf_{s})S(f-lf_{s})\right|^{2}}{\underset{l\in\mathbb{Z}}{\sum}\left|S(f-lf_{s})\right|^{2}\mathcal{S}_{\eta}(f-lf_{s})}.
\]

\end{theorem}

As expected, applying the prefilter modifies the channel gain and
colors the noise accordingly. The color of the noise is reflected
in the denominator term of the corresponding SNR in (\ref{eq:CapacityGeneralSamplingColorNoise})
at each $f\in[-f_{s}/2,f_{s}/2]$ within the sampling bandwidth. The
channel and prefilter response leads to an equivalent frequency-selective
channel, and the ideal uniform sampling that follows generates a folded
version of the non-sampled channel capacity. Specifically, this capacity
expression differs from the analog capacity given in Theorem \ref{thmGallagerChannelCapacity}
in that the SNR in the sampled scenario is $\gamma^{\text{s}}(f)$
in contrast to $\left|H(f)\right|^{2}/\mathcal{S}_{\eta}(f)$ for
the non-sampled scenario. $ $Water filling over $1/\gamma^{\text{s}}(f)$
determines the optimal power allocation.

\subsection{Approximate Analysis}

Rather than providing here a rigorous proof of Theorem \ref{thmPerfectCSIPrefilteredSamplerRigorous},
we first develop an approximate analysis by relating the aliased channel
to MISO channels, which allows for a communication theoretic interpretation.
The rigorous analysis, which is deferred to Appendix \ref{sec:Proof-of-Theorem-PerfectCSIPrefilteredSampler},
makes use of a discretization argument and asymptotic spectral properties
of Toeplitz matrices.

\begin{figure}[htbp]
\begin{centering}
\textsf{\includegraphics[scale=0.32]{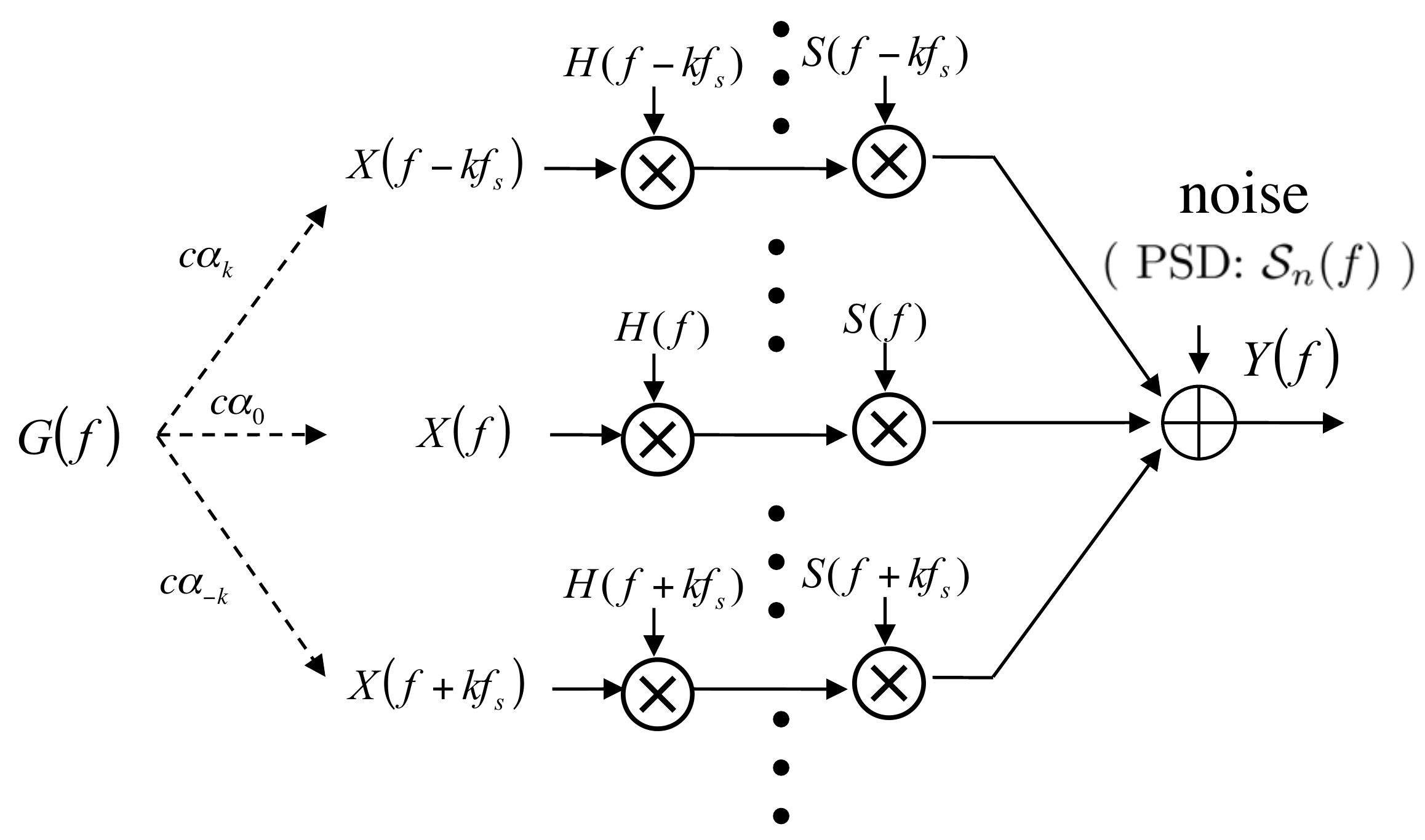}} 
\par\end{centering}

\caption{\label{fig:GeneralUniformSamplerEquivalent}Equivalent MISO Gaussian
channel for a given $f\in[-f_{s}/2,f_{s}/2]$ under filtering followed
by sampling. The additive noise has power spectral density $\mathcal{S}_{n}(f)=\sum_{l\in\mathbb{Z}}\mathcal{S}_{\eta}(f-lf_{s})\left|S(f-lf_{s})\right|^{2}$.}
\end{figure}

Consider first the equivalence between the sampled channel and a MISO
channel at a single frequency $f\in[-f_{s}/2,f_{s}/2]$. $ $As part
of the approximation, we suppose the Fourier transform $X(f)$ of
the transmitted signal exists%
\footnote{The Fourier transform of the input signal typically does not exist
since the input may be a stationary process. %
} . The Fourier transform of the sampled signal at any $f\in[-f_{s}/2,f_{s}/2]$
is given by
\begin{equation}
\frac{1}{T_{s}}\sum_{k\in\mathbb{Z}}H\left(f-kf_{s}\right)S(f-kf_{s})X\left(f-kf_{s}\right)\label{eq:AliasedSamplesFT}
\end{equation}
due to aliasing. The summing operation allows us to treat the aliased
channel at each $f$ within the sampling bandwidth as a separate MISO
channel with countably many input branches and a single output branch,
as illustrated in Fig. \ref{fig:GeneralUniformSamplerEquivalent}.

By assumption, the noise has spectral density $\mathcal{S}_{\eta}(f)$,
so that the filtered noise has power spectral density $\mathcal{S}_{\eta}(f)|S(f)|^{2}$.
The power spectral density of the sampled noise sequence at $f\in[-f_{s}/2,f_{s}/2]$
is then given by $\sum_{l\in\mathbb{Z}}\mathcal{S}_{\eta}(f-lf_{s})\left|S(f-lf_{s})\right|^{2}$.
If we term $\left\{ f-lf_{s}\mid\text{ }l\in\mathbb{Z}\right\} $
the \emph{aliased frequency set} for $f$, then the amount of power
allocated to $X(f-lf_{s})$ should ``match'' the corresponding channel
gain within each aliased set in order to achieve capacity. Specifically,
denote by $G(f)$ the transmitted signal for every $f\in[-f_{s}/2,f_{s}/2]$.
This signal is multiplied by a constant gain $c\alpha_{l}\text{ }(l\in\mathbb{Z})$,
and sent through the $l$th input branch, i.e.
\begin{equation}
X\left(f-lf_{s}\right)=c\alpha_{l}G(f),\quad\forall l\in\mathbb{Z},
\end{equation}
where $c$ is a normalizing constant, and 
\[
\alpha_{l}=\frac{H^{*}\left(f-lf_{s}\right)S^{*}\left(f-lf_{s}\right)}{\sum_{l}\left|H(f-lf_{s})S(f-lf_{s})\right|^{2}}.
\]
The resulting SNR can be expressed as the sum of SNRs (as shown in
\cite{Gold2005}) at each branch. Since the sampling operation combines
signal components at frequencies from each aliased set $\left\{ f-lf_{s}\mid l\in\mathbb{Z}\right\} $,
it is equivalent to having a set of parallel MISO channels, each indexed
by some $f\in[-f_{s}/2,f_{s}/2]$. The water-filling strategy is optimal
in allocating power among the set of parallel channels, which yields
the parametric equation (\ref{eq:WaterFillingConstraintColorNoise})
and completes our approximate analysis.

\subsection{Proof Sketch}

Since the Fourier transform is not well-defined for signals with infinite
energy, there exist technical flaws lurking in the approximate treatment
of the previous subsection. The key step to circumvent these issues
is to explore the asymptotic properties of Toeplitz matrices/operators.
This approach was used by Gallager \cite[Theorem 8.5.1]{Gallager68}
to prove the analog channel capacity theorem. Under uniform sampling,
however, the sampled channel no longer acts as a Toeplitz operator,
but instead becomes a block-Toeplitz operator. Since conventional
approaches \cite[Chapter 8.4]{Gallager68} do not accommodate for
block-Toeplitz matrices, a new analysis framework is needed. We provide
here a roadmap of our analysis framework, and defer the complete proof
to Appendix \ref{sec:Proof-of-Theorem-PerfectCSIPrefilteredSampler}.

\subsubsection{Discrete Approximation}

The channel response and the filter response are both assumed to be
continuous, which motivates us to use a discrete-time approximation
in order to transform the continuous-time operator into its discrete
counterpart. We discretize a time domain process by point-wise sampling
with period $\Delta$, e.g. $h(t)$ is transformed into $\left\{ h[n]\right\} $
by setting $h[n]=h(n\Delta).$ For any given $T$, this allows us
to use a finite-dimensional matrix to approximate the continuous-time
block-Toeplitz operator. Then, due to the continuity assumption, an
exact capacity expression can be obtained by letting $\Delta$ go
to zero.

\subsubsection{Spectral properties of block-Toeplitz matrices}

After discretization, the input-output relation is similar to a MIMO
discrete-time system. Applying MIMO channel capacity results leads
to the capacity for a given $T$ and $\Delta$. The channel capacity
is then obtained by taking $T$ to infinity and $\Delta$ to zero,
which can be related to the channel matrix's spectrum using Toeplitz
theory. Since the filtered noise is non-white and correlated across
time, we need to whiten it first. This, however, destroys the Toeplitz
properties of the original system matrix. In order to apply established
results in Toeplitz theory, we make use of the concept of \emph{asymptotic
equivalence} \cite{Gray06} that builds connections between Toeplitz
matrices and non-Toeplitz matrices. This allows us to relate the capacity
limit with spectral properties of the channel and filter response.

\subsection{Optimal Prefilters\label{sub:Optimal-Prefilters}}

\subsubsection{Derivation of optimal prefilters}

Since different prefilters lead to different channel capacities, a
natural question is how to choose $S(f)$ to maximize capacity. The
optimizing prefilter is given in the following theorem.

\begin{theorem}\label{Cor-OptimalPrefilter-GeneralUniformSampling}Consider
the system shown in Fig. \ref{fig:SingleFilter}, and define
\[
\gamma_{l}(f):=\frac{\left|H\left(f-lf_{s}\right)\right|^{2}}{\mathcal{S}_{\eta}\left(f-lf_{s}\right)}
\]
 for any integer $l$. Suppose that in each aliased set $\left\{ f-lf_{s}\mid l\in\mathbb{Z}\right\} $,
there exists $k$ such that 
\[
\gamma_{k}(f)=\sup_{l\in\mathbb{Z}}\gamma_{l}(f).
\]
Then the capacity in (\ref{eq:CapacityGeneralSamplingColorNoise})
is maximized by the filter with frequency response
\begin{equation}
S(f-kf_{s})=\begin{cases}
1, & \text{ }\mbox{if }\gamma_{k}(f)=\sup_{l\in\mathbb{Z}}\gamma_{l}(f),\\
0, & \text{ }\mbox{otherwise,}
\end{cases}\label{eq:OptimalPrefilterGeneralUniformSampling}
\end{equation}
for any $f\in\left[-f_{s}/2,f_{s}/2\right]$.

\end{theorem}

\begin{IEEEproof}It can be observed from (\ref{eq:CapacityGeneralSamplingColorNoise})
that the frequency response $S(f)$ at any $f$ can only affect the
SNR at $f\mbox{ mod }f_{s}$, indicating that we can optimize for
frequencies $f_{1}$ and $f_{2}$ $\left(f_{1}\neq f_{2};f_{1},f_{2}\in\left[-\frac{f_{s}}{2},\frac{f_{s}}{2}\right]\right)$
separately. Specifically, the SNR at each $f$ in the aliased channel
is given by
\begin{align*}
\gamma^{\text{s}}(f) & =\sum_{l\in\mathbb{Z}}\gamma_{l}(f)\lambda_{l}(f),
\end{align*}
 where
\[
\lambda_{l}(f)=\frac{\mathcal{S}_{\eta}(f-lf_{s})\left|S(f-lf_{s})\right|^{2}}{\sum_{l\in\mathbb{Z}}\left|S(f-lf_{s})\right|^{2}\mathcal{S}_{\eta}(f-lf_{s})},
\]
 and $\sum_{l}\lambda_{l}(f)=1$. That said, $\gamma^{\text{s}}(f)$
is a convex combination of $\left\{ \gamma_{l},l\in\mathbb{Z}\right\} $,
and is thus upper bounded by $\sup_{l\in\mathbb{Z}}\gamma_{l}$. This
bound can be attained by the filter given in (\ref{eq:OptimalPrefilterGeneralUniformSampling}).
\end{IEEEproof} 

The optimal prefilter puts all its mass in those frequencies with
the highest SNR within each aliased set $\left\{ f-lf_{s}\mid l\in\mathbb{Z}\right\} $.
Even if the optimal prefilter does not exist, we can find a prefilter
that achieves an information rate arbitrarily close to the maximum
capacity once $\sup_{l\in\mathbb{Z}}\gamma_{l}(f)$ exists. The existence
of the supremum is guaranteed under mild conditions, e.g. when $\gamma_{l}(f)$
is bounded.

\subsubsection{Interpretations}

Recall that $S(f)$ is applied after the noise is added. One distinguishing
feature in the subsampled channel is the non-invertibility of the
prefiltering operation, i.e. we cannot recover the analog channel
output from sub-Nyquist samples. As shown above, the aliased SNR is
a convex combination of SNRs at all aliased branches, indicating that
$S(f)$ plays the role of \textit{``weighting''} different branches.
As in maximum ratio combining (MRC), those frequencies with larger
SNRs should be given larger weight, while those that suffer from poor
channel gains should be suppressed.

The problem of finding optimal prefilters corresponds to \textit{joint}
optimization over all input and filter responses. Looking at the equivalent
aliased channel for a given frequency $f\in[-f_{s}/2,f_{s}/2]$ as
illustrated in Fig. \ref{fig:GeneralUniformSamplerEquivalent}, we
have full control over both $X(f)$ and $S(f)$. Although MRC at the
transmitter side maximizes the combiner SNR for a MISO channel \cite{Gold2005},
it turns out to be suboptimal for our joint optimization problem.
Rather, the optimal solution is to perform selection combining \cite{Gold2005}
by setting $S(f-lf_{s})$ to one for some $l=l_{0}$, as well as noise
suppression by setting $S(f-lf_{s})$ to zero for all other $l$s.
In fact, setting $S(f)$ to zero precludes the undesired effects of
noise from low SNR frequencies, which is crucial in maximizing data
rate.

Another interesting observation is that optimal prefiltering equivalently
generates an \textit{alias-free} channel. After passing through an
optimal prefilter, all frequencies modulo $f_{s}$ except the one
with the highest SNR are removed, and hence the optimal prefilter
suppresses aliasing and out-of-band noise. This alias-suppressing
phenomena, while different from many sub-Nyquist works that advocate
mixing instead of alias suppressing \cite{MisEld2010Theory2Practice},
arises from the fact that we have control over the input shape.

\subsection{Numerical examples}

\subsubsection{Additive Gaussian Noise Channel without Prefiltering}

\begin{figure}
\begin{centering}
\includegraphics[scale=0.9]{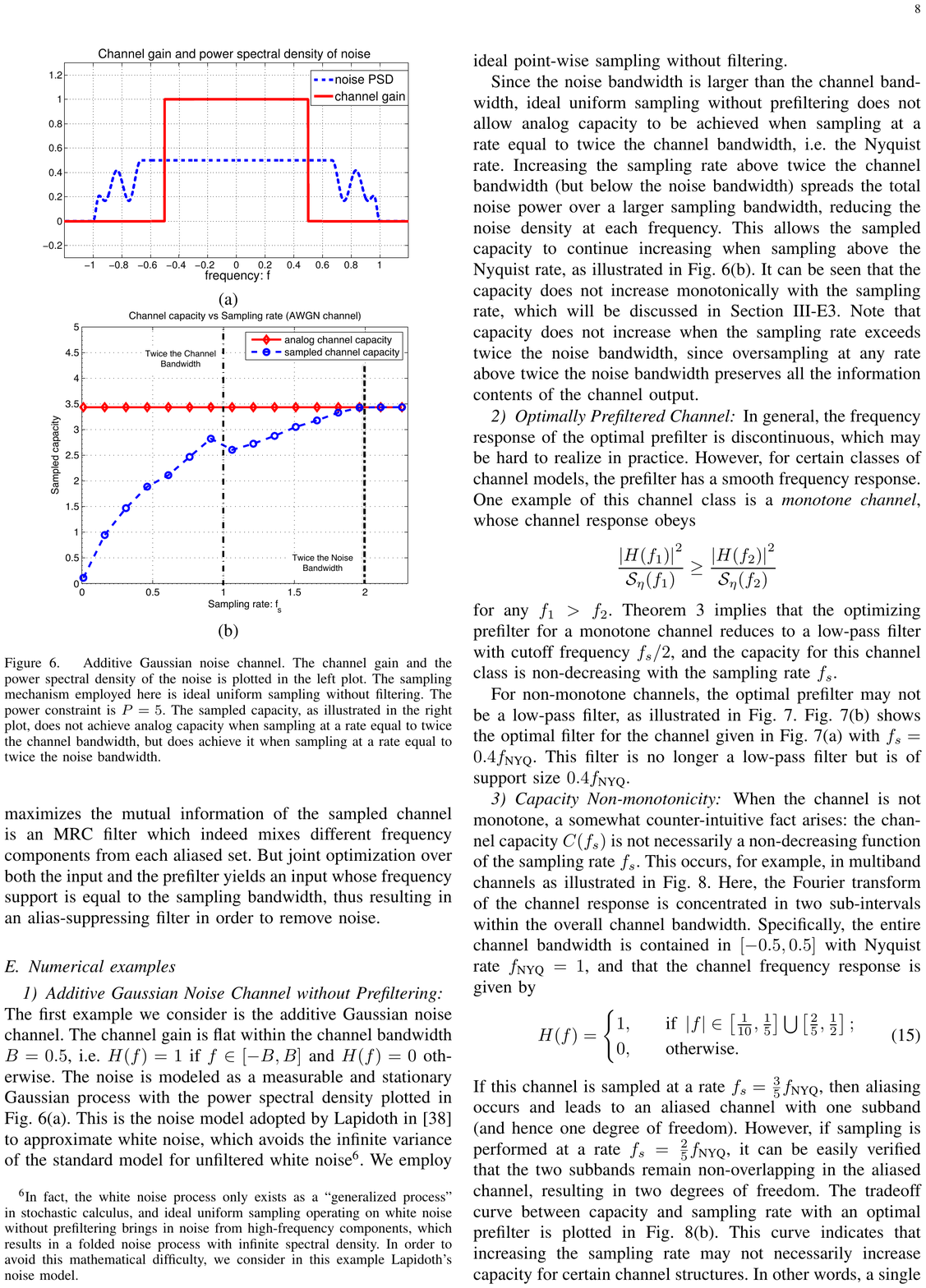}
\par\end{centering}

\begin{centering}
(a)
\par\end{centering}

\begin{centering}
\includegraphics[scale=0.9]{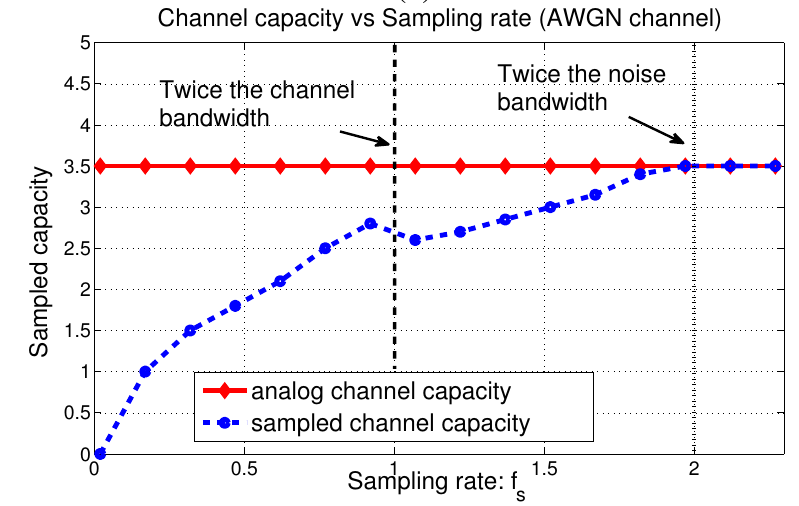}
\par\end{centering}

\centering{}(b)\caption{\label{fig:LapidothAWGN}Capacity of sampled additive Gaussian noise
channel under ideal uniform sampling without filtering. (a) The channel
gain and the PSD of the noise. (b) Sampled channel capacity v.s. analog
channel capacity under a power constraint $P=5$. }
\end{figure}

The first example we consider is the additive Gaussian noise channel.
The channel gain is flat within the channel bandwidth $B=0.5$, i.e.
$H(f)=1$ if $f\in\left[-B,B\right]$ and $H(f)=0$ otherwise. The
noise is modeled as a measurable and stationary Gaussian process with
the power spectral density plotted in Fig. \ref{fig:LapidothAWGN}(a).
This is the noise model adopted by Lapidoth in\cite{Lapidoth2009}
to approximate white noise, which avoids the infinite variance of
the standard model for unfiltered white noise. We employ ideal point-wise
sampling without filtering. 

Since the noise bandwidth is larger than the channel bandwidth, ideal
uniform sampling without prefiltering does not allow analog capacity
to be achieved when sampling at a rate equal to twice the channel
bandwidth, i.e. the Nyquist rate. Increasing the sampling rate above
twice the channel bandwidth (but below the noise bandwidth) spreads
the total noise power over a larger sampling bandwidth, reducing the
noise density at each frequency. This allows the sampled capacity
to continue increasing when sampling above the Nyquist rate, as illustrated
in Fig. \ref{fig:LapidothAWGN}(b). It can be seen that the capacity
does not increase monotonically with the sampling rate. We will discuss
this phenomena in more detail in Section \ref{sub:Capacity-Non-monotonicity-Single-Filter}.

\subsubsection{Optimally Filtered Channel}

In general, the frequency response of the optimal prefilter is discontinuous,
which may be hard to realize in practice. However, for certain classes
of channel models, the prefilter has a smooth frequency response.
One example of this channel class is a \textit{monotone channel},
whose channel response obeys $\left|H(f_{1})\right|^{2}/\mathcal{S}_{\eta}(f_{1})\geq\left|H(f_{2})\right|^{2}/\mathcal{S}_{\eta}(f_{2})$
for any $f_{1}>f_{2}$. Theorem \ref{Cor-OptimalPrefilter-GeneralUniformSampling}
implies that the optimizing prefilter for a monotone channel reduces
to a low-pass filter with cutoff frequency $f_{s}/2$.

\begin{figure}[htbp]
\begin{centering}
\textsf{\includegraphics{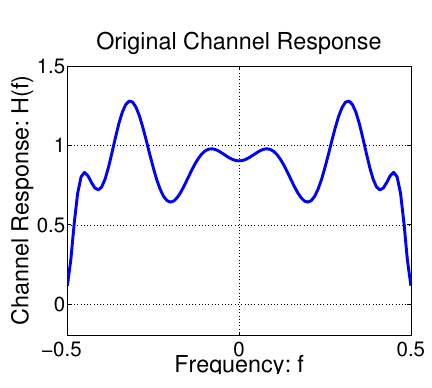}\includegraphics{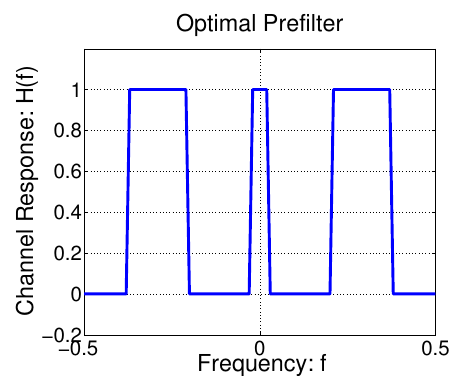}} 
\par\end{centering}

\centerline{$\quad$(a)$\hspace{12em}$ (b)}

\begin{centering}
\textsf{\includegraphics{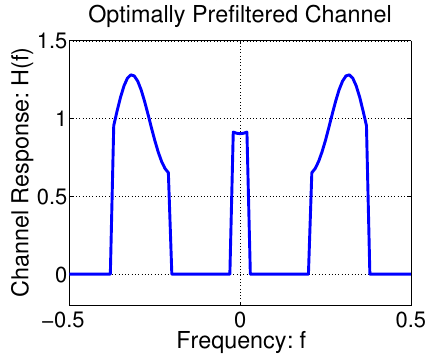}\includegraphics{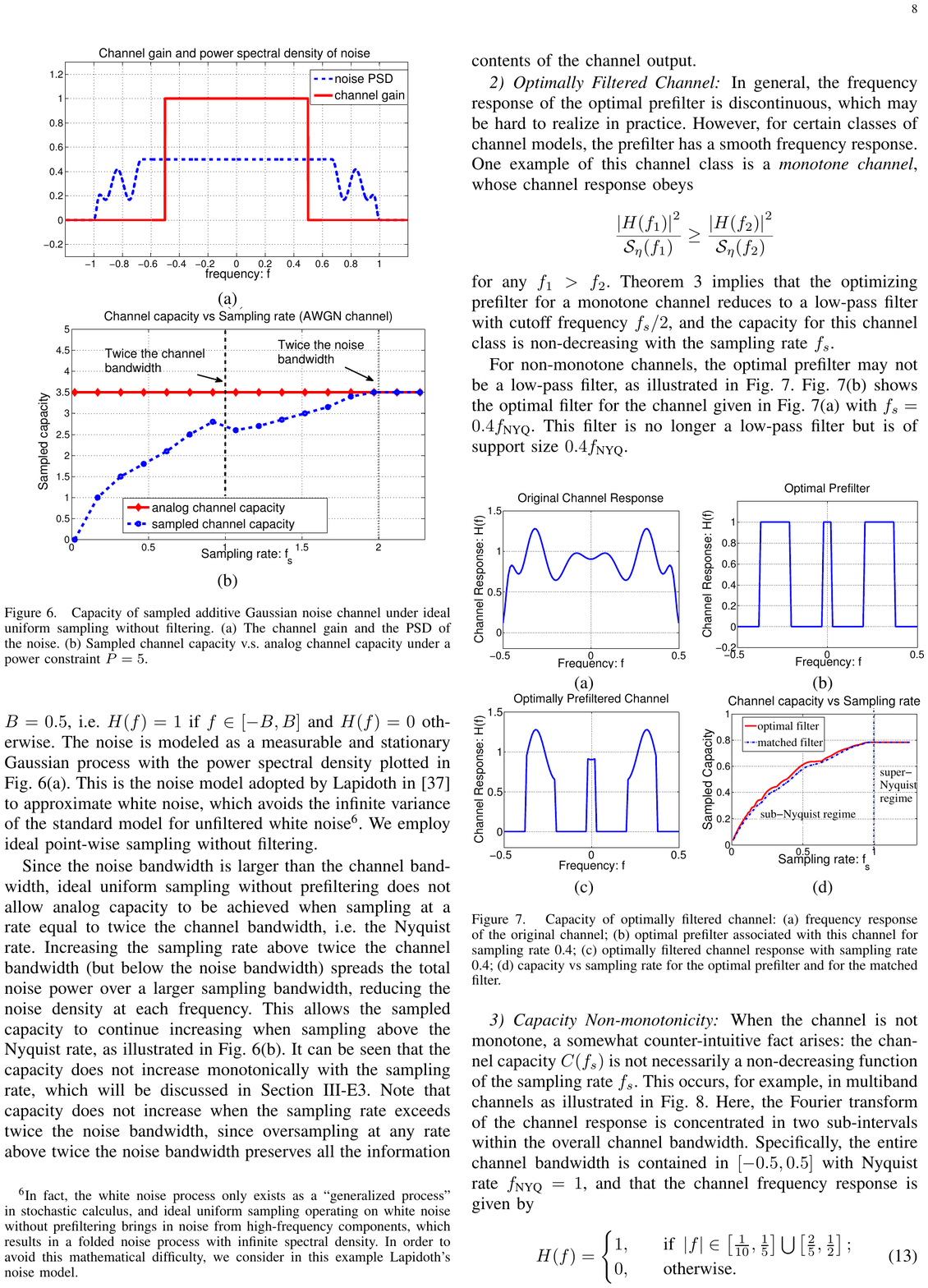}} 
\par\end{centering}

\centerline{$\quad$(c)$\hspace{12em}$ (d)}

\caption{\label{fig:PolynomialChannelSingleFilter}Capacity of optimally filtered
channel: (a) frequency response of the original channel; (b) optimal
prefilter associated with this channel for sampling rate 0.4; (c)
optimally filtered channel response with sampling rate 0.4; (d) capacity
vs sampling rate for the optimal prefilter and for the matched filter. }
\end{figure}

For non-monotone channels, the optimal prefilter may not be a low-pass
filter, as illustrated in Fig. \ref{fig:PolynomialChannelSingleFilter}.
Fig. \ref{fig:PolynomialChannelSingleFilter}(b) shows the optimal
filter for the channel given in Fig. \ref{fig:PolynomialChannelSingleFilter}(a)
with $f_{s}=0.4f_{\text{NYQ}}$, which is no longer a low-pass filter.

\begin{figure}[htbp]
\begin{centering}
\textsf{\includegraphics[scale=0.7]{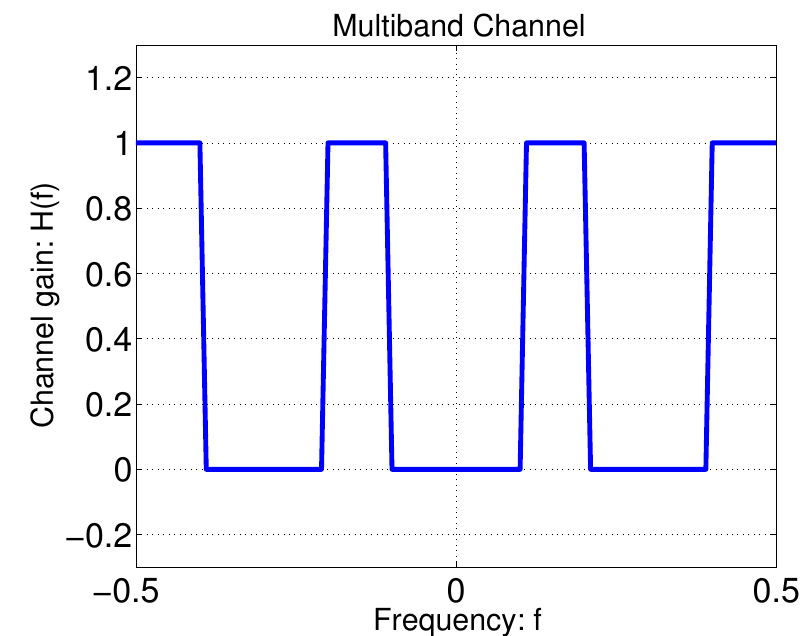}}
\par\end{centering}

\centerline{$\quad$(a)}

\begin{centering}
\textsf{\includegraphics[scale=0.9]{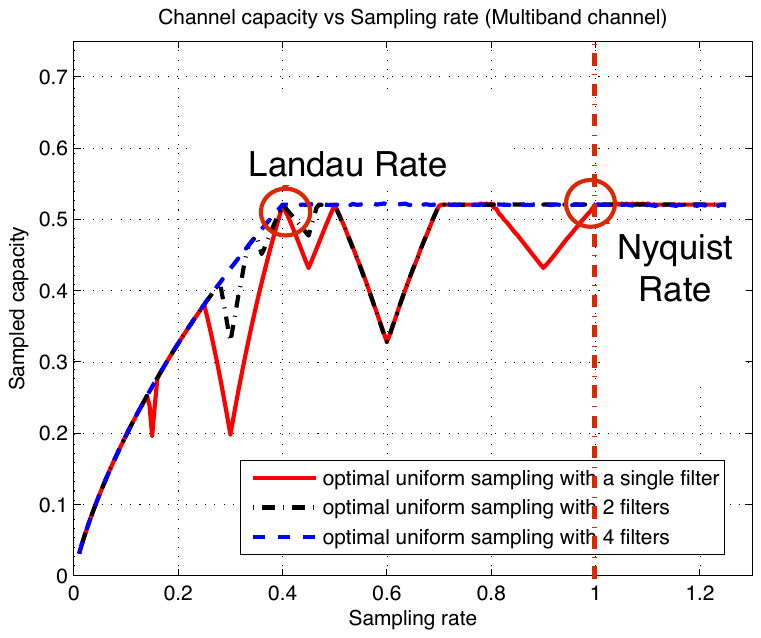}} 
\par\end{centering}

\centerline{$\quad$(b)}

\caption{\label{fig:UniformSamplerMultiband}Sampled channel capacity for a
multiband channel under filter-bank sampling. (a) Channel gain of
the multiband channel. The power constraint is $P=10$, and the noise
power is $\sigma_{\eta}^{2}=1$. (b) Sampled channel capacity for
a single filter followed by sampling and for a filter bank followed
by sampling for a bank of two filters and of four filters. }
\end{figure}

\subsubsection{Capacity Non-monotonicity\label{sub:Capacity-Non-monotonicity-Single-Filter}}

When the channel is not monotone, a somewhat counter-intuitive fact
arises: the channel capacity $C(f_{s})$ is not necessarily a non-decreasing
function of the sampling rate $f_{s}$. This occurs, for example,
in multiband channels as illustrated in Fig. \ref{fig:UniformSamplerMultiband}.
Here, the Fourier transform of the channel response is concentrated
in two sub-intervals within the overall channel bandwidth. Specifically,
the entire channel bandwidth is contained in $\left[-0.5,0.5\right]$
with Nyquist rate $f_{\text{NYQ}}=1$, and that the channel frequency
response is given by
\begin{equation}
H(f)=\begin{cases}
1, & \quad\mbox{if }\left|f\right|\in\left[\frac{1}{10},\frac{1}{5}\right]\bigcup\left[\frac{2}{5},\frac{1}{2}\right];\\
0, & \quad\mbox{otherwise.}
\end{cases}
\end{equation}
If this channel is sampled at a rate $f_{s}=\frac{3}{5}f_{\text{NYQ}}$,
then aliasing occurs and leads to an aliased channel with one subband
(and hence one degree of freedom). However, if sampling is performed
at a rate $f_{s}=\frac{2}{5}f_{\text{NYQ}}$. It can be easily verified
that the two subbands remain non-overlapping in the aliased channel,
resulting in two degrees of freedom. 

The tradeoff curve between capacity and sampling rate with an optimal
prefilter is plotted in Fig. \ref{fig:UniformSamplerMultiband}(b).
This curve indicates that increasing the sampling rate may not necessarily
increase capacity for certain channel structures. In other words,
a single filter followed by sampling largely constrains our ability
to exploit channel and signal structures. This is not the case for
more general sampling structures, as we show in the next section.

\section{A Bank of Filters Followed by Sampling\label{sec:Multi-channel-Prefiltered-Uniform}}

\subsection{Main Results}

We now treat filter-bank sampling, in which the channel output is
filtered and sampled through $M$ multiple branches as illustrated
in Fig. \ref{fig:SingleAntenna}. 

In order to state our capacity results, we introduce two matrices
${\bf F}_{s}$ and ${\bf F}_{h}$ defined in the Fourier domain. Here,
${\bf F}_{s}$ is an infinite matrix of $m$ rows and infinitely many
columns and ${\bf F}_{h}$ is a diagonal infinite matrix such that
for every $i$ ($1\leq i\leq k$) and every integer $l$: 
\begin{align*}
\left({\bf F}_{s}(f)\right)_{i,l} & =S_{i}\left(f-\frac{lf_{s}}{M}\right)\sqrt{\mathcal{S}_{\eta}\left(f-\frac{lf_{s}}{M}\right)},\\
\left({\bf F}_{h}(f)\right)_{l,l} & =H\left(f-\frac{lf_{s}}{M}\right)/\sqrt{\mathcal{S}_{\eta}\left(f-\frac{lf_{s}}{M}\right)}.
\end{align*}

\begin{theorem}\label{thmPerfectCSIFilterBankSingleAntenna}Consider
the system shown in Fig. \ref{fig:SingleAntenna}. Assume that $h(t)$
and $s_{i}(t)$ $(1\leq i\leq M)$ are all continuous, bounded and
absolutely Riemann integrable. Additionally, assume that $h_{\eta}(t):=\mathcal{F}^{-1}\left(\frac{H\left(f\right)}{\sqrt{\mathcal{S}_{\eta}\left(f\right)}}\right)$
satisfies $h_{\eta}(t)=o\left(t^{-\epsilon}\right)$ for some constant
$\epsilon>1$, and that ${\bf F}_{s}$ is right-invertible for every
$f$. Define $\tilde{{\bf F}}_{s}\overset{\Delta}{=}\left({\bf F}_{s}{\bf F}_{s}^{*}\right)^{-\frac{1}{2}}{\bf F}_{s}$.
The capacity $C(f_{s})$ of the sampled channel with a power constraint
$P$ is given as
\[
C(f_{s})={\displaystyle \int}_{-\frac{f_{s}}{2M}}^{\frac{f_{s}}{2M}}\frac{1}{2}\sum_{i=1}^{M}\log^{+}\left(\nu\lambda_{i}\left(\tilde{{\bf F}}_{s}{\bf F}_{h}{\bf F}_{h}^{*}\tilde{{\bf F}}_{s}^{*}\right)\right)\mathrm{d}f,
\]
where
\[
{\displaystyle \int}_{-\frac{f_{s}}{2M}}^{\frac{f_{s}}{2M}}\sum_{i=1}^{M}\left[\nu-\frac{1}{\lambda_{i}\left(\tilde{{\bf F}}_{s}{\bf F}_{h}{\bf F}_{h}^{*}\tilde{{\bf F}}_{s}^{*}\right)}\right]^{+}\mathrm{d}f=P.
\]
Here, $\lambda_{i}\left(\tilde{{\bf F}}_{s}{\bf F}_{h}{\bf F}_{h}^{*}\tilde{{\bf F}}_{s}^{*}\right)$
denotes the $i$th largest eigenvalue of $\tilde{{\bf F}}_{s}{\bf F}_{h}{\bf F}_{h}^{*}\tilde{{\bf F}}_{s}^{*}$.
\end{theorem}$ $\begin{remark}We can express this capacity in a
more traditional MIMO capacity form as
\begin{align}
C(f_{s}) & =\max_{\left\{ {\bf Q}(f)\right\} \in\mathcal{Q}}{\displaystyle \int}_{-\frac{f_{s}}{2M}}^{\frac{f_{s}}{2M}}\frac{1}{2}\log\det\left({\bf I}_{M}+\tilde{{\bf F}}_{s}{\bf F}_{h}{\bf Q}{\bf F}_{h}^{*}\tilde{{\bf F}}_{s}^{*}\right)\mathrm{d}f,\label{eq:ChannelCapacitySingleAntenna}
\end{align}
 where $\tilde{{\bf F}}_{s}\overset{\Delta}{=}\left({\bf F}_{s}{\bf F}_{s}^{*}\right)^{-\frac{1}{2}}{\bf F}_{s}$
and
\begin{align*}
\mathcal{Q} & =\left\{ {\bf Q}(f):\left|f\right|\leq\frac{f_{s}}{2M},{\bf Q}(f)\in\mathbb{S}_{+};\right.\\
 & \quad\quad\quad\quad\left.\int_{-\frac{f_{s}}{2M}}^{\frac{f_{s}}{2M}}\mathrm{Tr}\left({\bf Q}(f)\right)\mathrm{d}f=P.\right\} 
\end{align*}

\end{remark}

The optimal $\left\{ {\bf Q}(f)\right\} $ corresponds to a water-filling
power allocation strategy based on the singular values of the equivalent
channel matrix $\tilde{{\bf F}}_{s}{\bf F}_{h}$, where ${\bf F}_{h}$
is associated with the original channel and $\tilde{{\bf F}}_{s}$
arises from prefiltering and noise whitening. For each $f\in[-f_{s}/2M,f_{s}/2M]$,
the integrand in (\ref{eq:ChannelCapacitySingleAntenna}) can be interpreted
as a MIMO capacity formula. We have $M$ receive branches, and can
still optimize the transmitted signals $\left\{ X\left(f-\frac{lf_{s}}{M}\right)\mid l\in\mathbb{Z}\right\} $
at a countable number of input branches, but this time we have $M$
receive branches. The channel capacity is achieved when the transmit
signals are designed to decouple this MIMO channel into $M$ parallel
channels (and hence $M$ degrees of freedom), each associated with
one of its singular directions. 

\begin{figure}[htbp]
\begin{centering}
\textsf{\includegraphics[scale=0.33]{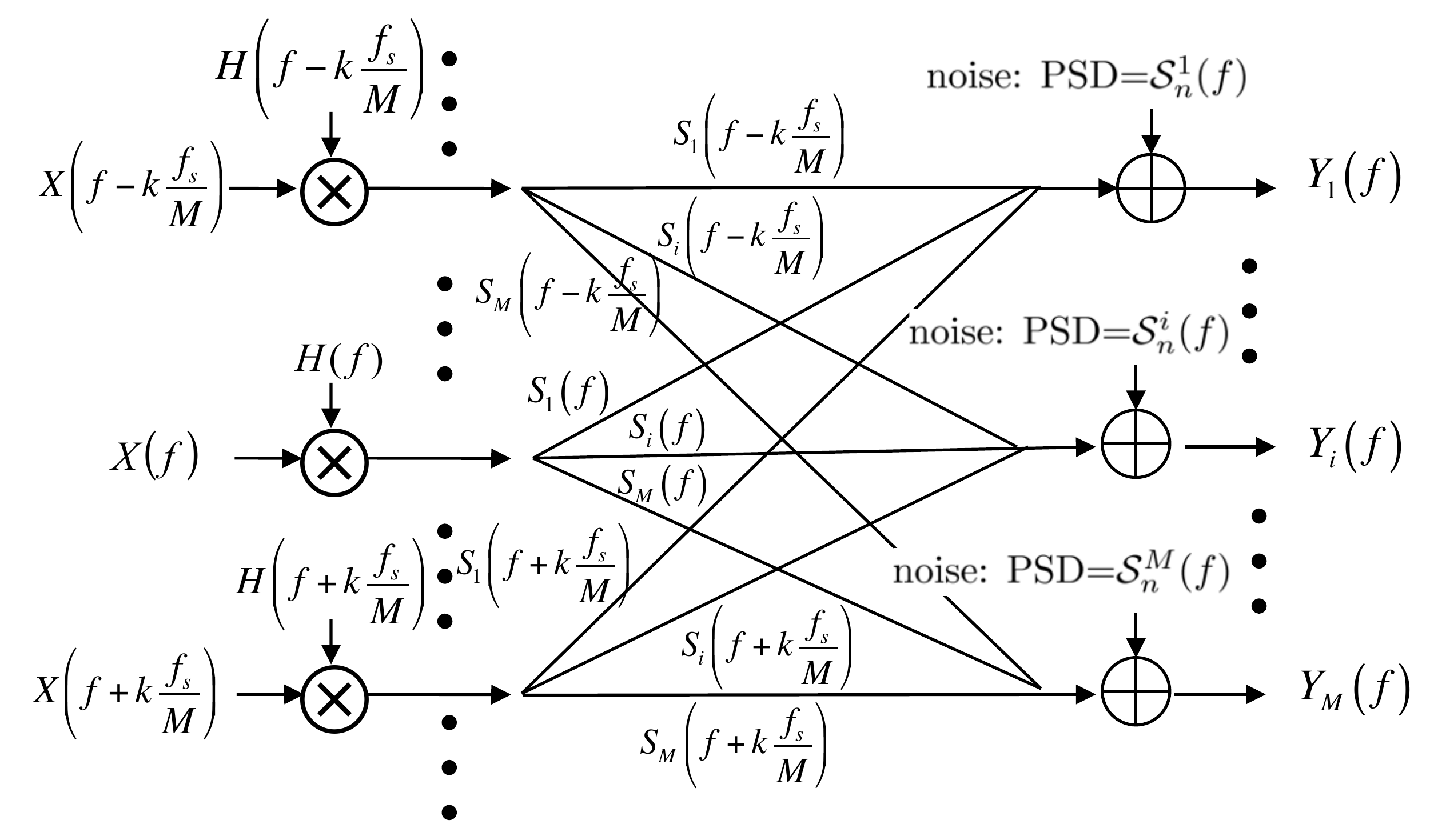}} 
\par\end{centering}

\caption{\label{fig:FilterBankSingleAntennaMIMO}Equivalent MIMO Gaussian channel
for a frequency $f\in[-f_{s}/2M,f_{s}/2M]$ under sampling with a
bank of $M$ filters. Here, $\mathcal{S}_{n}^{i}(f)=\sum_{l\in\mathbb{Z}}\mathcal{S}_{\eta}(f-lf_{s}/M)\left|S_{i}(f-lf_{s}/M)\right|^{2}$.}
\end{figure}

\subsection{Approximate Analysis}

The sampled analog channel under filter-bank sampling can be studied
through its connection with MIMO Gaussian channels (see Fig. \ref{fig:FilterBankSingleAntennaMIMO}).
Consider first a single frequency $f\in[-f_{s}/2M,f_{s}/2M]$. Since
we employ a bank of filters each followed by an ideal uniform sampler,
the equivalent channel has $M$ receive branches, each corresponding
to one branch of filtered sampling at rate $f_{s}/M$. The noise received
in the $i$th branch is zero-mean Gaussian with spectral density 
\begin{align*}
 & \sum_{l\in\mathbb{Z}}\left|S_{i}\left(f-\frac{lf_{s}}{M}\right)\right|^{2}\mathcal{S}_{\eta}\left(f-\frac{lf_{s}}{M}\right),\quad f\in\left[-\frac{f_{s}}{2M},\frac{f_{s}}{2M}\right],
\end{align*}
 indicating the mutual correlation of noise at different branches.
The received noise vector can be whitened by multiplying ${\bf Y}(f)=[\cdots,Y(f),Y(f-f_{s}),\cdots]^{T}$
by an $M\times M$ whitening matrix $\left({\bf F}_{s}(f){\bf F}_{s}^{*}(f)\right)^{-\frac{1}{2}}$.
Since the whitening operation is invertible, it preserves capacity.
After whitening, the channel of Fig. \ref{fig:FilterBankSingleAntennaMIMO}
at frequency $f$ has the following channel matrix
\begin{equation}
\left({\bf F}_{s}(f){\bf F}_{s}^{*}(f)\right)^{-\frac{1}{2}}{\bf F}_{s}(f){\bf F}_{h}(f)=\tilde{{\bf F}}_{s}(f){\bf F}_{h}(f).
\end{equation}

MIMO Gaussian channel capacity results \cite{Tel1999} immediately
imply that the capacity of the channel in Fig. \ref{fig:FilterBankSingleAntennaMIMO}
at any $f\in[-f_{s}/2M,f_{s}/2M]$ can be expressed as 
\begin{equation}
\max_{{\bf Q}}\frac{1}{2}\log\det\left[{\bf I}+\tilde{{\bf F}}_{s}(f){\bf F}_{h}(f){\bf Q}(f){\bf F}_{h}^{*}(f)\tilde{{\bf F}}_{s}^{*}(f)\right]
\end{equation}
 subject to the constraints that $\text{trace}\left({\bf Q}(f)\right)\leq P(f)$
and ${\bf Q}(f)\in\mathbb{S}_{+}$, where ${\bf Q}(f)$ denotes the
power allocation matrix. Performing water-filling power allocation
across all parallel channels leads to our capacity expression.

\subsection{Optimal Filter Bank\label{sub:Optimal-Filter-Banks}}

\subsubsection{Derivation of optimal filter banks}

In general, $\log\det[{\bf I}_{M}+\tilde{{\bf F}}_{s}{\bf F}_{h}{\bf Q}{\bf F}_{h}^{*}\tilde{{\bf F}}_{s}^{*}]$
is not perfectly determined by $\tilde{{\bf F}}_{s}(f)$ and ${\bf F}_{h}(f)$
at a single frequency $f$, but also depends on the water-level, since
the optimal power allocation strategy relies on the power constraint
$P/\sigma_{\eta}^{2}$ as well as ${\bf F}_{s}$ and ${\bf F}_{h}$
across all $f$. In other words, $\log\det[{\bf I}_{M}+\tilde{{\bf F}}_{s}{\bf F}_{h}{\bf Q}{\bf F}_{h}^{*}\tilde{{\bf F}}_{s}^{*}]$
is a function of all singular values of $\tilde{{\bf F}}_{s}{\bf F}_{h}$
and the universal water-level associated with optimal power allocation.
Given two sets of singular values, we cannot determine which set is
preferable without accounting for the water-level, unless one set
is element-wise larger than the other. That said, if there exists
a prefilter that maximizes all singular values simultaneously, then
this prefilter will be universally optimal regardless of the water-level.
Fortunately, such optimal schemes exist, as we characterize in Theorem
\ref{Cor:OptimalFilterBank}. 

Since ${\bf F}_{h}(f)$ is a diagonal matrix, $\lambda_{k}\left({\bf F}_{h}{\bf F}_{h}^{*}\right)$
denotes the $k$th largest entry of ${\bf F}_{h}{\bf F}_{h}^{*}$
. The optimal filter bank can then be given as follows.

\begin{theorem}\label{Cor:OptimalFilterBank}Consider the system
shown in Fig. \ref{fig:SingleAntenna}. Suppose that for each aliased
set $\left\{ f-\frac{if_{s}}{M}\mid i\in\mathbb{Z}\right\} $ and
each $k$ $(1\leq k\leq M)$, there exists an integer $l$ such that
$\frac{\left|H\left(f-\frac{lf_{s}}{M}\right)\right|^{2}}{\mathcal{S}_{\eta}\left(f-\frac{lf_{s}}{M}\right)}$
is equal to the $k^{\text{th}}$ largest element in $\left\{ \frac{\left|H\left(f-\frac{if_{s}}{M}\right)\right|^{2}}{\mathcal{S}_{\eta}\left(f-\frac{if_{s}}{M}\right)}\mid i\in\mathbb{Z}\right\} $.
The capacity (\ref{eq:ChannelCapacitySingleAntenna}) under filter-bank
sampling is then maximized by a bank of filters for which the frequency
response of the $k^{\text{th}}$ filter is given by 
\begin{equation}
S_{k}\left(f-\frac{lf_{s}}{M}\right)=\begin{cases}
1, & \quad\mbox{if }\frac{\left|H\left(f-\frac{lf_{s}}{M}\right)\right|^{2}}{\mathcal{S}_{\eta}\left(f-\frac{lf_{s}}{M}\right)}=\lambda_{k}\left({\bf F}_{h}(f){\bf F}_{h}^{*}(f)\right);\\
0, & \quad\mbox{otherwise},
\end{cases}\label{eq:optimalFilterBank}
\end{equation}
for all $l\in\mathbb{Z}$, $1\leq k\leq M$ and $f\in\left[-\frac{f_{s}}{2M},\frac{f_{s}}{2M}\right]$.
The resulting maximum channel capacity is given by
\begin{align}
C(f_{s}) & =\frac{1}{2}{\displaystyle \int}_{-f_{s}/2M}^{f_{s}/2M}\sum_{k=1}^{M}\log^{+}\left(\nu\cdot\lambda_{k}\left({\bf F}_{h}{\bf F}_{h}^{*}\right)\right)\mathrm{d}f,
\end{align}
 where $\nu$ is chosen such that 
\begin{equation}
{\displaystyle \int}_{-f_{s}/2M}^{f_{s}/2M}\sum_{k=1}^{M}\left[\nu-\frac{1}{\lambda_{k}\left({\bf F}_{h}{\bf F}_{h}^{*}\right)}\right]_{+}\mathrm{d}f=P.
\end{equation}

\end{theorem}

\begin{IEEEproof}See Appendix \ref{sec:Proof-of-Corollary-Optimal-Filter-Bank}.
\end{IEEEproof}

The choice of prefilters in (\ref{eq:optimalFilterBank}) achieves
the upper bounds on all singular values, and is hence universally
optimal regardless of the water level. Since $\tilde{{\bf F}}_{s}$
has orthonormal rows, it acts as an orthogonal projection and outputs
an $M$-dimensional subspace. The rows of the diagonal matrix ${\bf F}_{h}$
are orthogonal to each other. Therefore, the subspace closest to the
channel space spanned by ${\bf F}_{h}$ corresponds to the $M$ rows
of ${\bf F}_{h}$ containing the highest channel gains out of the
entire aliased frequency set $\left\{ f-\frac{lf_{s}}{M}\mid l\in\mathbb{Z}\right\} $.
The maximum data rate is then achieved when the filter bank outputs
$M$ frequencies with the highest SNR among the set of frequencies
equivalent modulo $\frac{f_{s}}{M}$ and suppresses noise from all
other branches. 

We note that if we consider the enlarged aliased set $\left\{ f-lf_{s}/M\mid l\in\mathbb{Z}\right\} $,
then the optimal filter bank is equivalent to generating an alias-free
channel over the frequency interval $\left[-f_{s}/2M,f_{s}/2M\right]$.
This again arises from the nature of the joint-optimization problem:
since we are allowed to control the input shape and sampling jointly,
we can adjust the input shape based on the channel structure in each
branch, which turn out to be alias-suppressing.

\subsection{Discussion and Numerical Examples }

In a \textit{monotone} channel, the optimal filter bank will sequentially
crop out the $M$ best frequency bands, each of bandwidth $f_{s}/M$.
Concatenating all of these frequency bands results in a low-pass filter
with cut-off frequency $f_{s}/2$, which is equivalent to single-branch
sampling with an optimal filter. In other words, for monotone channels,
using filter banks harvests no gain in capacity compared to a single
branch with a filter followed by sampling. 

For more general channels, the capacity is not necessarily a monotone
function of $f_{s}$. Consider again the multiband channel where the
channel response is concentrated in two sub-intervals, as illustrated
in Fig. \ref{fig:UniformSamplerMultiband}(a). As discussed above,
sampling following a single filter only allows us to select the best
single frequency with the highest SNR out of the set $\left\{ f-lf_{s}\mid l\in\mathbb{Z}\right\} $,
while sampling following filter banks allows us to select the best
$f$ out of the set $\left\{ f-l\frac{f_{s}}{M}\mid l\in\mathbb{Z}\right\} $.
Consequently, the channel capacity with filter-bank sampling exceeds
that of sampling with a single filter, but neither capacity is monotonically
increasing in $f_{s}$. This is shown in Fig. \ref{fig:UniformSamplerMultiband}(b).
Specifically, we see in this figure that when we apply a bank of two
filters prior to sampling, the capacity curve is still non-monotonic
but outperforms a single filter followed by sampling.

Another consequence of our results is that when the number of branches
is optimally chosen, the Nyquist-rate channel capacity can be achieved
by sampling at any rate above the Landau rate. In order to show this,
we introduce the following notion of a channel permutation. We call
$\tilde{H}(f)$ a \textit{permutation} of a channel response $H(f)$
at rate $f_{s}$ if, for any $f$, 
\[
\left\{ \frac{|\tilde{H}(f-lf_{s})|^{2}}{\mathcal{S}_{\eta}(f-lf_{s})}\mid l\in\mathbb{Z}\right\} =\left\{ \frac{\left|H(f-lf_{s})\right|^{2}}{\mathcal{S}_{\eta}(f-lf_{s})}\mid l\in\mathbb{Z}\right\} .
\]
The following proposition characterizes a sufficient condition that
allows the Nyquist-rate channel capacity to be achieved at any sampling
rate above the Landau rate. 

\begin{prop}\label{propLandauRateSampling} If there exists a permutation
$\tilde{H}(f)$ of $H(f)$ at rate $\frac{f_{s}}{M}$ such that the
support of $\tilde{H}(f)$ is $[-f_{L}/2,f_{L}/2]$, then optimal
sampling following a bank of $M$ filters achieves Nyquist-rate capacity
when $f_{s}\geq f_{L}$. \end{prop}

Examples of channels satisfying Proposition \ref{propLandauRateSampling}
include any multiband channel with $N$ subbands among which $K$
subbands have non-zero channel gain. For any $f_{s}\geq f_{L}=\frac{K}{N}f_{\text{NYQ}}$,
we are always able to permute the channel at rate $f_{s}/K$ to generate
a band-limited channel of spectral support size $f_{L}$. Hence, sampling
above the Landau rate following $K$ filters achieves the Nyquist-rate
channel capacity. This is illustrated in Fig. \ref{fig:UniformSamplerMultiband}(b)
where sampling with a four-branch filter bank has a higher capacity
than sampling with a single filter, and achieves the Nyquist-rate
capacity whenever $f_{s}\geq\frac{2}{5}f_{\text{NYQ}}$. The optimal
filter-bank sampling for most general channels is identified in \cite{ChenEldarGoldsmith2012},
where both the number of branches and per-branch sampling rate are
allowed to vary.

\section{Modulation and Filter Banks Followed by Sampling\label{sec:Multi-channel-Pre-modulated-Pre-filtered}}

\subsection{Main Results}

We now treat modulation and filter banks followed by sampling. Assume
that $\tilde{T}_{s}:=MT_{s}=\frac{b}{a}T_{q}$ where $a$ and $b$
are coprime integers, and that the Fourier transform of $q_{i}(t)$
is given as $\sum_{l}c_{i}^{l}\delta(f-lf_{q})$. Before stating our
theorem, we introduce the following two Fourier symbol matrices ${\bf F}^{\eta}$
and ${\bf F}^{h}$. The $aM\times\infty$-dimensional matrix ${\bf F}^{\eta}$
contains $M$ submatrices with the $\alpha$th submatrix given by
an $a\times\infty$-dimensional matrix ${\bf F}_{\alpha}^{\eta}{\bf F}_{\alpha}^{p}$.
Here, for any $v\in\mathbb{Z}$, $1\leq l\leq a$, and $1\leq\alpha\leq M$,
we have
\begin{align*}
\left({\bf F}_{\alpha}^{\eta}\right)_{l,v} & =\left({\bf F}_{\alpha}^{p}\right)_{v,v}\left[\sum_{u}c_{\alpha}^{u}S_{\alpha}\left(-f+uf_{q}+v\frac{f_{q}}{b}\right)\right.\\
 & \quad\quad\quad\quad\left.\cdot\exp\left(-j2\pi lMT_{s}\left(f-uf_{q}-v\frac{f_{q}}{b}\right)\right)\right].
\end{align*}
The matrices ${\bf F}_{\alpha}^{p}$ and ${\bf F}^{h}$ are infinite
diagonal matrices such that for every integer $l$:
\begin{align*}
\left({\bf F}_{\alpha}^{p}\right)_{l,l} & =P_{\alpha}\left(-f+l\frac{f_{q}}{b}\right)\sqrt{\mathcal{S}_{\eta}\left(-f+l\frac{f_{q}}{b}\right)},\\
\left({\bf F}^{h}\right)_{l,l} & =\frac{H\left(-f+l\frac{f_{q}}{b}\right)}{\sqrt{\mathcal{S}_{\eta}\left(-f+l\frac{f_{q}}{b}\right)}}.
\end{align*}

\begin{theorem}\label{thmPremodulatedFilterBank}Consider the system
shown in Fig. \ref{fig:PremodulatedPrefilteredSampler}. Assume that
$h(t)$, $p_{i}(t)$ and $s_{i}(t)$ $(1\leq i\leq M)$ are all continuous,
bounded and absolutely Riemann integrable, ${\bf F}^{\eta}$ is right
invertible, and that the Fourier transform of $q_{i}(t)$ is given
as $\sum_{l}c_{i}^{l}\delta(f-lf_{q})$. Additionally, suppose that
$h_{\eta}(t):=\mathcal{F}^{-1}\left(\frac{H\left(f\right)}{\sqrt{\mathcal{S}_{\eta}(f)}}\right)$
satisfies $h_{\eta}(t)=o\left(t^{-\epsilon}\right)$ for some constant
$\epsilon>1$. We further assume that $aMT_{s}=bT_{q}$ where $a$
and $b$ are coprime integers. The capacity $C(f_{s})$ of the sampled
channel with a power constraint $P$ is given by
\begin{align}
C(f_{s}) & ={\displaystyle \int}_{-\frac{f_{s}}{2aM}}^{\frac{f_{s}}{2aM}}\frac{1}{2}\sum_{i=1}^{aM}\log^{+}\left(\nu\lambda_{i}\left(\left({\bf F}^{\eta}{\bf F}^{\eta*}\right)^{-\frac{1}{2}}{\bf F}^{\eta}{\bf F}^{h}\cdot\right.\right.\nonumber \\
 & \quad\quad\quad\quad\quad\quad\left.\left.{\bf F}^{h*}{\bf F}^{\eta*}\left({\bf F}^{\eta}{\bf F}^{\eta*}\right)^{-\frac{1}{2}}\right)\right)\mathrm{d}f,\label{eq:CapacityModulationBank}
\end{align}
 where $\nu$ is chosen such that
\begin{align*}
P & ={\displaystyle \int}_{-\frac{f_{s}}{2aM}}^{\frac{f_{s}}{2aM}}\sum_{i=1}^{aM}\left[\nu-\lambda_{i}^{-1}\left(\left({\bf F}^{\eta}{\bf F}^{\eta*}\right)^{-\frac{1}{2}}{\bf F}^{\eta}{\bf F}^{h}\cdot\right.\right.\\
 & \quad\quad\quad\quad\quad\quad\left.\left.{\bf F}^{h*}{\bf F}^{\eta*}\left({\bf F}^{\eta}{\bf F}^{\eta*}\right)^{-\frac{1}{2}}\right)\right]^{+}\mathrm{d}f.
\end{align*}
\end{theorem}

\begin{remark}The right invertibility of ${\bf F}^{\eta}$ ensures
that the sampling method is non-degenerate, e.g. the modulation sequence
cannot be zero. \end{remark}

The optimal $\nu$ corresponds to a water-filling power allocation
strategy based on the singular values of the equivalent channel matrix
$\left({\bf F}^{\eta}{\bf F}^{\eta*}\right)^{-\frac{1}{2}}{\bf F}^{\eta}{\bf F}^{h}$,
where $\left({\bf F}^{\eta}{\bf F}^{\eta*}\right)^{-\frac{1}{2}}$
is due to noise prewhitening and ${\bf F}^{\eta}{\bf F}^{h}$ is the
equivalent channel matrix after modulation and filtering. This result
can again be interpreted by viewing (\ref{eq:CapacityModulationBank})
as the MIMO Gaussian channel capacity of the equivalent channel. We
note that a closed-form capacity expression may be hard to obtain
for general modulating sequences $q_{i}(t)$. This is because the
multiplication operation corresponds to convolution in the frequency
domain which does not preserve Toeplitz properties of the original
operator associated with the channel filter. When $q_{i}(t)$ is periodic,
however, it can be mapped to a spike train in the frequency domain,
which preserves block Toeplitz properties, as described in more detail
in Appendix \ref{sec:Proof-of-Theorem-Premodulated-Filter-Bank}.

\subsection{Approximate Analysis}

The Fourier transform of the signal prior to modulation in the $i$th
branch at a given frequency $f$ can be expressed as $P_{i}(f)R(f)$,
where $R(f)=H(f)X(f)+N(f)$. Multiplication of this pre-modulation
signal with the modulation sequence $q_{i}(t)=\sum_{l}c_{i}^{l}\delta\left(f-lf_{q}\right)$
corresponds to convolution in the frequency domain. 

Recall that $bT_{q}=aMT_{s}$ with integers $a$ and $b$. We therefore
divide all samples $\left\{ y_{i}[k]\mid k\in\mathbb{Z}\right\} $
in the $i$th branch into $a$ groups, where the $l$th ($0\leq l<a$)
group contains $\left\{ y_{i}[l+ka]\mid k\in\mathbb{Z}\right\} $.
Hence, each group is equivalent to the samples obtained by sampling
at rate $f_{s}/Ma=f_{q}/b$. The sampling system, when restricted
to the output on each group of the sampling set, can be treated as
LTI, thus justifying its equivalent representation in the spectral
domain. Specifically, for the $i$th branch, we denote by 
\[
g_{\eta}^{i}(t,\tau):=\int s_{i}(t-\tau_{1})q_{i}(\tau_{1})p(\tau_{1}-\tau)\mathrm{d}\tau_{1}
\]
the output response of the preprocessing system at time $t$ due to
an input impulse at time $\tau$. We then introduce a new LTI impulse
response $\tilde{g}_{l}^{i}(t)$ associated with the $l$th group
such that $\tilde{g}_{l}^{i}(t):=g_{\eta}^{i}(l\tilde{T}_{s},l\tilde{T}_{s}-t)$.
It can easily be shown that when the same sampling set $\left\{ \left(l+ka\right)\tilde{T}_{s}\mid k\in\mathbb{Z}\right\} $
is employed, the preprocessing system associated with $g_{\eta}^{i}(t,\tau)$
results in the same sampled output as the one associated with $\tilde{g}_{l}^{i}(t)$.
This allows us to treat the samples of each distinct group as the
ones obtained by an LTI preprocessing system followed by uniform sampling. 

\begin{figure}
\centering

\includegraphics[scale=0.35]{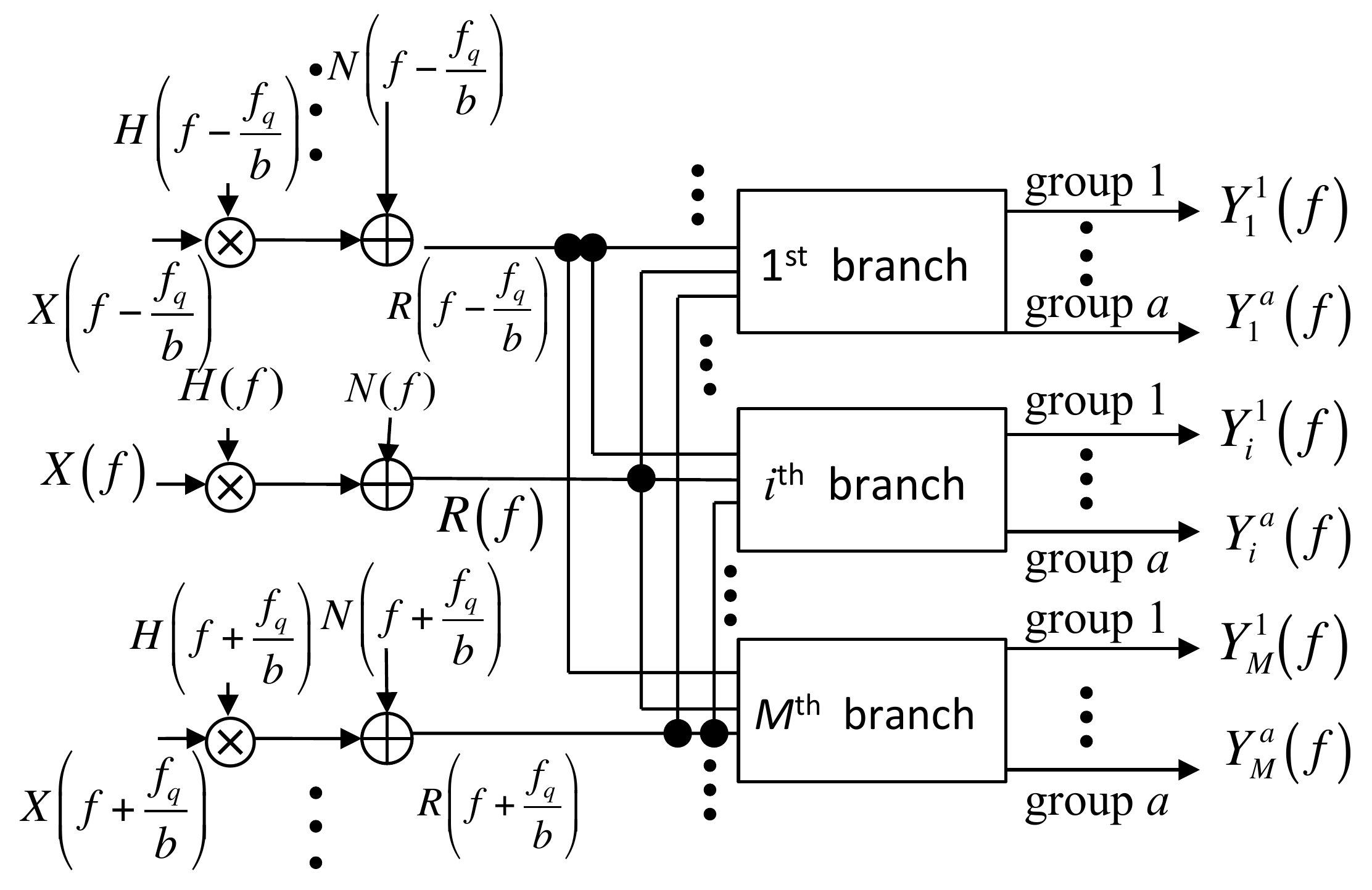}

\begin{centering}
(a)
\par\end{centering}

\includegraphics[scale=0.35]{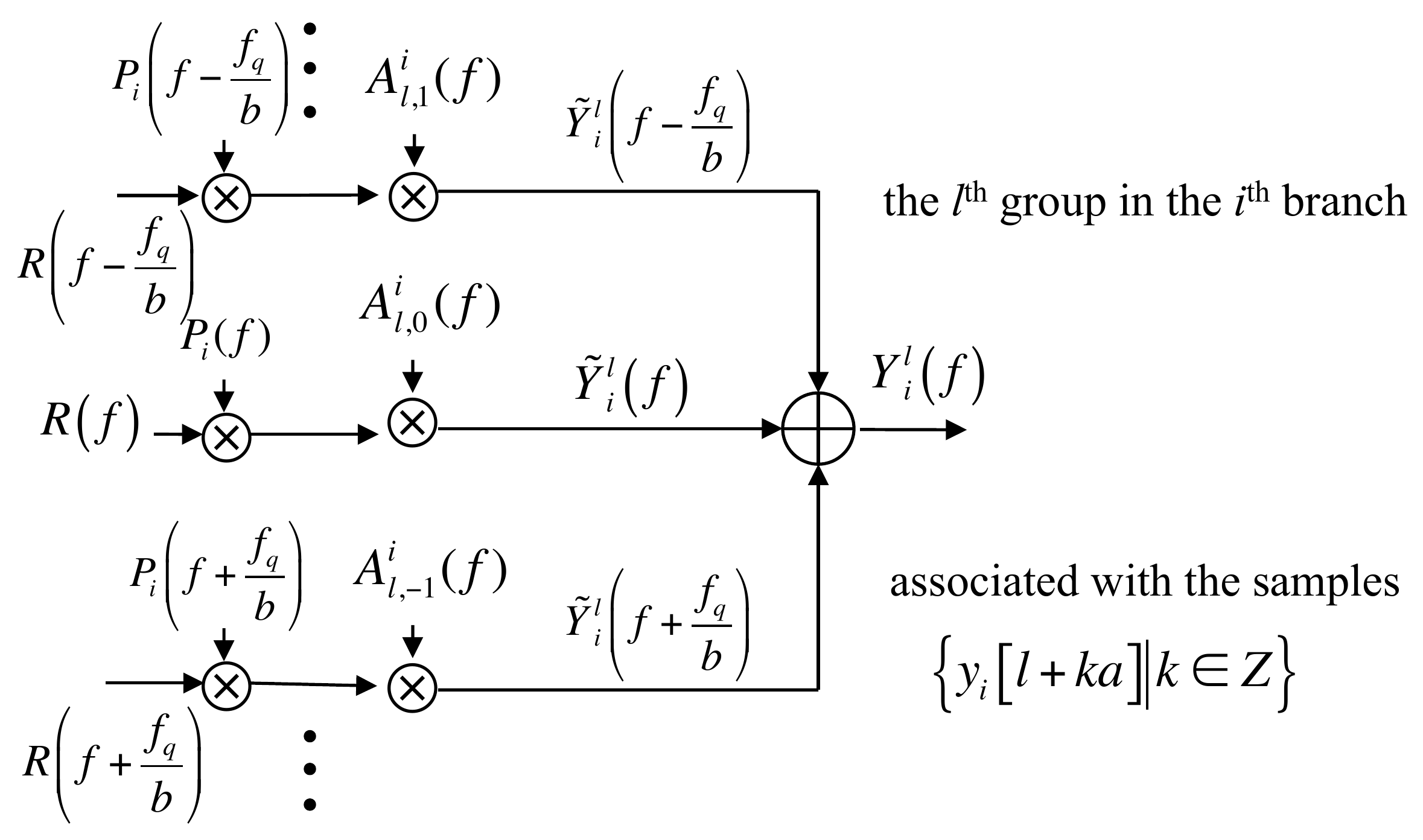}

\begin{centering}
(b)
\par\end{centering}

\caption{\label{fig:Equivalent-MIMO-Gaussian-Channel-Modulation}Equivalent
MIMO Gaussian channel for a given $f\in\left[0,f_{q}/b\right)$ under
sampling with modulation banks and filter banks. (a) The overall MIMO
representation, where each branch has $a$ output each corresponding
to a distinct group. (b) The MISO representation of the $l$th group
in the $i$th branch, where $A_{l,v}^{i}(f)$ is defined in (\ref{eq:AvlModulation}).
This is associated with the set of samples $\left\{ y_{i}[l+ka]\mid k\in\mathbb{Z}\right\} $.}
\end{figure}

Suppose the channel output $R(f)$ is passed through the LTI preprocessing
system associated with the $l$th group of the $i$th branch, i.e.
the one associated with $\tilde{g}_{l}^{i}(t)$. The Fourier transform
of the output of this LTI system prior to uniform sampling, as marked
in Fig \ref{fig:Equivalent-MIMO-Gaussian-Channel-Modulation}(b),
can be written as 
\begin{align*}
 & \tilde{Y}_{i}^{l}(f)\\
\overset{\Delta}{=} & P_{i}(f)R(f)\left(S_{i}(f)\exp\left(j2\pi fl\tilde{T}_{s}\right)*\sum_{u}c_{i}^{u}\delta\left(f-uf_{q}\right)\right)\\
= & P_{i}(f)R(f)\sum_{u}c_{i}^{u}S_{i}\left(f-uf_{q}\right)\exp\left(j2\pi l\tilde{T}_{s}\left(f-uf_{q}\right)\right).
\end{align*}
After uniform sampling at rate $f_{q}/b$, the Fourier transform of
the samples in the $l$th group can be expressed as 
\begin{align*}
 & Y_{i}^{l}(f)=\sum_{v}\tilde{Y}_{i}^{l}\left(f-\frac{vf_{q}}{b}\right)\\
= & \sum_{v}P_{i}\left(f-\frac{vf_{q}}{b}\right)R\left(f-\frac{vf_{q}}{b}\right)\sum_{u}c_{i}^{u}\cdot\\
 & S_{i}\left(f-uf_{q}-\frac{vf_{q}}{b}\right)\exp\left(j2\pi l\tilde{T}_{s}\left(f-uf_{q}-\frac{vf_{q}}{b}\right)\right)\\
= & \sum_{v}A_{l,v}^{i}(f)P_{i}\left(f-v\frac{f_{q}}{b}\right)R\left(f-v\frac{f_{q}}{b}\right),
\end{align*}
where 
\begin{align}
A_{l,v}^{i}(f) & \overset{\Delta}{=}\sum_{u}c_{i}^{u}S_{i}\left(f-uf_{q}-\frac{vf_{q}}{b}\right)\cdot\nonumber \\
 & \quad\quad\quad\quad\exp\left(j2\pi l\tilde{T}_{s}\left(f-uf_{q}-\frac{vf_{q}}{b}\right)\right).\label{eq:AvlModulation}
\end{align}

Since the sampled outputs of the original sampling system are equivalent
to the union of samples obtained by $Ma$ LTI systems each followed
by uniform sampling at rate $f_{q}/b$, we can transform the true
sampling system into a MIMO Gaussian channel with an infinite number
of input branches and finitely many output branches, as illustrated
in Fig. \ref{fig:Equivalent-MIMO-Gaussian-Channel-Modulation}. The
well-known formula for the capacity of a MIMO channel can now be used
to derive our capacity results.

We note that due to the convolution in the spectral domain, the frequency
response of the sampled output at frequency $f$ is a linear combination
of frequency components $\left\{ X(f)\right\} $ and $\left\{ N(f)\right\} $
from several different aliased frequency sets. We define the \emph{modulated
aliased frequency set} as a generalization of the aliased set. Specifically,
for each $f$, the modulated aliased set is given by%
\footnote{We note that although each modulated aliased set is countable, it
may be a dense set when $f_{q}/\tilde{f}_{s}$ is irrational. Under
the assumption in Theorem \ref{thmPremodulatedFilterBank}, however,
the elements in the set have a minimum spacing of $f_{q}/b$.%
} $\left\{ f-lf_{q}-k\tilde{f}_{s}\mid l,k\in\mathbb{Z}\right\} $.
By our assumption that $f_{q}=\frac{b}{a}\tilde{f}_{s}$ with $a$
and $b$ being relatively prime, simple results in number theory imply
that 
\begin{align*}
\left\{ f_{0}-lf_{q}-k\tilde{f}_{s}\mid l,k\in\mathbb{Z}\right\}  & =\left\{ f_{0}-lf_{q}/b\mid l\in\mathbb{Z}\right\} \\
 & =\left\{ f_{0}-l\tilde{f}_{s}/a\mid l\in\mathbb{Z}\right\} .
\end{align*}
In other words, for a given $f_{0}\in\left[-f_{q}/2b,f_{q}/2b\right]$,
the sampled output at $f_{0}$ depends on the input in the entire
modulated aliased set. Since the sampling bandwidth at each branch
is $\tilde{f}_{s}$, all outputs at frequencies $\left\{ f_{0}-lf_{q}/b\mid l\in\mathbb{Z};\text{ }-\tilde{f}_{s}/2\leq f_{0}-lf_{q}/b\leq\tilde{f}_{s}/2\right\} $
rely on the inputs in the same modulated aliased set. This can be
treated as a Gaussian MIMO channel with a countable number of input
branches at the frequency set $\left\{ f_{0}-l\tilde{f}_{s}/a\mid l\in\mathbb{Z}\right\} $
and $aM$ groups of output branches, each associated with one group
of sample sequences in one branch. As an example, we illustrate in
Fig. \ref{fig:Equivalent-MIMO-Gaussian-Channel-Modulation} the equivalent
MIMO Gaussian channel under sampling following a single branch of
modulation and filtering, when $S(f)=0$ for all $f\notin\left[-f_{s}/2,f_{s}/2\right]$.

The effective frequencies of this frequency-selective MIMO Gaussian
channel range from $-f_{q}/2b$ to $f_{q}/2b$, which gives us a set
of parallel channels each representing a single frequency $f$. The
water-filling power allocation strategy is then applied to achieve
capacity.

A rigorous proof of Theorem \ref{thmPremodulatedFilterBank} based
on Toeplitz properties is provided in Appendix \ref{sec:Proof-of-Theorem-Premodulated-Filter-Bank}.

\subsection{An Upper Bound on Sampled Capacity}

Following the same analysis of optimal filter-bank sampling developed
in Section \ref{sub:Optimal-Filter-Banks}, we can derive an upper
bound on the sampled channel capacity. 

\begin{corollary}\label{Cor:UpperBoundModulationBank}Consider the
system shown in Fig. \ref{fig:PremodulatedPrefilteredSampler}. Suppose
that for each aliased set $\left\{ f-if_{q}/b\mid i\in\mathbb{Z}\right\} $
and each $k$ $(1\leq k\leq aM)$, there exists an integer $l$ such
that $\frac{\left|H\left(f-lf_{q}/b\right)\right|^{2}}{\mathcal{S}_{\eta}\left(f-lf_{q}/b\right)}$
is equal to the $k^{\text{th}}$ largest element in $\left\{ \frac{\left|H\left(f-if_{q}/b\right)\right|^{2}}{\mathcal{S}_{\eta}\left(f-if_{q}/b\right)}\mid i\in\mathbb{Z}\right\} $.
The capacity (\ref{eq:CapacityModulationBank}) under sampling following
modulation and filter banks can be upper bounded by
\begin{align}
C^{\text{u}}(f_{s}) & \overset{\Delta}{=}\frac{1}{2}{\displaystyle \int}_{-f_{q}/2b}^{f_{q}/2b}\sum_{k=1}^{aM}\log^{+}\left(\nu\cdot\lambda_{k}\left({\bf F}_{h}{\bf F}_{h}^{*}\right)\right)\mathrm{d}f,
\end{align}
 where $\nu$ is chosen such that 
\begin{equation}
{\displaystyle \int}_{-f_{q}/2b}^{f_{q}/2b}\sum_{k=1}^{aM}\left[\nu-\frac{1}{\lambda_{k}\left({\bf F}_{h}{\bf F}_{h}^{*}\right)}\right]^{+}\mathrm{d}f=P.
\end{equation}

\end{corollary}

\begin{IEEEproof}By observing that $\left({\bf F}^{\eta}{\bf F}^{\eta*}\right)^{-\frac{1}{2}}{\bf F}^{\eta}$
has orthonormal rows, we can derive the result using Proposition \ref{lem_BoundsOnMSingularValues}
in Appendix \ref{sec:Proof-of-Corollary-Optimal-Filter-Bank}. \end{IEEEproof}

The upper bound of Corollary \ref{Cor:UpperBoundModulationBank} coincides
with the upper bound on sampled capacity under $aM$-branch filter-bank
sampling. This basically implies that for a given sampling rate $f_{s}$,
modulation and filter bank sampling does not outperform filter-bank
sampling in maximizing sampled channel capacity. In other words, we
can always achieve the same performance by adding more branches in
filter-bank sampling. 

Note however that this upper bound may not be tight, since we restrict
our analysis to periodic modulation sequences. General modulation
is not discussed here.

\subsection{Single-branch Sampling with Modulation and Filtering v.s. Filter-bank
Sampling}

Although the class of modulation and filter bank sampling does not
provide capacity gain compared with filter-bank sampling, it may potentially
provide implementation advantages, depending on the modulation period
$T_{q}$. Specifically, modulation-bank sampling may achieve a larger
capacity region than that achievable by filter-bank sampling with
the same number of branches. We consider here two special cases of
single-branch modulation sampling, and investigate whether any hardware
benefit can be harvested.

\subsubsection{$f_{s}/M=f_{q}/a$ for some integer $a$}

In this case, the modulated aliased set is $\left\{ f-kf_{s}/M-lf_{q}\mid k,l\in\mathbb{Z}\right\} =\left\{ f-kf_{s}/M\mid k\in\mathbb{Z}\right\} $,
which is equivalent to the original aliased frequency set. That said,
the sampled output $Y\left(f\right)$ is still a linear combination
of $\left\{ R\left(f-kf_{s}/M\right)\mid k\in\mathbb{Z}\right\} $.
But since linear combinations of these components can be attained
by simply adjusting the prefilter response $S(f)$, the modulation
bank does not provide any further design degrees of freedom, and hence
does not improve the capacity region achievable by sampling with a
bank of $M$ filters.

\subsubsection{$f_{s}/M=bf_{q}$ for some integer $b$}

In this case, the modulated aliased set is enlarged to $\left\{ f-kf_{s}/M-lf_{q}\mid k,l\in\mathbb{Z}\right\} =\left\{ f-lf_{q}\mid l\in\mathbb{Z}\right\} $,
which may potentially provide implementation gain compared with filter-bank
sampling with the same number of branches. We illustrate this in the
following example.

\begin{example}\label{Example-Modulation}Suppose that the channel
contains $3$ subbands with channel gains as plotted in Fig. \ref{fig:ModulationBankExampleChannelGain},
and that the noise is of unit spectral density within these 3 subbands
and 0 otherwise. 

(i) Let us first consider single-branch sampling with filtering with
$f_{s}=2$. As illustrated in Fig. \ref{fig:ModulationBankExampleChannelGain},
Subband 1 and 3 are mixed together due to aliasing. According to Section
\ref{sub:Optimal-Prefilters}, the optimal prefilter without modulation
would be a band-pass filter with passband $[-1.5,0.5]$, resulting
in a channel containing 2 subbands with respective channel gains $2$
and $ $$1$. 

\begin{figure}
\centering\includegraphics{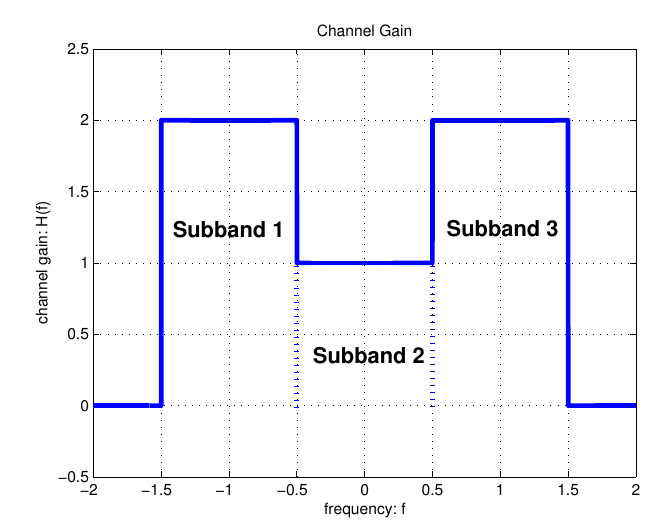}\caption{\label{fig:ModulationBankExampleChannelGain}The channel gain of Example
\ref{Example-Modulation}. The noise is of unit power spectral density. }
\end{figure}

(ii) If we add a modulation sequence with period $T_{q}=2T_{s}$,
then the channel structure can be better exploited. Specifically,
suppose that the modulation sequence obeys $c^{0}=1$, $c^{3}=1$,
and $c^{i}=0$ for all other $i$'s, and that the post-modulation
filter is a band-pass filter with passbands $[-1.5,-0.5]\cup[3.5,4.5]$.
We can see that this moves spectral contents of Subband 1 and Subband
3 to frequency bands $[-1.5,-0.5]$ and $[3.5,4.5]$, respectively,
which are alias-free. Therefore, we obtain a two-subband channel with
respective channel gains both equal to 2, thus outperforming a single
branch of sampling with filtering. \end{example}

More generally, let us consider the following scenario. Suppose that
the channel of bandwidth $W=\frac{2L}{K}f_{s}$ is equally divided
into $2L$ subbands each of bandwidth $f_{q}=f_{s}/K$ for some integers
$K$ and $L$. The SNR $ $$\left|H\left(f\right)\right|^{2}/\mathcal{S}_{\eta}(f)$
within each subband is assumed to be flat. For instance, in the presence
of white noise, if $f_{q}\ll B_{c}$ with $B_{c}$ being the coherence
bandwidth \cite{Gold2005}, the channel gain (and hence the SNR) is
roughly equal across the subband. Algorithm 1 given below generates
an alias-free sampled analog channel, which is achieved by moving
the $K$ subbands with the highest SNRs to alias-free locations. By
Corollary \ref{Cor:UpperBoundModulationBank}, this algorithm determines
an optimal sampling mechanism that maximizes capacity under a single
branch of sampling with modulation and filtering. Specifically, take
any $f\in[-f_{q}/2,f_{q}/2]$, and the algorithm works as follows.\vspace{10pt}

\begin{center}
\begin{tabular}{>{\raggedright}p{3.2in}}
\hline 
\textbf{Algorithm 1}\tabularnewline
\hline 
1. \quad{}\textbf{Initialize.} Find the $K$ largest elements in
$\left\{ \frac{\left|H\left(f-lf_{q}\right)\right|^{2}}{\mathcal{S}_{\eta}\left(f-lf_{q}\right)}\mid l\in\mathbb{Z},-L\leq l\leq L-1\right\} $.
Denote by $\left\{ l_{i}\mid1\leq i\leq K\right\} $ the index set
of these $K$ elements such that $l_{1}>l_{2}>\cdots>l_{K}$. Set
$L^{*}:=\min\left\{ k\mid k\in\mathbb{Z},k\geq2L,k\text{ mod }K=0\right\} $
.\tabularnewline
2. \quad{}For $i=1:K$

\hspace{2.5em}Let $ $$\alpha:=i\cdot L^{*}+i-l_{i}$.

\hspace{2.5em}Set $c^{\alpha}=1$, and $S(f+\alpha f_{p})=1$.\tabularnewline
\hline 
\end{tabular}
\par\end{center}

Algorithm 1 first selects the $K$ subbands with the highest SNR,
and then moves each of the selected subbands to a new location by
appropriately setting $\left\{ c^{i}\right\} $, which guarantees
that (1) the movement does not corrupt any of the previously chosen
locations; (2) the contents in the newly chosen locations will be
alias-free. The post-modulation filter is applied to suppress the
frequency contents outside the set of newly chosen subband locations.
One drawback of this algorithm is that we need to preserve as many
as $2LK$ subbands in order to make it work. 

The performance of Algorithm 1 is equivalent to the one using an optimal
filter bank followed by sampling with sampling rate $f_{q}$ at each
branch. Hence, single-branch sampling effectively achieves the same
performance as multi-branch filter-bank sampling. This approach may
be preferred since building multiple analog filters is often expensive
(in terms of power consumption, size, or cost). We note, however,
that for a given overall sampling rate, modulation-bank sampling does
not outperform filter-bank sampling with an arbitrary number of branches.
The result is formally stated as follows.

\begin{prop}\label{prop-ModulationUpperBound}Consider the setup
in Theorem \ref{thmPremodulatedFilterBank}. For a given overall sampling
rate $f_{s}$, sampling with $M$ branches of optimal modulation and
filter banks does not achieve higher sampled capacity compared to
sampling with an optimal bank of $aM$ filters.

\end{prop}

Hence, the main advantage of applying a modulation bank is a hardware
benefit, namely, using fewer branches and hence less analog circuitry
to achieve the same capacity.

\section{Connections between Capacity and MMSE\label{sec:ConnectionCapacityMMSE}}

In Sections \ref{sub:Optimal-Prefilters} and \ref{sub:Optimal-Filter-Banks},
we derived respectively the optimal prefilter and the optimal filter
bank that maximize capacity. It turns out that such choices of sampling
methods coincide with the optimal prefilter / filter bank that minimize
the MSE between the Gaussian channel input and the signal reconstructed
from sampling the channel output, as detailed below.

Consider the following sampling problem. Let $x(t)$ be a zero-mean
wide-sense stationary (WSS) stochastic signal whose power spectral
density (PSD) $\mathcal{S}_{X}(f)$ satisfies a power constraint %
\footnote{We restrict our attention to WSS input signals. This restriction,
while falling short of generality, allows us to derive sampling results
in a simple way. %
} $\int_{-\infty}^{\infty}\mathcal{S}_{X}(f)\mathrm{d}f=P$. This input
is passed through a channel consisting of an LTI filter and additive
stationary Gaussian noise. We sample the channel output using a filter
bank at a fixed rate $f_{s}/M$ in each branch, and recover a \emph{linear}
MMSE estimate $\hat{x}(t)$ of $x(t)$ from its samples in the sense
of minimizing $\mathbb{E}(\left|x(t)-\hat{x}(t)\right|^{2})$ for
$t\in\mathbb{R}$. We propose to jointly optimize $x(t)$ and the
sampling method. Specifically, our joint optimization problem can
now be posed as follows: for which input process $x(t)$ and for which
filter bank is the estimation error $\mathbb{E}(\left|x(t)-\hat{x}(t)\right|^{2})$
minimized for $t\in\mathbb{R}$.

It turns out that the optimal input and the optimal filter bank coincide
with those maximizing channel capacity, which is captured in the following
proposition. 

\begin{prop}\label{lem-optimal-filter-bank-sampling-theoretic} Suppose
the channel input $x(t)$ is any WSS signal. For a given sampling
system, let $\hat{x}(t)$ denote the optimal linear estimate of $x(t)$
from the digital sequence $\left\{ {\bf y}[n]\right\} $. Then the
capacity-optimizing filter bank given in (\ref{eq:optimalFilterBank})
and its corresponding optimal input $x(t)$ minimize the linear MSE
$\mathbb{E}(\left|x(t)-\hat{x}(t)\right|^{2})$ over all possible
LTI filter banks. 

\end{prop}

\begin{IEEEproof}See Appendix \ref{sec:Proof-of-Lemma-optimal-filter-bank-sampling-theoretic}.\end{IEEEproof}

Proposition \ref{lem-optimal-filter-bank-sampling-theoretic} implies
that the input signal and the filter bank optimizing channel capacity
also minimize the MSE between the original input signal and its reconstructed
output. We note that if the samples $\left\{ {\bf y}[n]\right\} $
and $x(t)$ are jointly Gaussian random variables, then the MMSE estimate
$\hat{x}(t)$ for a given input process $x(t)$ is linear in $\left\{ {\bf y}[n]\right\} $.
That said, for Gaussian inputs passed through Gaussian channels, the
capacity-maximizing filter bank also minimizes the MSE even if we
take into account nonlinear estimation. Thus, under sampling with
filter-banks for Gaussian channels, information theory reconciles
with sampling theory through the SNR metric when determining optimal
systems. Intuitively, high SNR typically leads to large capacity and
small MSE.

Proposition \ref{lem-optimal-filter-bank-sampling-theoretic} includes
the optimal prefilter under single-prefilter sampling as a special
case. We note that a similar MSE minimization problem was investigated
decades ago with applications in PAM \cite{ChaDon1971,Eri1973}: a
given random input $x(t)$ is prefiltered, corrupted by noise, uniformly
sampled, and then postfiltered to yield a linear estimate $\hat{x}(t)$.
The goal in that work was to minimize the MSE between $x(t)$ and
$\hat{x}(t)$ over all prefiltering (or pulse shaping) and postfiltering
mechanisms. While our problem differs from this PAM design problem
by optimizing directly over the random input instead of the pulse
shape, the two problems are similar in spirit and result in the same
alias-suppressing filter. However, earlier work did not account for
filter-bank sampling or make connections between minimizing MSE and
maximizing capacity.

\section{Conclusions and Future Work}

We have characterized sampled channel capacity as a function of sampling
rate for different sampling methods, thereby forming a new connection
between sampling theory and information theory. We show how the capacity
of a sampled analog channel is affected by reduced sampling rate and
identify optimal sampling structures for several classes of sampling
methods, which exploit structure in the sampling design. These results
also indicate that capacity is not always monotonic in sampling rate,
and illuminate an intriguing connection between MIMO channel capacity
and capacity of undersampled analog channels. The capacity optimizing
sampling structures are shown to extract the frequency components
with highest SNRs from each aliased set, and hence suppress aliasing
and out-of-band noise. We also show that for Gaussian inputs over
Gaussian channels, the optimal filter / filter bank also minimizes
the MSE between the channel input and the reconstructed signal. Our
work establishes a framework for using the information-theoretic metric
of capacity to optimize sampling structures, offering a different
angle from traditional design of sampling methods based on other performance
metrics.

Our work uncovers additional questions at the intersection of sampling
theory and information theory. For instance, an upper bound on sampled
capacity under sampling rate constraints for more general nonuniform
sampling methods would allow us to evaluate which sampling mechanisms
are capacity-achieving for any channel. Moreover, for channels where
there is a gap between achievable rates and the capacity upper bound,
these results might provide insight into new sampling mechanisms that
might close the gap to capacity. Investigation of capacity under more
general nonuniform sampling techniques is an interesting topic that
is studied in our companion paper \cite{ChenEldarGoldsmith2012}.
In addition, the optimal sampling structure for time-varying channels
will require different analysis than used in the time-invariant case.
It is also interesting to investigate what sampling mechanisms are
optimal for channels when the channel state is partially or fully
unknown. A deeper understanding of how to exploit channel structure
may also guide the design of sampling mechanisms for multiuser channels
that require more sophisticated cooperation schemes among users and
are impacted in a more complex way by subsampling.

\appendices

\section{Proof of Theorem \ref{thmPerfectCSIPrefilteredSamplerRigorous}\label{sec:Proof-of-Theorem-PerfectCSIPrefilteredSampler}}

We begin by an outline of the proof. A discretization argument is
first used to approximate arbitrarily well the analog signals by discrete-time
signals, which allows us to make use of the properties of Toeplitz
matrices instead of the more general Toeplitz operators. By noise
whitening, we effectively convert the sampled channel to a MIMO channel
with i.i.d. noise for any finite time interval. Finally, the asymptotic
properties of Toeplitz matrices are exploited in order to relate the
eigenvalue distribution of the equivalent channel matrix with the
Fourier representation of both channel filters and prefilters. The
proofs of several auxiliary lemmas are deferred to Appendix \ref{sec:Proofs-of-Auxiliary-Lemmas}.

Instead of directly proving Theorem \ref{thmPerfectCSIPrefilteredSamplerRigorous},
we prove the theorem for a simpler scenario where the noise $\eta(t)$
is of \emph{unit spectral density}. In this case, our goal is to prove
that the capacity is equivalent to
\begin{align*}
C(f_{s}) & =\frac{1}{2}{\displaystyle \int}_{-\frac{f_{s}}{2}}^{\frac{f_{s}}{2}}\log^{+}\left(\nu\frac{\overset{}{\underset{l\in\mathbb{Z}}{\sum}}\left|H(f-lf_{s})S(f-lf_{s})\right|^{2}}{\overset{}{\underset{l\in\mathbb{Z}}{\sum}}\left|S(f-lf_{s})\right|^{2}}\right)\mathrm{d}f
\end{align*}
where the water level $\nu$ can be calculated through the following
equation
\[
{\displaystyle \int}_{-\frac{f_{s}}{2}}^{\frac{f_{s}}{2}}\left(\nu-\frac{\sum_{l}\left|S(f-lf_{s})\right|^{2}}{\sum_{l}\left|H(f-lf_{s})S(f-lf_{s})\right|^{2}}\right)^{+}\mathrm{d}f=P.
\]
This capacity result under white noise can then be immediately extended
to accommodate for colored noise. Suppose the additive noise is of
power spectral density $\mathcal{S}_{\eta}(f)$. We can then split
the channel filter $H\left(f\right)$ into two parts with respective
frequency response $H\left(f\right)/\sqrt{\mathcal{S}_{\eta}(f)}$
and $\sqrt{\mathcal{S}_{\eta}(f)}$. Equivalently, the channel input
is passed through an LTI filter with frequency response $H\left(f\right)/\sqrt{\mathcal{S}_{\eta}(f)}$,
contaminated by white noise, and then passed through a filter with
transfer function $\sqrt{\mathcal{S}_{\eta}(f)}S(f)$ followed by
an ideal sampler with rate $f_{s}$. This equivalent representation
immediately leads to the capacity in the presence of colored noise
by substituting corresponding terms into the capacity with white noise.

\subsection{Channel Discretization and Diagonalization}

Given that $h(t)$ is continuous and Riemann integrable, one approach
to study the continuous-time problem is via reduction to an equivalent
discrete-time problem \cite[Chapter 16]{KaiSayHas2000}. In this subsection,
we describe the method of obtaining our equivspace discretization
approximations to the continuous-time problems, which will allow us
to exploit the properties of block-Toeplitz matrices instead of the
more complicated block-Toeplitz operators.

For notational simplicity, we define
\[
g_{u,v}=\frac{1}{\Delta}\int_{0}^{\Delta}g\left(uT_{s}-v\Delta+\tau\right)\mathrm{d}\tau
\]
for any function $g(t)$. If $g(t)$ is a continuous function, then
$\lim_{\Delta\rightarrow0}g_{u,v}=g\left(uT_{s}-v\Delta\right)$,
where $v$ may be a function of $\Delta$. We also define $\tilde{h}(t):=h(t)*s(t)$.
Set $T=nT_{s}$ and $T_{s}=k\Delta$ for some integers $n$ and $k$,
and define
\begin{align*}
\tilde{{\bf h}}_{i} & :=\Delta\cdot\left[\tilde{h}_{i,0},\tilde{h}_{i,1},\cdots,\tilde{h}_{i,k-1}\right],\\
{\bf s}_{i} & :=\Delta\cdot\left[s_{i,0},s_{i,1},\cdots,s_{i,k-1}\right],\\
\left({\bf x}^{n}\right)_{i} & :=\frac{1}{\Delta}\int_{0}^{\Delta}x\left(i\Delta+\tau\right)\mathrm{d}\tau\text{ }\left(0\leq i<nk\right),\\
\left({\bf \eta}\right)_{i} & :=\frac{1}{\Delta}\int_{0}^{\Delta}{\bf \eta}\left(i\Delta+\tau\right)\mathrm{d}\tau\text{ }\left(i\in\mathbb{Z}\right).
\end{align*}
 We also define 
\begin{align*}
\tilde{{\bf H}}^{n} & :=\left[\begin{array}{cccc}
\tilde{{\bf h}}_{0} & \tilde{{\bf h}}_{-1} & \cdots & \tilde{{\bf h}}_{-n+1}\\
\tilde{{\bf h}}_{1} & \tilde{{\bf h}}_{0} & \cdots & \tilde{{\bf h}}_{-n+2}\\
\vdots & \vdots & \cdots & \vdots\\
\tilde{{\bf h}}_{n-1} & \tilde{{\bf h}}_{n-2} & \cdots & \tilde{{\bf h}}_{0}
\end{array}\right],
\end{align*}
\[
{\bf S}^{n}:=\left[\begin{array}{cccc}
\cdots & {\bf s}_{0} & {\bf s}_{-1} & \cdots\\
\cdots & {\bf s}_{1} & {\bf s}_{0} & \cdots\\
\cdots & \vdots & \vdots & \cdots\\
\cdots & {\bf s}_{n-1} & {\bf s}_{n-2} & \cdots
\end{array}\right].
\]
With these definitions, the original channel model can be approximated
with the following discretized channel:
\begin{equation}
{\bf y}^{n}=\tilde{{\bf H}}^{n}{\bf x}^{n}+{\bf S}^{n}{\bf \eta}.
\end{equation}
As can be seen, $\tilde{{\bf H}}^{n}$ is a fat \textit{block Toeplitz}
matrix. Moreover, ${\bf S}^{n}{\bf S}^{n*}$ is asymptotically equivalent
to a Toeplitz matrix, as will be shown in Appendix \ref{sub:ProofTheorem2PartC}.
We note that each element $\eta_{i}$ is a zero-mean Gaussian variable
with variance $\mathbb{E}(\left|\eta_{i}\right|^{2})=1/\Delta$. In
addition, $\mathbb{E}\left({\bf \eta}_{i}{\bf \eta}_{l}^{*}\right)=0$
for any $i\neq l$, implying that ${\bf \eta}$ is an i.i.d. Gaussian
vector. The filtered noise ${\bf S}^{n}{\bf \eta}$ is no longer i.i.d.
Gaussian, which motivates us to whiten the noise first.

The prewhitening matrix is given by ${\bf S}_{\text{w}}^{n}:=\left({\bf S}^{n}{\bf S}^{n*}\right)^{-\frac{1}{2}}$,
which follows from the fact that
\begin{align*}
 & \mathbb{E}{\bf S}_{\text{w}}^{n}{\bf S}^{n}{\bf \eta}\left({\bf S}_{\text{w}}^{n}{\bf S}^{n}{\bf \eta}\right)^{*}={\bf S}_{\text{w}}^{n}{\bf S}^{n}\mathbb{E}\left({\bf \eta}{\bf \eta}^{*}\right){\bf S}^{n*}{\bf S}_{\text{w}}^{n*}\\
= & \frac{1}{\Delta}\left({\bf S}^{n}{\bf S}^{n*}\right)^{-\frac{1}{2}}{\bf S}^{n}{\bf S}^{n*}\left({\bf S}^{n}{\bf S}^{n*}\right)^{-\frac{1}{2}}=\frac{1}{\Delta}{\bf I}^{n}.
\end{align*}
This basically implies that ${\bf S}_{\text{w}}^{n}{\bf S}^{n}$ projects
the i.i.d. Gaussian noise $\eta$ onto an $n$-dimensional subspace,
and that ${\bf S}_{\text{w}}^{n}\left({\bf S}^{n}\eta\right)$ is
now $n$-dimensional i.i.d. Gaussian noise. Left-multiplication with
this whitening matrix yields a new output
\begin{align*}
\tilde{{\bf y}}^{n}: & =\left({\bf S}^{n}{\bf S}^{n*}\right)^{-\frac{1}{2}}\left(\tilde{{\bf H}}^{n}{\bf x}^{n}+{\bf S}^{n}{\bf \eta}\right)\\
 & =\left({\bf S}^{n}{\bf S}^{n*}\right)^{-\frac{1}{2}}\tilde{{\bf H}}^{n}{\bf x}^{n}+\tilde{{\bf \eta}}^{n}.
\end{align*}
Here, $\tilde{\eta}^{n}$ consists of independent zero-mean Gaussian
elements with variance $1/\Delta$. Since the prewhitening operation
${\bf S}_{\text{w}}^{n}$ is invertible, we have
\begin{equation}
I\left({\bf x}^{n};\tilde{{\bf y}}^{n}\right)=I\left({\bf x}^{n};{\bf y}^{n}\right).
\end{equation}
In this paper, we will use $I_{k,\Delta}\left({\bf x}^{n};{\bf y}^{n}\right)$
and $I\left({\bf x}^{n};{\bf y}^{n}\right)$ interchangeably to denote
the mutual information between the $nk$-dimensional vector ${\bf x}^{n}$
and ${\bf y}^{n}$.

Moreover, when $x(t)$ is of bounded variance (i.e. $\sup_{t}\mathbb{E}\left|x(t)\right|^{2}\leq\infty$)
and the additive noise is Gaussian, it has been shown \cite{wu2012functional}
that the mutual information is weakly continuous in the input distribution.
Therefore, $\lim_{k\rightarrow\infty}I_{k,\Delta}\left({\bf x}^{n};{\bf y}^{n}\right)\rightarrow I\left(\left\{ x\left(t\right)\right\} _{t=0}^{T};\left\{ {\bf y}\left[n\right]\right\} _{t=0}^{T}\right).$
As $k$ increases, the discretized sequence becomes a finer approximation
to the continuous-time signal. The uniform continuity of the probability
measure of $x(t)$ and the continuity of mutual information immediately
imply that $\lim_{n\rightarrow\infty}\frac{1}{nT_{\text{s}}}I_{k,\Delta}\left({\bf x}^{n};{\bf y}^{n}\right)$
converges uniformly in $k$. We also observe that for every given
$n$, $\lim_{k\rightarrow\infty}I_{k,\Delta}\left({\bf x}^{n};{\bf y}^{n}\right)$
exists due to the continuity condition of the mutual information.
Therefore, applying the Moore-Osgood theorem in real analysis allows
us to exchange the order of limits. 

Based on the above arguments, the capacity of the sampled analog channel
can be expressed as the following limit
\begin{align*}
C(f_{s}) & =\lim_{k\rightarrow\infty}\lim_{n\rightarrow\infty}\frac{1}{nT_{s}}\sup_{p(x):\frac{1}{nk}\mathbb{E}\left(\left\Vert {\bf x}^{n}\right\Vert _{2}^{2}\right)\leq P}I_{k,\Delta}\left({\bf x}^{n},{\bf y}^{n}\right)\\
 & =\lim_{k\rightarrow\infty}\lim_{n\rightarrow\infty}\frac{f_{s}}{n}\sup_{p(x):\frac{1}{nk}\mathbb{E}\left(\left\Vert {\bf x}^{n}\right\Vert _{2}^{2}\right)\leq P}I_{k,\Delta}\left({\bf x}^{n},\tilde{{\bf y}}^{n}\right).
\end{align*}
Note that it suffices to investigate the case where $T$ is an integer
multiple of $T_{s}$ since $\lim_{T\rightarrow\infty}\frac{1}{T}\sup I\left(x(0,T];\left\{ y[n]\right\} \right)=\lim_{n\rightarrow\infty}\frac{1}{nT_{s}}\sup I\left(x(0,nT_{s}];\left\{ y[n]\right\} \right)$. 

\subsection{Preliminaries on Toeplitz Matrices}

Before proceeding to the proof of the theorem, we briefly introduce
several basic definitions and properties related to Toeplitz matrices.
Interested readers are referred to \cite{Gray06,GreSze1984} for more
details.

A Toeplitz matrix is an $n\times n$ matrix ${\bf T}^{n}$ where $\left({\bf T}^{n}\right)_{k,l}=t_{k-l}$,
which implies that a Toeplitz matrix ${\bf T}^{n}$ is uniquely defined
by the sequence $\left\{ t_{k}\right\} $. A special case of Toeplitz
matrices is circulant matrices where every row of the matrix ${\bf C}^{n}$
is a right cyclic shift of the row above it. The Fourier series (or
symbol) with respect to the sequence of Toeplitz matrices $\left\{ {\bf T}^{n}:=\left[t_{k-l};k,l=0,1,\cdots,n-1\right]:n\in\mathbb{Z}\right\} $
is given by 
\begin{equation}
F(\omega)=\sum_{k=-\infty}^{+\infty}t_{k}\exp\left(jk\omega\right),\quad\omega\in\left[-\pi,\pi\right].\label{eq:FourierSeries-1}
\end{equation}
Since the sequence $\left\{ t_{k}\right\} $ uniquely determines $F(\omega)$
and vice versa, we denote by ${\bf T}^{n}(F)$ the Toeplitz matrix
generated by $F$ (and hence $\left\{ t_{k}\right\} $). We also define
a related circulant matrix ${\bf C}^{n}(F)$ with top row $(c_{0}^{(n)},c_{1}^{(n)},\cdots,c_{n-1}^{(n)})$,
where
\begin{equation}
c_{k}^{(n)}=\frac{1}{n}\sum_{i=0}^{n-1}F\left(\frac{2\pi i}{n}\right)\exp\left(\frac{2\pi jik}{n}\right).\label{eq:circulantConstruction}
\end{equation}

One key concept in our proof is asymptotic equivalence, which is formally
defined as follows \cite{Gray06}.

\begin{definition}[{\bf Asymptotic Equivalence}] Two sequences of
$n\times n$ matrices $\left\{ {\bf A}^{n}\right\} $ and $\left\{ {\bf B}^{n}\right\} $
are said to be asymptotically equivalent if

(1) ${\bf A}^{n}$ and ${\bf B}^{n}$ are uniformly bounded, i.e.
there exists a constant $c$ independent of $n$ such that
\begin{equation}
\left\Vert {\bf A}^{n}\right\Vert _{2},\left\Vert {\bf B}^{n}\right\Vert _{2}\leq c<\infty,\quad n=1,2,\cdots
\end{equation}

(2) $\lim_{n\rightarrow\infty}\frac{1}{\sqrt{n}}\left\Vert {\bf A}^{n}-{\bf B}^{n}\right\Vert _{\text{F}}=0$.

\end{definition}

We will abbreviate asymptotic equivalence of $\left\{ {\bf A}^{n}\right\} $
and $\left\{ {\bf B}^{n}\right\} $ by ${\bf A}^{n}\sim{\bf B}^{n}$.
Two important results regarding asymptotic equivalence are given in
the following lemmas \cite{Gray06}.

\begin{lem}\label{lemmaAsymptoticEquivalenceEigenvalues}Suppose
${\bf A}^{n}\sim{\bf B}^{n}$ with eigenvalues $\left\{ \alpha_{n,k}\right\} $
and $\left\{ \beta_{n,k}\right\} $, respectively. Let $g(x)$ be
an arbitrary continuous function. Then if the limits $\lim_{n\rightarrow\infty}\frac{1}{n}\sum_{k=0}^{n-1}g\left(\alpha_{n,k}\right)$
and $\lim_{n\rightarrow\infty}\frac{1}{n}\sum_{k=0}^{n-1}g\left(\beta_{n,k}\right)$
exist, we have 
\begin{equation}
\lim_{n\rightarrow\infty}\frac{1}{n}\sum_{k=0}^{n-1}g\left(\alpha_{n,k}\right)=\lim_{n\rightarrow\infty}\frac{1}{n}\sum_{k=0}^{n-1}g\left(\beta_{n,k}\right).
\end{equation}
 \end{lem}

\begin{lem} \label{lemmaMultiplicationInverse}(a) Suppose a sequence
of Toeplitz matrices ${\bf T}^{n}$ where $\left({\bf T}^{n}\right)_{ij}=t_{i-j}$
satisfies that $\left\{ t_{i}\right\} $ is absolutely summable. Suppose
the Fourier series $F(\omega)$ related to ${\bf T}^{n}$ is positive
and ${\bf T}^{n}$ is Hermitian. Then we have
\begin{equation}
{\bf T}^{n}(F)\sim{\bf C}^{n}(F).
\end{equation}
If we further assume that there exists a constant $\epsilon>0$ such
that $F\left(\omega\right)\geq\epsilon>0$ for all $\omega\in\left[0,2\pi\right]$,
then we have 
\begin{equation}
{\bf T}^{n}(F)^{-1}\sim{\bf C}^{n}(F)^{-1}={\bf C}^{n}(1/F)\sim{\bf T}^{n}\left(1/F\right).
\end{equation}

(b) Suppose ${\bf A}^{n}\sim{\bf B}^{n}$ and ${\bf C}^{n}\sim{\bf D}^{n}$,
then ${\bf A}^{n}{\bf C}^{n}\sim{\bf B}^{n}{\bf D}^{n}$.\end{lem}

Toeplitz or block Toeplitz matrices have well-known asymptotic spectral
properties \cite{GreSze1984,Tilli98}. The notion of asymptotic equivalence
allows us to approximate non-Toeplitz matrices by Toeplitz matrices,
which will be used in the next subsection to analyze the spectral
properties of the channel matrix.

\subsection{Capacity via Convergence of the Discrete Model\label{sub:ProofTheorem2PartC}}

After channel discretization, we can calculate the capacity for each
finite duration $T$ using well-known MIMO Gaussian channel capacity,
which, however, depends on the spectrum of the truncated channel and
may vary dramatically for different $T$. By our definition of capacity,
we will pass $T$ to infinty and see whether the finite-duration capacity
converges, and if so, whether there is a closed-form expression for
the limit. Fortunately, the beautiful asymptotic properties of block-Toeplitz
matrices guarantees the existance of the limit and allows for a closed-form
solution using the frequency response of $h(t)$ and $s(t)$. 

To see this, we first construct a new channel whose capacity is easier
to obtain, and will show that the new channel has asymptotically equivalent
channel capacity as the original channel. As detailed below, each
key matrix associated with the new channel is a Toeplitz matrix, whose
spectrum can be well approximated in the asymptotic regime \cite{Gray06}. 

Consider the spectral properties of the Hermitian matrices $\tilde{{\bf H}}^{n}\tilde{{\bf H}}^{n*}$
and ${\bf S}^{n}{\bf S}^{n*}$ . We can see that
\begin{equation}
\left(\tilde{{\bf H}}^{n}\tilde{{\bf H}}^{n*}\right)_{ij}=\left(\tilde{{\bf H}}^{n}\tilde{{\bf H}}^{n*}\right)_{ji}^{*}=\sum_{t=-j+1}^{n-j}\tilde{{\bf h}}_{j-i+t}\tilde{{\bf h}}_{t}^{*}.
\end{equation}
 Obviously, $\tilde{{\bf H}}^{n}\tilde{{\bf H}}^{n*}$ is not a Toeplitz
matrix. Instead of investigating the eigenvalue distribution of $\tilde{{\bf H}}^{n}\tilde{{\bf H}}^{n*}$
directly, we look at a new \textit{Hermitian Toeplitz} matrix $\hat{{\bf H}}^{n}$
associated with $\tilde{{\bf H}}^{n}\tilde{{\bf H}}^{n*}$ such that
for any $i\leq j$:
\begin{equation}
\left(\hat{{\bf H}}^{n}\right)_{ij}=\left(\hat{{\bf H}}^{n}\right)_{ij}^{*}=\sum_{t=-\infty}^{\infty}\tilde{{\bf h}}_{j-i+t}\tilde{{\bf h}}_{t}^{*}.
\end{equation}

\begin{lem}\label{lemmaAsymptoticEquivalenceSH}The above definition
of $\hat{{\bf H}}^{n}$ implies that
\begin{equation}
\hat{{\bf H}}^{n}\sim\tilde{{\bf H}}^{n}\tilde{{\bf H}}^{n*}.
\end{equation}
 \end{lem}

\begin{IEEEproof}See Appendix \ref{sec:Proof-of-Lemma-Asymptotic-SH}.\end{IEEEproof}

On the other hand, for any $1\leq i\leq j\leq n$, we have 
\begin{equation}
\left({\bf S}^{n}{\bf S}^{n*}\right)_{ij}=\left({\bf S}^{n}{\bf S}^{n*}\right)_{ji}^{*}=\sum_{t=-\infty}^{\infty}{\bf s}_{j-i+t}{\bf s}_{t}^{*}.
\end{equation}
Hence, the Hermitian matrix $\hat{{\bf S}}^{n}:={\bf S}^{n}{\bf S}^{n*}$
is still Toeplitz. However, the matrix of interest in the capacity
will be $\left({\bf S}^{n}{\bf S}^{n*}\right)^{-1/2}$ instead. We
therefore construct an asymptotically equivalent circulant matrix
${\bf C}^{n}$ as defined in (\ref{eq:circulantConstruction}), which
will preserves the Toeplitz property when we take $\left({\bf C}^{n}\right)^{-1/2}$
\cite{Gray06}. Formally speaking, $\left({\bf S}^{n}{\bf S}^{n*}\right)^{-1}$
can be related to $\left({\bf C}^{n}\right)^{-1}$ as follows.

\begin{lem}\label{lem-S-Inverse-Asymptotic}If there exists some
constant $\epsilon_{s}>0$ such that for all $f\in\left[-\frac{f_{s}}{2},\frac{f_{s}}{2}\right]$,
\begin{equation}
\sum_{l\in\mathbb{Z}}\left|S\left(f-lf_{s}\right)\right|^{2}\geq\epsilon_{s}>0
\end{equation}
 holds, then $\left({\bf C}^{n}\right)^{-1}\sim\left({\bf S}^{n}{\bf S}^{n*}\right)^{-1}$.

\end{lem}

\begin{IEEEproof}See Appendix \ref{sec:Proof-of-Lemma-S-Inverse-Asymptotic}.\end{IEEEproof}

One of the most useful properties of a circulant matrix ${\bf C}^{n}$
is that its eigenvectors $\left\{ {\bf u}_{c}^{(m)}\right\} $ are
\begin{equation}
{\bf u}_{c}^{(m)}=\frac{1}{\sqrt{n}}\left(1,e^{-2\pi jm/n},\cdots,e^{-2\pi j(n-1)m/n}\right).
\end{equation}
 Suppose the eigenvalue decomposition of ${\bf C}^{n}$ is given as
\begin{equation}
{\bf C}^{n}={\bf U}_{c}{\bf \Lambda}_{c}{\bf U}_{c}^{*},
\end{equation}
 where ${\bf U}_{c}$ is a Fourier coefficient matrix, and ${\bf \Lambda}_{c}$
is a diagonal matrix where each element in the diagonal is positive.

The concept of asymptotic equivalence allows us to explicitly relate
our matrices of interest to both circulant matrices and Toeplitz matrices,
whose asymptotic spectral densities have been well studied.

\begin{lem}\label{lem-asymptoticSpectralPropertyGeneralSampling}For
any continuous function $g(x)$, we have
\begin{align*}
 & \lim_{n\rightarrow\infty}\frac{1}{n}\sum_{i=1}^{n}g\left(\lambda_{i}\right)\\
= & T_{s}{\displaystyle \int}_{-\frac{f_{s}}{2}}^{\frac{f_{s}}{2}}g\left(\frac{\sum_{l\in\mathbb{Z}}\left|H(f-lf_{s})S(f-lf_{s})\right|^{2}}{\sum_{l\in\mathbb{Z}}\left|S(f-lf_{s})\right|^{2}}\right)\mathrm{d}f,
\end{align*}
 where $\lambda_{i}$ denotes the $i$th eigenvalue of $\left({\bf S}^{n}{\bf S}^{n*}\right)^{-\frac{1}{2}}\tilde{{\bf H}}^{n}\tilde{{\bf H}}^{n*}\left({\bf S}^{n}{\bf S}^{n*}\right)^{-\frac{1}{2}}$.

\end{lem}

\begin{IEEEproof}See Appendix \ref{sec:Proof-of-Lem-Asymptotic-Spectral-General-Sampling}.\end{IEEEproof}

We can now prove the capacity result. The standard capacity results
for parallel channels \cite[Theorem 7.5.1]{Gallager68} implies that
the capacity of the discretized sampled analog channel is given by
the parametric equations
\begin{align}
C_{T} & =\frac{1}{2T}\sum_{i}\log^{\text{+}}\left(\nu\lambda_{i}\right),\\
\frac{Pnk}{1/\Delta} & =\sum_{i}\left[\nu-1/\lambda_{i}\right]^{\text{+}},\label{eq:PowerConstraint}
\end{align}
 where $\nu$ is the water level of the optimal power allocation over
this discrete model, as can be calculated through (\ref{eq:PowerConstraint}).
Since this capacity depends on the eigenvalues of $\left({\bf S}^{n}{\bf S}^{n*}\right)^{-\frac{1}{2}}\tilde{{\bf H}}^{n}\tilde{{\bf H}}^{n*}\left({\bf S}^{n}{\bf S}^{n*}\right)^{-\frac{1}{2}}$,
then by Lemma \ref{lem-asymptoticSpectralPropertyGeneralSampling},
the convergence as $T\rightarrow\infty$ is guaranteed and the capacity
$C=\lim_{T\rightarrow\infty}C_{T}$ can be expressed using $H(f)$
and $S(f)$. Specifically,
\begin{align*}
 & \lim_{T\rightarrow\infty}C_{T}\left(\nu\right)=\lim_{T\rightarrow\infty}\frac{1}{T}\sum_{i}\frac{1}{2}\log^{+}\left[\nu\lambda_{i}\right]\\
= & \frac{1}{2}{\displaystyle \int}_{-f_{s}/2}^{f_{s}/2}\log^{+}\left(\nu\frac{\sum_{l\in\mathbb{Z}}\left|H(f-lf_{s})S(f-lf_{s})\right|^{2}}{\sum_{l\in\mathbb{Z}}\left|S(f-lf_{s})\right|^{2}}\right)\mathrm{d}f.
\end{align*}
 Similarly, (\ref{eq:PowerConstraint}) can be transformed into
\begin{align*}
 & PT_{s}=\frac{Pk}{1/\Delta}=\frac{1}{n}\sum_{i}\left[\nu-1/\lambda_{i}\right]^{+}\\
= & T_{s}{\displaystyle \int}_{-f_{s}/2}^{f_{s}/2}\left[\nu-\frac{\sum_{l\in\mathbb{Z}}\left|S(f-lf_{s})\right|^{2}}{\sum_{l\in\mathbb{Z}}\left|H(f-lf_{s})S(f-lf_{s})\right|^{2}}\right]^{+}\mathrm{d}f,
\end{align*}
 which completes the proof.

\section{Proof of Theorem \ref{thmPerfectCSIFilterBankSingleAntenna}\label{sec:Proof-of-Theorem-PerfectCSIFilterBank}}

We follow similar steps as in the proof of Theorem \ref{thmPerfectCSIPrefilteredSamplerRigorous}:
we approximate the sampled channel using a discretized model first,
whiten the noise, and then find capacity of the equivalent channel
matrix. Due to the use of filter banks, the equivalent channel matrix
is no longer asymptotically equivalent to a Toeplitz matrix, but instead
a block-Toeplitz matrix. This motivates us to exploit the asymptotic
properties of block-Toeplitz matrices.

\subsection{Channel Discretization and Diagonalization}

Let $\tilde{T}_{s}=MT_{s}$, and suppose we have $T=n\tilde{T}_{s}$
and $\tilde{T}_{s}=k\Delta$ with integers $n$ and $k$. Similarly,
we can define
\[
\tilde{h}_{i}(t):=h(t)*s_{i}(t),\quad\text{and}
\]
\[
\tilde{{\bf h}}_{i}^{l}=\left[\tilde{h}_{i}\left(l\tilde{T}_{s}\right),\tilde{h}_{i}\left(l\tilde{T}_{s}-\Delta\right),\cdots,\tilde{h}_{i}\left(l\tilde{T}_{s}-(k-1)\Delta\right)\right].
\]
We introduce the following two matrices as
\[
\tilde{{\bf H}}_{i}^{n}=\left[\begin{array}{cccc}
\tilde{{\bf h}}_{i}^{0} & \tilde{{\bf h}}_{i}^{-1} & \cdots & \tilde{{\bf h}}_{i}^{-n+1}\\
\tilde{{\bf h}}_{i}^{1} & \tilde{{\bf h}}_{i}^{0} & \cdots & \tilde{{\bf h}}_{i}^{-n+2}\\
\vdots & \vdots & \vdots & \vdots\\
\tilde{{\bf h}}_{i}^{n-1} & \tilde{{\bf h}}_{i}^{n-2} & \cdots & \tilde{{\bf h}}_{i}^{0}
\end{array}\right]
\]
\[
\text{and}\quad{\bf S}_{i}^{n}=\left[\begin{array}{cccc}
\cdots & {\bf s}_{i}^{0} & {\bf s}_{i}^{-1} & \cdots\\
\cdots & {\bf s}_{i}^{1} & {\bf s}_{i}^{0} & \cdots\\
\vdots & \vdots & \vdots & \vdots\\
\cdots & {\bf s}_{i}^{n-1} & {\bf s}_{i}^{n-2} & \cdots
\end{array}\right].
\]
We also set $\left({\bf x}^{n}\right)_{i}=\frac{1}{\Delta}\int_{0}^{\Delta}x\left(i\Delta+\tau\right)\mathrm{d}\tau\left(0\leq i<nk\right)$,
and $\left({\bf \eta}\right)_{i}=\frac{1}{\Delta}\int_{0}^{\Delta}{\bf \eta}\left(i\Delta+\tau\right)\mathrm{d}\tau$
$\left(i\in\mathbb{Z}\right)$. Defining ${\bf y}^{n}=\left[y_{1}[0],\cdots,y_{1}[n-1],y_{2}[0],\cdots,y_{2}[n-1],\cdots,y_{M}[n-1]\right]^{T}$
leads to the discretized channel model
\[
{\bf y}^{n}=\left[\begin{array}{c}
\tilde{{\bf H}}_{1}^{n}\\
\tilde{{\bf H}}_{2}^{n}\\
\vdots\\
\tilde{{\bf H}}_{M}^{n}
\end{array}\right]{\bf x}^{n}+\left[\begin{array}{c}
{\bf S}_{1}^{n}\\
{\bf S}_{2}^{n}\\
\vdots\\
{\bf S}_{M}^{n}
\end{array}\right]{\bf \eta}.
\]
 Whitening the noise gives us
\[
\tilde{{\bf y}}^{n}=\left(\left[\begin{array}{c}
{\bf S}_{1}^{n}\\
{\bf S}_{2}^{n}\\
\vdots\\
{\bf S}_{M}^{n}
\end{array}\right]\left[\begin{array}{ccc}
{\bf S}_{1}^{n*} & \cdots & {\bf S}_{M}^{n*}\end{array}\right]\right)^{-\frac{1}{2}}\left[\begin{array}{c}
\tilde{{\bf H}}_{1}^{n}\\
\tilde{{\bf H}}_{2}^{n}\\
\vdots\\
\tilde{{\bf H}}_{M}^{n}
\end{array}\right]{\bf x}_{n}+\tilde{\eta},
\]
 where $\tilde{{\bf \eta}}$ is i.i.d. Gaussian variable with variance
$1/\Delta$. We can express capacity of the sampled analog channel
under filter-bank sampling as the following limit
\begin{align*}
C(f_{s}) & =\lim_{k\rightarrow\infty}\lim_{n\rightarrow\infty}\frac{f_{s}}{Mn}\sup I\left({\bf x}^{n};\tilde{{\bf y}}^{n}\right),
\end{align*}
 Here, the supremum is taken over all distribution of ${\bf x}^{n}$
subject to a power constraint $\frac{1}{nk}\mathbb{E}\left(\left\Vert x_{n}\right\Vert _{2}^{2}\right)\leq P$.

\subsection{Capacity via Convergence of the Discrete Model}

We can see that for any $1\leq u,v\leq m$,
\begin{equation}
{\bf S}_{u}^{n}{\bf S}_{v}^{n*}=\tilde{{\bf S}}_{u,v}^{n},
\end{equation}
 where the Toeplitz matrix $\tilde{{\bf S}}_{u,v}^{n}$ is defined
such that for any $1\leq i\leq j\leq n$
\begin{equation}
\left(\tilde{{\bf S}}_{u,v}^{n}\right)_{i,j}=\sum_{t=-\infty}^{\infty}{\bf s}_{u}^{j-i+t}\left({\bf s}_{v}^{t}\right)^{*}.
\end{equation}
 Let ${\bf S}^{n}=\left[{\bf S}_{1}^{n*},{\bf S}_{2}^{n*},\cdots,{\bf S}_{M}^{n*}\right]^{*}$.
Then the Hermitian block Toeplitz matrix 
\[
\tilde{{\bf S}}^{n}:=\left[\begin{array}{cccc}
\tilde{{\bf S}}_{1,1}^{n} & \tilde{{\bf S}}_{1,2}^{n} & \cdots & \tilde{{\bf S}}_{1,M}^{n}\\
\tilde{{\bf S}}_{2,1}^{n} & \tilde{{\bf S}}_{2,2}^{n} & \cdots & \tilde{{\bf S}}_{2,M}^{n}\\
\vdots & \vdots & \vdots & \vdots\\
\tilde{{\bf S}}_{M,1}^{n} & \tilde{{\bf S}}_{M,2}^{n} & \cdots & \tilde{{\bf S}}_{M,M}^{n}
\end{array}\right]
\]
 satisfies $\tilde{{\bf S}}^{n}={\bf S}^{n}{\bf S}^{n*}$. Additionally,
we define $\hat{{\bf H}}_{u,v}^{n}\text{ }\left(1\leq u,v\leq M\right)$
, where
\begin{equation}
\left(\hat{{\bf H}}_{u,v}^{n}\right)_{i,j}=\sum_{t=-\infty}^{\infty}\tilde{{\bf h}}_{u}^{j-i+t}\left(\tilde{{\bf h}}_{v}^{t}\right)^{*},
\end{equation}
and we let $\tilde{{\bf H}}^{n}=\left[\tilde{{\bf H}}_{1}^{n*},\tilde{{\bf H}}_{2}^{n*},\cdots,\tilde{{\bf H}}_{M}^{n*}\right]^{*}$.
The block Toeplitz matrix
\[
\hat{{\bf H}}^{n}:=\left[\begin{array}{cccc}
\hat{{\bf H}}_{1,1}^{n} & \hat{{\bf H}}_{1,2}^{n} & \cdots & \hat{{\bf H}}_{1,M}^{n}\\
\hat{{\bf H}}_{2,1}^{n} & \hat{{\bf H}}_{2,2}^{n} & \cdots & \hat{{\bf H}}_{2,M}^{n}\\
\vdots & \vdots & \vdots & \vdots\\
\hat{{\bf H}}_{M,1}^{n} & \hat{{\bf H}}_{M,2}^{n} & \cdots & \hat{{\bf H}}_{M,M}^{n}
\end{array}\right]
\]
satisfies
\begin{align*}
 & \lim_{n\rightarrow\infty}\frac{1}{\sqrt{nM}}\left\Vert \hat{{\bf H}}^{n}-\tilde{{\bf H}}^{n}\tilde{{\bf H}}^{n*}\right\Vert _{\text{F}}\\
\leq & \lim_{n\rightarrow\infty}\frac{1}{\sqrt{M}}\sum_{1\leq u,v\leq M}\frac{1}{\sqrt{n}}\left\Vert \hat{{\bf H}}_{u,v}^{n}-\tilde{{\bf H}}_{u}^{n}\tilde{{\bf H}}_{v}^{n*}\right\Vert =0.
\end{align*}
\begin{equation}
\Longrightarrow\quad\hat{{\bf H}}^{n}\sim\tilde{{\bf H}}^{n}\tilde{{\bf H}}^{n*}.
\end{equation}

The $M\times M$ Fourier symbol matrix ${\bf F}_{\tilde{s}}(f)$ associated
with $\tilde{{\bf S}}^{n}$ has elements $\left[{\bf F}_{\tilde{s}}\left(f\right)\right]_{u,v}$
given by
\begin{align*}
 & \left[{\bf F}_{\tilde{s}}\left(f\right)\right]_{u,v}\\
= & \frac{\Delta^{2}}{\tilde{T}_{s}^{2}}\sum_{i=0}^{k-1}\left(\sum_{l_{1}}S_{u}\left(-f+l_{1}\tilde{f}_{s}\right)\exp\left(-j2\pi\left(f-l_{1}\tilde{f}_{s}\right)i\Delta\right)\right)\\
 & \quad\left(\sum_{l_{2}}S_{v}\left(-f+l_{2}\tilde{f}_{s}\right)\exp\left(-j2\pi\left(f-l_{2}\tilde{f}_{s}\right)i\Delta\right)\right)^{*}\\
= & \frac{\Delta^{2}}{\tilde{T}_{s}^{2}}\sum_{i=0}^{k-1}\left(\sum_{l_{1},l_{2}}S_{u}\left(-f+l_{1}\tilde{f}_{s}\right)S_{v}^{*}\left(-f+l_{2}\tilde{f}_{s}\right)\right.\\
 & \quad\quad\quad\left.\exp\left(-j2\pi\left(l_{2}-l_{1}\right)\tilde{f}_{s}i\Delta\right)\right)\\
= & \frac{\Delta}{\tilde{T}_{s}}\sum_{l\in\mathbb{Z}}S_{u}\left(-f+l\tilde{f}_{s}\right)S_{v}^{*}\left(-f+l\tilde{f}_{s}\right).
\end{align*}
 Denote by $\left\{ {\bf T}^{n}\left({\bf F}_{\tilde{s}}^{-1}\right)\right\} $
the sequence of block Toeplitz matrices generated by ${\bf F}_{\tilde{s}}^{-1}(f)$,
and denote by ${\bf T}_{l_{1},l_{2}}^{n}\left({\bf F}_{\tilde{s}}^{-1}\right)$
the $(l_{1},l_{2})$ Toeplitz block of ${\bf T}^{n}\left({\bf F}_{\tilde{s}}^{-1}\right)$.
It can be verified that
\begin{align*}
\sum_{l_{2}=1}^{M}{\bf T}_{l_{1},l_{2}}^{n}\left({\bf F}_{\tilde{s}}^{-1}\right)\cdot\tilde{{\bf S}}_{l_{2},l_{3}}^{n} & \sim{\bf T}_{n}\left(\sum_{l_{2}=1}^{M}\left[{\bf F}_{\tilde{s}}^{-1}\right]_{l_{1},l_{2}}\left[{\bf F}_{\tilde{s}}\right]_{l_{2},l_{3}}\right)\\
 & ={\bf T}^{n}\left(\delta[l_{1}-l_{3}]\right),
\end{align*}
 which immediately yields
\begin{equation}
{\bf T}^{n}\left({\bf F}_{\tilde{s}}^{-1}\right)\tilde{{\bf S}}^{n}\sim{\bf I}\text{ }\Longrightarrow\text{ }{\bf T}^{n}\left({\bf F}_{\tilde{s}}^{-1}\right)\sim\left(\tilde{{\bf S}}^{n}\right)^{-1}\sim\left({\bf S}^{n}{\bf S}^{n*}\right)^{-1}.
\end{equation}
 Therefore, for any continuous function $g(x)$, \cite[Theorem 5.4]{Tilli98}
implies that
\begin{align*}
 & \lim_{n\rightarrow\infty}\frac{1}{nM}\sum_{i=1}^{nM}g\left(\lambda_{i}\left\{ \left({\bf S}^{n}{\bf S}^{n*}\right)^{-\frac{1}{2}}\tilde{{\bf H}}^{n}\tilde{{\bf H}}^{n*}\left({\bf S}^{n}{\bf S}^{n*}\right)^{-\frac{1}{2}}\right\} \right)\\
= & {\displaystyle \int}_{-\frac{f_{s}}{2M}}^{\frac{f_{s}}{2M}}\sum_{i=1}^{M}g\left(\lambda_{i}\left({\bf F}_{\tilde{s}}^{-\frac{1}{2}}{\bf F}_{\tilde{h}}{\bf F}_{\tilde{s}}^{-\frac{1}{2}}\right)\right)\mathrm{d}f.
\end{align*}
Denote ${\bf F}_{s}^{\ddagger}=\left({\bf F}_{s}{\bf F}_{s}^{*}\right)^{-\frac{1}{2}}{\bf F}_{s}$,
then the capacity of parallel channels \cite{Gallager68}, which is
achieved via water filling power allocation, yields
\begin{align*}
 & C(f_{s})\\
= & \lim_{n\rightarrow\infty}\frac{\sum_{i=1}^{nM}\log^{+}\left(\nu\lambda_{i}\left\{ \left({\bf S}^{n}{\bf S}^{n*}\right)^{-\frac{1}{2}}\tilde{{\bf H}}^{n}\tilde{{\bf H}}^{n*}\left({\bf S}^{n}{\bf S}^{n*}\right)^{-\frac{1}{2}}\right\} \right)}{2nMT_{s}}\\
= & {\displaystyle \int}_{-\frac{f_{s}}{2M}}^{\frac{f_{s}}{2M}}\frac{1}{2}\sum_{i=1}^{M}\log^{+}\left(\nu\lambda_{i}\left({\bf F}_{\tilde{s}}^{-\frac{1}{2}}{\bf F}_{\tilde{h}}{\bf F}_{\tilde{s}}^{-\frac{1}{2}}\right)\right)\mathrm{d}f\\
= & \frac{1}{2}{\displaystyle \int}_{-\frac{f_{s}}{2M}}^{\frac{f_{s}}{2M}}\underset{i=1}{\overset{M}{\sum}}\log^{+}\left(\nu\lambda_{i}\left({\bf F}_{s}^{\ddagger}{\bf F}_{h}{\bf F}_{h}^{*}{\bf F}_{s}^{\ddagger*}\right)\right)\mathrm{d}f,
\end{align*}
 where
\begin{align*}
P & ={\displaystyle \int}_{-\frac{f_{s}}{2M}}^{\frac{f_{s}}{2M}}\sum_{i=1}^{M}\left[\nu-\frac{1}{\lambda_{i}\left({\bf F}_{\tilde{s}}^{-\frac{1}{2}}{\bf F}_{\tilde{h}}{\bf F}_{\tilde{s}}^{-\frac{1}{2}}\right)}\right]^{+}\mathrm{d}f\\
 & ={\displaystyle \int}_{-\frac{f_{s}}{2M}}^{\frac{f_{s}}{2M}}\sum_{i=1}^{M}\left[\nu-\frac{1}{\lambda_{i}{\bf F}_{s}^{\ddagger}{\bf F}_{h}{\bf F}_{h}^{*}{\bf F}_{s}^{\ddagger*}}\right]^{+}\mathrm{d}f.
\end{align*}
This completes the proof.

\section{Proof of Theorem \ref{Cor:OptimalFilterBank}\label{sec:Proof-of-Corollary-Optimal-Filter-Bank} }

Theorem \ref{Cor:OptimalFilterBank} immediately follows from the
following proposition.

\begin{prop}\label{lem_BoundsOnMSingularValues}The $k$th largest
eigenvalue $\lambda_{k}$ of the positive semidefinite matrix $\tilde{{\bf F}}_{s}{\bf F}_{h}{\bf F}_{h}^{*}\tilde{{\bf F}}_{s}^{*}$
is bounded by
\begin{equation}
0\leq\lambda_{k}\leq\lambda_{k}\left({\bf F}_{h}{\bf F}_{h}^{*}\right),\quad1\leq k\leq M.
\end{equation}
These upper bounds can be attained simultaneously by the filter (\ref{eq:optimalFilterBank}). 

\end{prop}

\begin{IEEEproof}Recall that at a given $f$, ${\bf F}_{h}$ is an
infinite diagonal matrix satisfying $\left({\bf F}_{h}\right)_{l,l}=H\left(f-\frac{lf_{s}}{M}\right)$
for all $l\in\mathbb{Z}$, and that $\tilde{{\bf F}}_{s}=\left({\bf F}_{s}{\bf F}_{s}^{*}\right)^{-\frac{1}{2}}{\bf F}_{s}$.
Hence, $\tilde{{\bf F}}_{s}{\bf F}_{h}{\bf F}_{h}^{*}\tilde{{\bf F}}_{s}^{*}$
is an $M\times M$ dimensional matrix. We observe that
\begin{equation}
\tilde{{\bf F}}_{s}\left(\tilde{{\bf F}}_{s}\right)^{*}=\left({\bf F}_{s}{\bf F}_{s}^{*}\right)^{-\frac{1}{2}}{\bf F}_{s}{\bf F}_{s}^{*}\left({\bf F}_{s}{\bf F}_{s}^{*}\right)^{-\frac{1}{2}}={\bf I},
\end{equation}
 which indicates that the rows of $\tilde{{\bf F}}_{s}$ are orthonormal.
Hence, the operator norm of $\tilde{{\bf F}}_{s}$ is no larger than
$1$, which leads to
\[
\lambda_{1}\left(\tilde{{\bf F}}_{s}{\bf F}_{h}{\bf F}_{h}^{*}\tilde{{\bf F}}_{s}^{*}\right)=\left\Vert \tilde{{\bf F}}_{s}{\bf F}_{h}\right\Vert _{2}^{2}\leq\left\Vert {\bf F}_{h}\right\Vert _{2}^{2}=\lambda_{1}\left({\bf F}_{h}{\bf F}_{h}^{*}\right).
\]

Denote by $\left\{ {\bf e}_{k},\mbox{ }k\geq1\right\} $ the standard
basis where ${\bf e}_{k}$ is a vector with a $1$ in the $k$th coordinate
and $0$ otherwise. We introduce the index set $\left\{ i_{1},i_{2},\cdots,i_{M}\right\} $
such that ${\bf {\bf e}}_{i_{k}}$ $\left(1\leq k\leq M\right)$ is
the eigenvector associated with the $k$th largest eigenvalues of
the diagonal matrix ${\bf F}_{h}{\bf F}_{h}^{*}$. 

Suppose that ${\bf v}_{k}$ is the eigenvector associated with the
$k$th largest eigenvalue $\lambda_{k}$ of $\tilde{{\bf F}}_{s}{\bf F}_{h}{\bf F}_{h}^{*}\tilde{{\bf F}}_{s}^{*}$,
and denote by $\left(\tilde{{\bf F}}_{s}\right)_{k}$ the $k$th \emph{column}
of $\tilde{{\bf F}}_{s}$. Since $\tilde{{\bf F}}_{s}{\bf F}_{h}{\bf F}_{h}^{*}\tilde{{\bf F}}_{s}^{*}$
is Hermitian positive semidefinite, its eigendecomposition yields
an orthogonal basis of eigenvectors. Observe that $\left\{ {\bf v}_{1},\cdots,{\bf v}_{k}\right\} $
spans a $k$-dimensional space and that $\left\{ \left(\tilde{{\bf F}}_{s}\right)_{j},1\leq j\leq k-1\right\} $
spans a subspace of dimension no more than $k-1$. For any $k\geq2$,
there exists $k$ scalars $a_{1},\cdots,a_{k}$ such that
\begin{equation}
\sum_{i=1}^{k}a_{i}{\bf v}_{i}\perp\left\{ \left(\tilde{{\bf F}}_{s}\right)_{i_{j}},1\leq j\leq k-1\right\} \text{ and }\sum_{i=1}^{k}a_{i}{\bf v}_{i}\neq0.\label{eq:VEigenvectorsOrthogonal}
\end{equation}
This allows us to define the following unit vector 
\begin{equation}
{\bf \tilde{v}}_{k}\overset{\Delta}{=}\sum_{i=1}^{k}\frac{a_{i}}{\sqrt{\sum_{j=1}^{k}\left|a_{j}\right|^{2}}}{\bf v}_{i},
\end{equation}
 which is orthogonal to $\left\{ \left(\tilde{{\bf F}}_{s}\right)_{j},1\leq j\leq k-1\right\} $.
We observe that
\begin{align}
\left\Vert \tilde{{\bf F}}_{s}{\bf F}_{h}{\bf F}_{h}^{*}\tilde{{\bf F}}_{s}^{*}{\bf \tilde{v}}_{k}\right\Vert _{2}^{2} & =\left\Vert \sum_{i=1}^{k}\frac{a_{i}}{\sqrt{\sum_{j=1}^{k}\left|a_{j}\right|^{2}}}\tilde{{\bf F}}_{s}{\bf F}_{h}{\bf F}_{h}^{*}\tilde{{\bf F}}_{s}^{*}{\bf v}_{i}\right\Vert _{2}^{2}\nonumber \\
 & =\left\Vert \sum_{i=1}^{k}\frac{a_{i}\lambda_{i}}{\sqrt{\sum_{j=1}^{k}\left|a_{j}\right|^{2}}}{\bf v}_{i}\right\Vert _{2}^{2}\nonumber \\
 & =\sum_{i=1}^{k}\frac{\lambda_{i}^{2}\left|a_{i}\right|^{2}}{\sum_{j=1}^{k}\left|a_{j}\right|^{2}}\geq\lambda_{k}^{2}.
\end{align}
 Define ${\bf u}_{k}:=\tilde{{\bf F}}_{s}^{*}\tilde{{\bf v}}_{k}$.
From (\ref{eq:VEigenvectorsOrthogonal}) we can see that $\left({\bf u}_{k}\right)_{i}=\left\langle \left(\tilde{{\bf F}}_{s}\right)_{i},\tilde{{\bf v}}_{i}\right\rangle =0$
holds for all $i\in\left\{ i_{1},i_{2},\cdots,i_{k-1}\right\} $.
In other words, ${\bf u}_{k}\perp\left\{ {\bf e}_{i_{1}},\cdots,{\bf e}_{i_{k-1}}\right\} $.
This further implies that
\begin{align}
\lambda_{k}^{2} & \leq\left\Vert \tilde{{\bf F}}_{s}{\bf F}_{h}{\bf F}_{h}^{*}\tilde{{\bf F}}_{s}^{*}{\bf \tilde{v}}_{k}\right\Vert _{2}^{2}\leq\left\Vert \tilde{{\bf F}}_{s}\right\Vert _{2}^{2}\left\Vert {\bf F}_{h}{\bf F}_{h}^{*}\tilde{{\bf F}}_{s}^{*}{\bf \tilde{v}}_{k}\right\Vert _{2}^{2}\nonumber \\
 & \leq\left\Vert {\bf F}_{h}{\bf F}_{h}^{*}{\bf u}_{k}\right\Vert _{2}^{2}\\
 & \leq\sup_{{\bf x}\perp\text{span}\left\{ {\bf e}_{i_{1}},\cdots,{\bf e}_{i_{k-1}}\right\} }\left\Vert {\bf F}_{h}{\bf F}_{h}^{*}{\bf x}\right\Vert _{2}^{2}\\
 & =\lambda_{k}^{2}\left({\bf F}_{h}{\bf F}_{h}^{*}\right)
\end{align}
 by observing that ${\bf F}_{h}{\bf F}_{h}^{*}$ is a diagonal matrix.

Setting
\begin{align*}
 & S_{k}\left(f-\frac{lf_{s}}{M}\right)\\
= & \begin{cases}
1, & \quad\mbox{if }\left|H\left(f-\frac{lf_{s}}{M}\right)\right|^{2}=\lambda_{k}\left({\bf F}_{h}(f){\bf F}_{h}^{*}(f)\right),\\
0, & \quad\mbox{otherwise},
\end{cases}
\end{align*}
 yields $\tilde{{\bf F}}_{s}={\bf F}_{s}$ and hence $\tilde{{\bf F}}_{s}{\bf F}_{h}{\bf F}_{h}^{*}\tilde{{\bf F}}_{s}$
is a diagonal matrix such that 
\begin{equation}
\left(\tilde{{\bf F}}_{s}{\bf F}_{h}{\bf F}_{h}^{*}\tilde{{\bf F}}_{s}^{*}\right)_{k,k}=\lambda_{k}\left({\bf F}_{h}{\bf F}_{h}^{*}\right).
\end{equation}
 Apparently, this choice of $S_{k}(f)$ allows the upper bounds
\begin{equation}
\lambda_{k}\left(\tilde{{\bf F}}_{s}{\bf F}_{h}{\bf F}_{h}^{*}\tilde{{\bf F}}_{s}^{*}\right)=\lambda_{k}\left({\bf F}_{h}{\bf F}_{h}^{*}\right),\quad\forall1\leq k\leq M
\end{equation}
to be attained simultaneously.\end{IEEEproof}

By extracting out the $M$ frequencies with the highest SNR from each
aliased set $\left\{ f-lf_{s}/M\mid l\in\mathbb{Z}\right\} $, we
achieve $\lambda_{k}=\lambda_{k}\left({\bf F}_{h}{\bf F}_{h}^{*}\right)$,
thus achieving the maximum capacity.

\section{Proof of Theorem \ref{thmPremodulatedFilterBank}\label{sec:Proof-of-Theorem-Premodulated-Filter-Bank}}

Following similar steps as in the proof of Theorem \ref{thmPerfectCSIFilterBankSingleAntenna},
we approximately convert the sampled channel into its discrete counterpart,
and calculate the capacity of the discretized channel model after
noise whitening. We note that the impulse response of the sampled
channel is no longer LTI due to the use of modulation banks. But the
periodicity assumption of the modulation sequences allows us to treat
the channel matrix as blockwise LTI, which provides a way to exploit
the properties of block-Toeplitz matrices. 

Again, we give a proof for the scenario where noise is white Gaussian
with unit spectral density. The capacity expression in the presence
of colored noise can immediately be derived by replacing $P_{i}(f)$
with $P_{i}(f)\sqrt{\mathcal{S}_{\eta}(f)}$ and $H(f)$ with $H(f)/\sqrt{\mathcal{S}_{\eta}(f)}$.

In the $i$th branch, the noise component at time $t$ is given by
\begin{align*}
 & s_{i}\left(t\right)*\left(q_{i}(t)\cdot\left(p_{i}(t)*\eta(t)\right)\right)\\
= & \int_{\tau_{1}}\mathrm{d}\tau_{1}s_{i}\left(t-\tau_{1}\right)\int_{\tau_{2}}q_{i}\left(\tau_{1}\right)p_{i}\left(\tau_{1}-\tau_{2}\right)\eta\left(\tau_{2}\right)\mathrm{d}\tau_{2}\\
= & \int_{\tau_{2}}\left({\displaystyle \int}_{\tau_{1}}s_{i}\left(t-\tau_{1}\right)q_{i}\left(\tau_{1}\right)p_{i}\left(\tau_{1}-\tau_{2}\right)\mathrm{d}\tau_{1}\right)\eta\left(\tau_{2}\right)\mathrm{d}\tau_{2}\\
= & \int_{\tau_{2}}g_{i}^{\eta}(t,\tau_{2})\eta(\tau_{2})\mathrm{d}\tau_{2},
\end{align*}
where $g_{i}^{\eta}(t,\tau_{2})\overset{\Delta}{=}\underset{\tau_{1}}{{\displaystyle \int}}s_{i}\left(t-\tau_{1}\right)q_{i}\left(\tau_{1}\right)p_{i}\left(\tau_{1}-\tau_{2}\right)\mathrm{d}\tau_{1}.$

Let $\tilde{T}_{s}=MT_{s}$. Our assumption $bT_{q}=a\tilde{T}_{s}$
immediately leads to
\begin{align*}
 & g_{i}^{\eta}\left(t+a\tilde{T}_{s},\tau+bT_{q}\right)\\
= & \int_{\tau_{1}}s_{i}\left(t+a\tilde{T}_{s}-\tau_{1}\right)q_{i}\left(\tau_{1}\right)p_{i}\left(\tau_{1}-\tau-a\tilde{T}_{s}\right)\mathrm{d}\tau_{1}\\
= & \int_{\tau_{1}}s_{i}\left(t-\tau_{1}\right)q_{i}\left(\tau_{1}+bT_{q}\right)p_{i}\left(\tau_{1}-\tau\right)\mathrm{d}\tau_{1}\\
= & \int_{\tau_{1}}s_{i}\left(t-\tau_{1}\right)q_{i}\left(\tau_{1}\right)p_{i}\left(\tau_{1}-\tau\right)\mathrm{d}\tau_{1}=g_{i}^{\eta}\left(t,\tau\right),
\end{align*}
implying that $g_{i}^{\eta}\left(t,\tau\right)$ is a block-Toeplitz
function. 

Similarly, the signal component
\[
s_{i}\left(t\right)*\left(q_{i}(t)\cdot\left(p_{i}(t)*h(t)*x(t)\right)\right)=\underset{\tau_{2}}{{\displaystyle \int}}g_{i}^{h}(t,\tau_{2})x(\tau_{2})\mathrm{d}\tau_{2},
\]
where
\[
g_{i}^{h}(t,\tau_{2})\overset{\Delta}{=}\underset{\tau_{1}}{{\displaystyle \int}}s_{i}\left(t-\tau_{1}\right)q_{i}\left(\tau_{1}\right)\underset{\tau_{3}}{{\displaystyle \int}}p_{i}\left(\tau_{1}-\tau_{2}-\tau_{3}\right)h(\tau_{3})\mathrm{d}\tau_{3}\mathrm{d}\tau_{1},
\]
which also satisfies the block-Toeplitz property $g_{i}^{h}\left(t+a\tilde{T}_{s},\tau+bT_{q}\right)=g_{i}^{h}\left(t,\tau\right)$.

Suppose that $T=n\tilde{T}_{s}$ and $\tilde{T}_{s}=k\Delta$ hold
for some integers $n$ and $k$. We can introduce two matrices ${\bf G}_{i}^{\eta}$
and ${\bf G}_{i}^{h}$ such that $\forall m\in\mathbb{Z},0\leq l<n$
\[
\begin{cases}
\left({\bf G}_{i}^{\eta}\right)_{l,m} & =g_{i}^{\eta}\left(l\tilde{T}_{s},m\Delta\right),\\
\left({\bf G}_{i}^{h}\right)_{l,m} & =g_{i}^{h}\left(l\tilde{T}_{s},m\Delta\right).
\end{cases}
\]
 Setting ${\bf y}_{i}^{n}=\left[y_{i}[0],y_{i}[1],\cdots,y_{i}[n-1]\right]^{T}$
leads to similar discretized approximation as in the proof of Theorem
\ref{thmPerfectCSIPrefilteredSamplerRigorous}:
\begin{equation}
{\bf y}_{i}^{n}={\bf G}_{i}^{h}{\bf x}^{n}+{\bf G}_{i}^{\eta}{\bf \eta}.
\end{equation}
Here, ${\bf \eta}$ is a i.i.d. zero-mean Gaussian vector where each
entry is of variance $1/\Delta$.

Hence, ${\bf G}_{i}^{h}$ and ${\bf G}_{i}^{\eta}$ are block Toeplitz
matrices satisfying $\left({\bf G}_{i}^{h}\right)_{l+a,m+ak}=\left({\bf G}_{i}^{h}\right)_{l,m}$
and $\left({\bf G}_{i}^{\eta}\right)_{l+a,m+ak}=\left({\bf G}_{i}^{\eta}\right)_{l,m}$.
Using the same definition of ${\bf x}^{n}$ and ${\bf \eta}$ as in
Appendix \ref{sec:Proof-of-Theorem-PerfectCSIFilterBank}, we can
express the system equation as
\begin{equation}
{\bf y}^{n}=\left[\begin{array}{c}
{\bf G}_{1}^{h}\\
{\bf G}_{2}^{h}\\
\vdots\\
{\bf G}_{M}^{h}
\end{array}\right]{\bf x}^{n}+\left[\begin{array}{c}
{\bf G}_{1}^{\eta}\\
{\bf G}_{2}^{\eta}\\
\vdots\\
{\bf G}_{M}^{\eta}
\end{array}\right]{\bf \eta}.
\end{equation}
Whitening the noise component yields
\begin{equation}
\tilde{{\bf y}}_{n}=\left(\left[\begin{array}{c}
{\bf G}_{1}^{\eta}\\
{\bf G}_{2}^{\eta}\\
\vdots\\
{\bf G}_{M}^{\eta}
\end{array}\right]\left[\begin{array}{c}
{\bf G}_{1}^{\eta}\\
{\bf G}_{2}^{\eta}\\
\vdots\\
{\bf G}_{M}^{\eta}
\end{array}\right]^{*}\right)^{-\frac{1}{2}}\left[\begin{array}{c}
{\bf G}_{1}^{h}\\
{\bf G}_{2}^{h}\\
\vdots\\
{\bf G}_{M}^{h}
\end{array}\right]{\bf x}_{n}+\tilde{\eta},
\end{equation}
 where $\tilde{{\bf \eta}}$ is i.i.d. Gaussian noise with variance
$1/\Delta$.

In order to calculate the capacity limit, we need to investigate the
Fourier symbols associated with these block Toeplitz matrices.

$ $\begin{lem}\label{lem: FourierSymbolModulationBank}At a given
frequency $f$, the Fourier symbol with respect to ${\bf G}_{\alpha}^{\eta}{\bf G}_{\beta}^{\eta}$
is given by $ak{\bf F}_{\alpha}^{\eta}{\bf F}_{\alpha}^{p}{\bf F}_{\beta}^{p*}{\bf F}_{\beta}^{\eta*}$,
and the Fourier symbol with respect to ${\bf G}_{\alpha}^{h}{\bf G}_{\beta}^{h}$
is given by $ak{\bf F}_{\alpha}^{\eta}{\bf F}_{\alpha}^{p}{\bf F}^{h}{\bf F}^{h*}{\bf F}_{\beta}^{p*}{\bf F}_{\beta}^{\eta*}$.
Here for any $\left(l,v\right)$ such that $v\in\mathbb{Z}$ and $1\leq l\leq a$,
we have
\begin{align*}
\left({\bf F}_{\alpha}^{\eta}\right)_{l,v} & =\sum_{u}c_{\alpha}^{u}S_{\alpha}\left(-f+uf_{q}+v\frac{f_{q}}{b}\right)\cdot\\
 & \quad\quad\quad\quad\exp\left(-j2\pi l\frac{T_{s}}{M}\left(f-uf_{q}-v\frac{f_{q}}{b}\right)\right).
\end{align*}
Also, ${\bf F}_{\alpha}^{p}$ and ${\bf F}^{h}$ are infinite diagonal
matrices such that for all $l\in\mathbb{Z}$
\[
\begin{cases}
\left({\bf F}_{\alpha}^{p}\right)_{l,l} & =P_{\alpha}\left(-f+l\frac{f_{q}}{b}\right),\\
\left({\bf F}^{h}\right)_{l,l} & =H\left(-f+l\frac{f_{q}}{b}\right).
\end{cases}
\]
\end{lem}

\begin{IEEEproof}See Appendix \ref{sec:Proof-of-Lemma-Fourier-Symbol-Modulation}.\end{IEEEproof}

Define ${\bf G}^{\eta}$ such that its $\left(\alpha,\beta\right)$
subblock is ${\bf G}_{\alpha}^{\eta}{\bf G}_{\beta}^{\eta*}$, and
${\bf G}^{h}$ such that its $\left(\alpha,\beta\right)$ subblock
is ${\bf G}_{\alpha}^{h}{\bf G}_{\beta}^{h*}$. Proceeding similarly
as in the proof of Theorem \ref{thmPerfectCSIFilterBankSingleAntenna},
we obtain
\[
\mathcal{F}\left({\bf G}^{\eta}\right)=ak{\bf F}^{\eta}{\bf F}^{\eta*}\quad\text{and}\quad\mathcal{F}\left({\bf G}^{h}\right)=ak{\bf F}^{\eta}{\bf F}^{h}{\bf F}^{h*}{\bf F}^{\eta*},
\]
where ${\bf F}^{\eta}$ contain $M\times1$ submatrices. The $\left(\alpha,1\right)$
submatrix of ${\bf F}^{\eta}$ is given by ${\bf F}_{\alpha}^{\eta}{\bf F}_{\alpha}^{p}$. 

Denote ${\bf F}^{\eta\ddagger}\overset{\Delta}{=}\left({\bf F}^{\eta}{\bf F}^{\eta*}\right)^{-\frac{1}{2}}{\bf F}^{\eta}$.
For any continuous function $g(x)$, \cite[Theorem 5.4]{Tilli98}
implies that
\begin{align*}
 & \lim_{n\rightarrow\infty}\frac{1}{naM}\sum_{i=1}^{naM}g\left(\lambda_{i}\left\{ \left({\bf G}^{\eta}\right)^{-\frac{1}{2}}{\bf G}^{h}\left({\bf G}^{\eta}\right)^{-\frac{1}{2}}\right\} \right)\\
= & {\displaystyle \int}_{-\frac{\tilde{f}_{s}}{2a}}^{\frac{\tilde{f}_{s}}{2a}}\sum_{i=1}^{aM}g\left(\lambda_{i}\left({\bf F}^{\eta\ddagger}{\bf F}^{h}{\bf F}^{h*}{\bf F}^{\eta\ddagger*}\right)\right)\mathrm{d}f.
\end{align*}
Then capacity of parallel channels, achieved via water-filling power
allocation, yields
\begin{align*}
C(f_{s})= & \lim_{n\rightarrow\infty}\sum_{i=1}^{naM}\frac{\log^{+}\left(\nu\lambda_{i}\left\{ \left({\bf G}^{\eta}\right)^{-\frac{1}{2}}{\bf G}^{h}\left({\bf G}^{\eta}\right)^{-\frac{1}{2}}\right\} \right)}{naM}\\
= & {\displaystyle \int}_{-\frac{\tilde{f}_{s}}{2a}}^{\frac{\tilde{f}_{s}}{2a}}\frac{1}{2}\sum_{i=1}^{aM}\log^{+}\left(\nu\lambda_{i}\left({\bf F}^{\eta\ddagger}{\bf F}^{h}{\bf F}^{h*}{\bf F}^{\eta\ddagger*}\right)\right)\mathrm{d}f,
\end{align*}
 where the water level $\nu$ can be computed through the following
parametric equation
\begin{align*}
P & =\lim_{n\rightarrow\infty}\frac{1}{naM}\sum_{i=1}^{naM}\left[\nu-\frac{1}{\lambda_{i}\left\{ \left({\bf G}^{\eta}\right)^{-\frac{1}{2}}{\bf G}^{h}\left({\bf G}^{\eta}\right)^{-\frac{1}{2}}\right\} }\right]^{+}\\
 & ={\displaystyle \int}_{-\frac{\tilde{f}_{s}}{2a}}^{\frac{\tilde{f}_{s}}{2a}}\sum_{i=1}^{aM}\left[\nu-\frac{1}{\lambda_{i}\left\{ \left({\bf G}^{\eta}\right)^{-\frac{1}{2}}{\bf G}^{h}\left({\bf G}^{\eta}\right)^{-\frac{1}{2}}\right\} }\right]^{+}\mathrm{d}f.
\end{align*}

\section{Proof of Proposition \ref{lem-optimal-filter-bank-sampling-theoretic}\label{sec:Proof-of-Lemma-optimal-filter-bank-sampling-theoretic}}

Denote by $y^{k}(t)$ the analog signal after passing through the
$k^{\text{th}}$ prefilter prior to ideal sampling. When both the
input signal $x(t)$ and the noise $\eta(t)$ are Gaussian, the MMSE
estimator of $x(t)$ from samples $\left\{ y^{k}[n]\mid1\leq k\leq M\right\} $
is linear. Recall that $\tilde{T}_{s}=MT_{s}$ and $\tilde{f}_{s}=f_{s}/M$.
A linear estimator of $x(t)$ from ${\bf y}[n]$ can be given as
\begin{equation}
\hat{x}(t)=\sum_{k\in\mathbb{Z}}{\bf {\bf g}}^{T}(t-k\tilde{T}_{s})\cdot{\bf y}(k\tilde{T}_{s}),\label{eq:xLinearEstimator}
\end{equation}
where we use the vector form ${\bf g}(t)=[g^{1}(t),\cdots,g^{M}(t)]^{T}$
and ${\bf y}(t)=[y^{1}(t),\cdots,y^{M}(t)]^{T}$ for notational simplicity.
Here, $g^{l}(t)$ denotes the interpolation function operating upon
the samples in the $l^{\text{th}}$ branch. We propose to find the
optimal estimator ${\bf g}(t)$ that minimizes the mean square estimation
error $\mathbb{E}\left(\left|x(t)-\hat{x}(t)\right|^{2}\right)$ for
some $t$.

From the orthogonality principle, the MMSE estimate $\hat{x}(t)$
obeys
\begin{equation}
\mathbb{E}\left(x(t){\bf y}^{*}(l\tilde{T}_{s})\right)=\mathbb{E}\left(\hat{x}(t){\bf y}^{*}(l\tilde{T}_{s})\right),\quad\forall l\in\mathbb{Z}.\label{eq:OrthogonalityPrinciple}
\end{equation}
Since $x(t)$ and $\eta(t)$ are both stationary Gaussian processes,
we can define ${\bf R}_{XY}(\tau):=\mathbb{E}\left(x(t){\bf y}^{*}(t-\tau)\right)$
to be the cross correlation function between $x(t)$ and ${\bf y}(t)$,
and ${\bf R}_{Y}(\tau):=\mathbb{E}\left({\bf y}(t){\bf y}^{*}(t-\tau)\right)$
the autocorrelation function of ${\bf y}(t)$. Plugging (\ref{eq:xLinearEstimator})
into (\ref{eq:OrthogonalityPrinciple}) leads to the following relation
\begin{align*}
{\bf R}_{XY}\left(t-l\tilde{T}_{s}\right) & =\sum_{k\in\mathbb{Z}}{\bf g}^{T}\left(t-k\tilde{T}_{s}\right){\bf R}_{Y}\left(k\tilde{T}_{s}-l\tilde{T}_{s}\right).
\end{align*}
Replacing $t$ by $t+l\tilde{T}_{s}$ , we can equivalently express
it as
\begin{align}
{\bf R}_{XY}(t) & =\sum_{k\in\mathbb{Z}}{\bf g}^{T}\left(t+l\tilde{T}_{s}-k\tilde{T}_{s}\right){\bf R}_{Y}\left(k\tilde{T}_{s}-l\tilde{T}_{s}\right)\nonumber \\
 & =\sum_{l\in\mathbb{Z}}{\bf g}^{T}\left(t-l\tilde{T}_{s}\right){\bf R}_{Y}\left(l\tilde{T}_{s}\right),\label{eq:CrossCorrelationRelation}
\end{align}
which is equivalent to the convolution of ${\bf g}(t)$ and ${\bf R}_{Y}(t)\cdot\sum_{l\in\mathbb{Z}}\delta\left(t-l\tilde{T}_{s}\right)$.

Let $\mathcal{F}(\cdot)$ denote Fourier transform operator. Define
the cross spectral density ${\bf S}_{XY}(f):=\mathcal{F}\left({\bf R}_{XY}(t)\right)$
and ${\bf S}_{Y}(f)=\mathcal{F}\left({\bf R}_{Y}\left(t\right)\right)$.
By taking the Fourier transform on both sides of (\ref{eq:CrossCorrelationRelation})
, we have
\[
{\bf S}_{XY}(f)={\bf G}(f)\mathcal{F}\left({\bf R}_{Y}(\tau)\sum_{l\in\mathbb{Z}}\delta(\tau-l\tilde{T}_{s})\right),
\]
which immediately yields that $\forall f\in\left[-\tilde{f}_{s}/2,\tilde{f}_{s}/2\right]$
\begin{align*}
{\bf G}(f) & ={\bf S}_{XY}(f)\left[\mathcal{F}\left({\bf R}_{Y}(\tau)\sum_{l\in\mathbb{Z}}\delta(\tau-l\tilde{T}_{s})\right)\right]^{-1}\\
 & ={\bf S}_{XY}(f)\left(\sum_{l\in\mathbb{Z}}{\bf S}_{Y}\left(f-l\tilde{f}_{s}\right)\right)^{-1}.
\end{align*}
 Since the noise $\eta(t)$ is independent of $x(t)$, the cross correlation
function ${\bf R}_{XY}(t)$ is
\begin{align*}
{\bf R}_{XY}(\tau) & =\mathbb{E}\left(x(t+\tau)\cdot\right.\\
 & \quad\quad\left.\left[\left(s_{1}*h*x\right)^{*}(t),\cdots,\left(s_{M}*h*x\right)^{*}(t)\right]\right).
\end{align*}
 which allows the cross spectral density to be derived as 
\begin{align}
{\bf S}_{XY}(f) & =H^{*}(f)\mathcal{S}_{X}(f)\left[S_{1}^{*}(f),\cdots,S_{M}^{*}(f)\right].
\end{align}
 Additionally, the spectral density of ${\bf y}(t)$ can be given
as the following $M\times M$ matrix 
\begin{equation}
{\bf S}_{Y}(f)=\left(\left|H(f)\right|^{2}\mathcal{S}{}_{X}(f)+\mathcal{S}_{\eta}(f)\right){\bf S}(f){\bf S}^{*}(f),
\end{equation}
 with $\mathcal{S}_{\eta}(f)$ denoting the spectral density of the
noise $\eta(t)$, and ${\bf S}(f)=\left[S_{1}(f),\cdots,S_{m}(f)\right]^{T}$.

Define 
\begin{align*}
{\bf K}(f): & =\sum_{l\in\mathbb{Z}}\left(\left|H(f-lf_{s})\right|^{2}\mathcal{S}_{X}(f-lf_{s})+\mathcal{N}(f-lf_{s})\right)\\
 & \quad\quad\quad\quad{\bf S}(f-lf_{s}){\bf S}^{*}(f-lf_{s}).
\end{align*}
The Wiener-Hopf linear reconstruction filter can now be written as
\[
{\bf G}(f)=H^{*}(f)\mathcal{S}_{X}(f){\bf S}^{*}(f){\bf K}^{-1}(f)
\]
Define $R_{X}(\tau)=\mathbb{E}\left(x(t)x^{*}(t-\tau)\right)$. Since
$\int_{-\infty}^{\infty}\mathcal{S}_{X}(f)\mathrm{d}f=R_{X}(0)$,
the resulting MSE is
\begin{align*}
\xi(t) & =\mathbb{E}\left(\left|x(t)\right|^{2}\right)-\mathbb{E}\left(\left|\hat{x}(t)\right|^{2}\right)\\
 & =\mathbb{E}\left(\left|x(t)\right|^{2}\right)-\mathbb{E}\left(x(t)\hat{x}^{*}(t)\right)\\
 & =R_{X}(0)-\mathbb{E}\left(x(t)\left(\sum_{l\in\mathbb{Z}}{\bf g}^{T}(t-lT_{s}){\bf y}(lT_{s})\right)^{*}\right)\\
 & =R_{X}\left(0\right)-\sum_{l\in\mathbb{Z}}{\bf R}_{XY}(t-lT_{s}){\bf g}(t-lT_{s}).
\end{align*}
Since $\mathcal{F}\left({\bf g}(-t)\right)=\left({\bf G}^{*}(f)\right)^{T}$
and ${\bf S}_{XY}=H^{*}(f)\mathcal{S}_{X}(f){\bf S}^{*}(f)$, Parseval's
identity implies that
\begin{align*}
\xi(t) & =\int_{-\infty}^{\infty}\left[\mathcal{S}_{X}(f)-{\bf G}^{*}(f){\bf S}_{XY}^{T}\right]\text{d}f\\
 & =\int_{-\infty}^{\infty}\left[\mathcal{S}_{X}(f)-\left|H(f)\mathcal{S}_{X}(f)\right|^{2}{\bf S}^{*}(f){\bf K}^{-1}(f){\bf S}(f)\right]\mathrm{d}f\\
 & ={\displaystyle \int}_{-\tilde{f}_{s}/2}^{\tilde{f}_{s}/2}\left[\sum_{l=-\infty}^{\infty}\mathcal{S}_{X}(f-l\tilde{f}_{s})-\tilde{T}_{s}{\bf V}_{\zeta}^{T}(f,\tilde{f}_{s})\cdot{\bf 1}\right]\mathrm{d}f.
\end{align*}
Suppose that we impose power constraints $\sum_{l\in\mathbb{Z}}\mathcal{S}_{X}(f-l\tilde{f}_{s})=P(f)$,
and define $\zeta(f):=\left|H(f)\mathcal{S}_{X}(f)\right|^{2}{\bf S}^{*}(f){\bf K}^{-1}(f){\bf S}(f)$.
For a given input process $x(t)$, the problem of finding the optimal
prefilter ${\bf S}(f)$ that minimizes MSE then becomes
\[
\underset{\left\{ S(f-lf_{s}),l\in\mathbb{Z}\right\} }{\mbox{maximize}}\quad{\bf V}_{\zeta}^{T}(f,\tilde{f}_{s})\cdot{\bf 1},
\]
 where the objective function can be alternatively rewritten in matrix
form
\begin{equation}
\mbox{trace}\left\{ {\bf F}_{X}^{\frac{1}{2}}{\bf F}_{h}^{*}{\bf F}_{s}^{*}\left({\bf F}_{s}\left({\bf F}_{h}{\bf F}_{h}^{*}+{\bf F}_{\eta}\right){\bf F}_{s}^{*}\right)^{-1}{\bf F}_{s}{\bf F}_{h}{\bf F}_{X}^{\frac{1}{2}}\right\} 
\end{equation}
 Here ${\bf F}_{X}$ and ${\bf F}_{\eta}$ are diagonal matrices such
that $\left({\bf F}_{X}\right)_{l,l}=\mathcal{S}_{X}(f-lf_{s})$ and
$\left({\bf F}_{\eta}\right)_{l,l}=\mathcal{S}_{\eta}(f+kf_{s})$.
We observe that
\begin{align}
 & \mbox{trace}\left\{ {\bf F}_{X}^{\frac{1}{2}}{\bf F}_{h}^{*}{\bf F}_{s}^{*}\left({\bf F}_{s}\left({\bf F}_{h}{\bf F}_{h}^{*}+{\bf F}_{\eta}\right){\bf F}_{s}^{*}\right)^{-1}{\bf F}_{s}{\bf F}_{h}{\bf F}_{X}^{\frac{1}{2}}\right\} \nonumber \\
= & \mbox{trace}\left\{ \left({\bf F}_{s}\left({\bf F}_{h}{\bf F}_{h}^{*}+{\bf F}_{\eta}\right){\bf F}_{s}^{*}\right)^{-1}{\bf F}_{s}{\bf F}_{h}{\bf F}_{X}{\bf F}_{h}^{*}{\bf F}_{s}^{*}\right\} \nonumber \\
\overset{(\text{a})}{=} & \mbox{trace}\left\{ \left({\bf Y}{\bf Y}^{*}\right)^{-1}{\bf Y}\left({\bf F}_{h}{\bf F}_{h}^{*}+{\bf F}_{\eta}\right)^{-\frac{1}{2}}{\bf F}_{h}{\bf F}_{X}{\bf F}_{h}^{*}\right.\nonumber \\
 & \quad\quad\quad\quad\quad\left.\left({\bf F}_{h}{\bf F}_{h}^{*}+{\bf F}_{\eta}\right)^{-\frac{1}{2}}{\bf Y}^{*}\right\} \nonumber \\
\overset{(\text{b})}{=} & \mbox{trace}\left\{ \left({\bf F}_{h}{\bf F}_{h}^{*}+{\bf F}_{\eta}\right)^{-1}{\bf F}_{h}{\bf F}_{X}{\bf F}_{h}^{*}{\bf Y}^{*}\left({\bf Y}{\bf Y}^{*}\right)^{-1}{\bf Y}\right\} \nonumber \\
\overset{(\text{c})}{=} & \mbox{trace}\left\{ \left({\bf F}_{h}{\bf F}_{h}^{*}+{\bf F}_{\eta}\right)^{-1}{\bf F}_{h}{\bf F}_{X}{\bf F}_{h}^{*}\tilde{{\bf Y}}^{*}\tilde{{\bf Y}}\right\} \nonumber \\
\overset{(\text{d})}{\leq} & \sup_{{\bf Z}\cdot{\bf Z}^{*}={\bf I}_{M}}\mbox{trace}\left\{ {\bf Z}\left({\bf F}_{h}{\bf F}_{h}^{*}+{\bf F}_{\eta}\right)^{-1}{\bf F}_{h}{\bf F}_{X}{\bf F}_{h}^{*}{\bf Z}^{*}\right\} \nonumber \\
= & \sum_{i=1}^{M}\lambda_{i}({\bf D}),
\end{align}
where (a) follows by introducing ${\bf Y}:={\bf F}_{s}\left({\bf F}_{h}{\bf F}_{h}^{*}+{\bf F}_{\eta}\right)^{\frac{1}{2}}$,
(b) follows from the fact that ${\bf F}_{h}$, ${\bf F}_{X}$, ${\bf F}_{\eta}$
are all diagonal matrices, (c) follows by introducing $\tilde{{\bf Y}}=\left({\bf Y}{\bf Y}^{*}\right)^{-\frac{1}{2}}{\bf Y}$,
and $ $(d) follows by observing that $\tilde{{\bf Y}}\tilde{{\bf Y}}^{*}=\left({\bf Y}{\bf Y}^{*}\right)^{-\frac{1}{2}}{\bf Y}{\bf Y}^{*}\left({\bf Y}{\bf Y}^{*}\right)^{-\frac{1}{2}}={\bf I}$.
Here, ${\bf D}$ is an infinite diagonal matrix such that ${\bf D}_{l,l}=\frac{\left|H(f-lf_{s})\right|^{2}\mathcal{S}_{X}(f-lf_{s})}{\left|H(f-lf_{s})\right|^{2}\mathcal{S}_{X}(f-lf_{s})+\mathcal{S}_{\eta}(f-lf_{s})}$.
In other words, the upper bound is the sum of the $M$ largest ${\bf D}_{i,i}$
which are associated with $M$ frequency points of highest SNR $\frac{\left|H(f+lf_{s})\right|^{2}\mathcal{S}_{X}(f+lf_{s})}{\mathcal{S}_{\eta}(f+lf_{s})}$.

Therefore, when restricted to the set of all permutations of $\left\{ \mathcal{S}_{X}(f),\mathcal{S}_{X}(f\pm f_{s}),\cdots\right\} $,
the minimum MSE is achieved when assigning the $M$ largest $\mathcal{S}_{X}(f+lf_{s})$
to $M$ branches with the largest SNR. In this case, the corresponding
optimal filter can be chosen such that
\begin{equation}
S_{k}(f-lf_{s})=\begin{cases}
1, & \quad\mbox{if }l=\hat{k}\\
0, & \quad\mbox{otherwise.}
\end{cases}
\end{equation}
 where $\hat{k}$ is the index of the $k^{\text{th}}$ largest element
in $\left\{ \left|H(f-lf_{s})\right|^{2}/\mathcal{S}_{\eta}(f-lf_{s}):l\in\mathbb{Z}\right\} $.

\section{Proofs of Auxiliary Lemmas \label{sec:Proofs-of-Auxiliary-Lemmas}}

\subsection{Proof of Lemma \ref{lemmaAsymptoticEquivalenceSH}\label{sec:Proof-of-Lemma-Asymptotic-SH}}

For any $i\leq j$, we have
\begin{align}
 & \left|\left(\tilde{{\bf H}}^{n}\tilde{{\bf H}}^{n*}-\hat{{\bf H}}^{n}\right)_{ij}\right|\nonumber \\
\leq & \left|\sum_{t=-\infty}^{-j}\tilde{{\bf h}}_{j-i+t}\tilde{{\bf h}}_{t}^{*}\right|+\left|\sum_{t=n-j+1}^{\infty}\tilde{{\bf h}}_{j-i+t}\tilde{{\bf h}}_{t}^{*}\right|.\label{eq:ResidualTerms}
\end{align}
 Since $h(t)$ is absolutely summable and Riemann integrable, for
sufficiently small $\Delta$, there exists a constant $c$ such that
$\sum_{i=-\infty}^{\infty}\left\Vert \tilde{{\bf h}}_{i}\right\Vert _{1}\leq c$.
In the following analysis, we define ${\bf R}^{1}$ and ${\bf R}^{2}$
to capture the two residual terms respectively, i.e.
\[
{\bf R}_{ij}^{1}=\sum_{t=-\infty}^{-j}\tilde{{\bf h}}_{j-i+t}\tilde{{\bf h}}_{t}^{*},\quad\text{and}\quad{\bf R}_{ij}^{2}=\sum_{t=n-j+1}^{\infty}\tilde{{\bf h}}_{j-i+t}\tilde{{\bf h}}_{t}^{*}.
\]
In order to prove that $\tilde{{\bf H}}^{n}\tilde{{\bf H}}^{n*}\sim\hat{{\bf H}}^{n}$,
we need to prove (1) $\lim_{n\rightarrow\infty}\frac{1}{n}\left\Vert \tilde{{\bf H}}^{n}\tilde{{\bf H}}^{n*}-\hat{{\bf H}}^{n}\right\Vert _{\text{F}}^{2}=0$,
or equivalently, $\lim_{n\rightarrow\infty}\frac{1}{n}\left\Vert {\bf R}^{2}\right\Vert _{\text{F}}^{2}=0$
and $\lim_{n\rightarrow\infty}\frac{1}{n}\left\Vert {\bf R}^{1}\right\Vert _{\text{F}}^{2}=0$;
(2) the $\ell_{2}$ norms of both $\tilde{{\bf H}}^{n}\tilde{{\bf H}}^{n*}$
and $\hat{{\bf H}}^{n}$ are uniformly bounded, i.e. $\exists M_{\text{u}}$
such that $\left\Vert \tilde{{\bf H}}^{n}\tilde{{\bf H}}^{n*}\right\Vert _{2}\leq M_{\text{u}}<\infty$
and $\left\Vert \hat{{\bf H}}^{n}\right\Vert _{2}\leq M_{\text{u}}<\infty$
for all $n$. 

(1) We first prove that $\lim_{n\rightarrow\infty}\frac{1}{n}\left\Vert \tilde{{\bf H}}^{n}\tilde{{\bf H}}^{n*}-\hat{{\bf H}}^{n}\right\Vert _{\text{F}}^{2}=0$.
By our assumptions, we have $h(t)=o\left(t^{-\epsilon}\right)$ for
some $\epsilon>1$. Since $s(t)$ is absolutely integrable, $\tilde{h}(t)=o\left(t^{-\epsilon}\right)$
also holds. Without loss of generality, we suppose that $j\geq i$.

(a) if $i\geq n^{\frac{1}{2\epsilon}}$, by the assumption $\tilde{h}(t)=o\left(\frac{1}{t^{\epsilon}}\right)$
for some $\epsilon>1$, one has
\begin{align}
\left|{\bf R}_{ij}^{1}\right|\leq & \sum_{t=-\infty}^{-j}\left\Vert \tilde{{\bf h}}_{j-i+t}\right\Vert _{1}\left\Vert \tilde{{\bf h}}_{t}\right\Vert _{\infty}\nonumber \\
\leq & \left(\max_{\tau\geq n^{\frac{1}{2\epsilon}}}\left\Vert \tilde{{\bf h}}_{-\tau}\right\Vert _{1}\right)\sum_{t=-\infty}^{-j}\left\Vert \tilde{{\bf h}}_{t}\right\Vert _{\infty}\nonumber \\
\leq & \left(\max_{\tau\geq n^{\frac{1}{2\epsilon}}}\left\Vert \tilde{{\bf h}}_{-\tau}\right\Vert _{1}\right)\sum_{t=-\infty}^{-j}\left\Vert \tilde{{\bf h}}_{t}\right\Vert _{1}\leq\mbox{ }c\max_{\tau\geq n^{\frac{1}{2\epsilon}}}\left\Vert \tilde{{\bf h}}_{-\tau}\right\Vert _{1}\nonumber \\
= & \mbox{ }kc\cdot o\left(\frac{1}{\sqrt{n}}\right)=o\left(\frac{1}{\sqrt{n}}\right).\label{eq:Rij_1_a}
\end{align}

(b) if $j\geq n^{\frac{1}{2\epsilon}}$, 
\begin{align}
\left|{\bf R}_{ij}^{1}\right|\leq & \sum_{t=-\infty}^{-j}\left\Vert \tilde{{\bf h}}_{j-i+t}\right\Vert _{1}\left\Vert \tilde{{\bf h}}_{t}\right\Vert _{\infty}\nonumber \\
\leq & \left(\sum_{t=-\infty}^{-j}\left\Vert \tilde{{\bf h}}_{j-i+t}\right\Vert _{1}\right)\max_{\tau\leq-j}\left\Vert \tilde{{\bf h}}_{\tau}\right\Vert _{\infty}\nonumber \\
\leq & \mbox{ }c\max_{\tau\geq n^{\frac{1}{2\epsilon}}}\left\Vert \tilde{{\bf h}}_{-\tau}\right\Vert _{\infty}\nonumber \\
= & \mbox{ }c\cdot o\left(\frac{1}{\sqrt{n}}\right)=o\left(\frac{1}{\sqrt{n}}\right).\label{eq:Rij_1_b}
\end{align}

(c) if $j<n^{\frac{1}{2\epsilon}}$ and $i<n^{\frac{1}{2\epsilon}}$,
we have
\begin{align}
\left|{\bf R}_{ij}^{1}\right|^{2} & \leq\left(\sum_{t=-\infty}^{\infty}\left\Vert \tilde{{\bf h}}_{j-i+t}\right\Vert _{1}\left\Vert \tilde{{\bf h}}_{t}\right\Vert _{\infty}\right)^{2}\nonumber \\
 & \leq\left(\sum_{t=-\infty}^{\infty}\left\Vert \tilde{{\bf h}}_{j-i+t}\right\Vert _{1}\right)^{2}\left(\max_{t}\left\Vert \tilde{{\bf h}}_{t}\right\Vert _{\infty}\right)^{2}\nonumber \\
 & \leq\left(\sum_{t=-\infty}^{\infty}\left\Vert \tilde{{\bf h}}_{j-i+t}\right\Vert _{1}\right)^{2}\left(\sum_{t=-\infty}^{\infty}\left\Vert \tilde{{\bf h}}_{t}\right\Vert _{1}\right)^{2}\nonumber \\
 & \leq c^{4}.\label{eq:Rij_1_c}
\end{align}
 By combining inequality (\ref{eq:Rij_1_a}), (\ref{eq:Rij_1_b})
and (\ref{eq:Rij_1_c}), we can obtain
\begin{align*}
 & \lim_{n\rightarrow\infty}\frac{1}{n}\left\Vert {\bf R}^{1}\right\Vert _{\text{F}}^{2}\\
= & \lim_{n\rightarrow\infty}\frac{1}{n}\left(\sum_{i,j<n^{\frac{1}{2\epsilon}}}\left|{\bf R}_{ij}^{1}\right|^{2}+\sum_{i\geq n^{\frac{1}{2\epsilon}}\text{ or }j\geq n^{\frac{1}{2\epsilon}}}\left|{\bf R}_{ij}^{1}\right|^{2}\right)\\
\leq & \lim_{n\rightarrow\infty}\frac{1}{n}\left[n^{\frac{1}{\epsilon}}\max_{i,j<n^{\frac{1}{2\epsilon}}}\left|{\bf R}_{ij}^{1}\right|^{2}+2n^{1+\frac{1}{2\epsilon}}\max_{i\text{ or }j\geq n^{\frac{1}{2\epsilon}}}\left|{\bf R}_{ij}^{1}\right|^{2}\right]\\
= & \lim_{n\rightarrow\infty}\frac{1}{n}\left[n^{\frac{1}{\epsilon}}c^{4}+2n^{1+\frac{1}{2\epsilon}}o\left(\frac{1}{n}\right)\right]=0
\end{align*}
Similarly, we can show that
\[
\lim_{n\rightarrow\infty}\frac{1}{n}\left\Vert {\bf R}^{2}\right\Vert _{\text{F}}^{2}=0,
\]
which immediately implies that
\[
\lim_{n\rightarrow\infty}\frac{1}{n}\left\Vert \hat{{\bf H}}^{n}-\tilde{{\bf H}}^{n}\tilde{{\bf H}}^{n*}\right\Vert _{\text{F}}^{2}=0.
\]

(2) We now proceed to show that $\left\Vert \tilde{{\bf H}}^{n}\tilde{{\bf H}}^{n*}\right\Vert _{2}$
and $\left\Vert \hat{{\bf H}}^{n}\right\Vert _{2}$ are uniformly
bounded. Since $\hat{{\bf H}}^{n}$ is a Toeplitz matrix, applying
\cite[Lemma 6]{Gray06} and \cite[Section 4.1]{Gray06} yields
\begin{align*}
\left\Vert \hat{{\bf H}}^{n}\right\Vert _{2} & \leq2\sum_{i=0}^{\infty}\sum_{t=-\infty}^{\infty}\left|\tilde{{\bf h}}_{i+t}\tilde{{\bf h}}_{t}^{*}\right|\\
 & \leq2\sum_{t=-\infty}^{+\infty}\left\Vert \tilde{{\bf h}}_{t}\right\Vert _{\infty}\sum_{i=0}^{\infty}\left\Vert \tilde{{\bf h}}_{i+t}\right\Vert _{1}\leq2c^{2}.
\end{align*}
 Additionally, since $\tilde{{\bf H}}^{n}$ is a block Toeplitz matrix,
\cite[Corollary 4.2]{Tilli98} allows us to bound the norm as
\begin{align*}
\left\Vert \tilde{{\bf H}}^{n}\tilde{{\bf H}}^{n*}\right\Vert _{2} & =\left\Vert \hat{{\bf H}}^{n}\right\Vert _{2}^{2}\leq\left\Vert {\bf F}_{\tilde{h}}(\omega)\right\Vert _{\infty}^{2}=\sup_{\omega}\sum_{i=0}^{k-1}\left|{\bf F}_{\tilde{h},i}(\omega)\right|^{2}\\
 & \leq\sum_{j=0}^{\infty}\left(\sum_{i=0}^{k-1}\left|\left(\tilde{{\bf h}}_{j}\right)_{i}\right|\right)^{2}\leq\left(\sum_{j=0}^{\infty}\left\Vert \tilde{{\bf h}}_{j}\right\Vert _{1}\right)^{2}\leq c^{2}.
\end{align*}
Hence, by definition of asymptotic equivalence, we have $\hat{{\bf H}}^{n}\sim\tilde{{\bf H}}^{n}\tilde{{\bf H}}^{n*}$.

\subsection{Proof of Lemma \ref{lem-S-Inverse-Asymptotic}\label{sec:Proof-of-Lemma-S-Inverse-Asymptotic}}

We know that ${\bf S}^{n}{\bf S}^{n*}=\hat{{\bf S}}^{n}$, hence,
${\bf C}^{n}\sim\hat{{\bf S}}^{n}={\bf S}^{n}{\bf S}^{n*}$. Recall
that $\left(\hat{{\bf S}}^{n}\right)_{1i}=\sum_{t=-\infty}^{\infty}{\bf s}_{i-1+t}{\bf s}_{t}^{*}$.
For a given $k$, the Fourier series related to $\left\{ {\bf C}^{n}\right\} $
can be given as
\begin{equation}
F_{c}^{k}(\omega)=\sum_{i=-\infty}^{\infty}\left(\sum_{t=-\infty}^{\infty}{\bf s}_{i+t}{\bf s}_{t}^{*}\right)\exp(ji\omega).
\end{equation}
By Lemma \ref{lemmaMultiplicationInverse}, in order to show $\left({\bf C}^{n}\right)^{-1}\sim\left({\bf S}^{n}{\bf S}^{n*}\right)^{-1}$,
we will need to show that $F_{c}^{k}(\omega)$ is uniformly bounded
away from 0.

When $k$ is sufficiently large, the Riemann integrability of $s(t)$
implies that
\begin{align*}
F_{c}^{k}(\omega) & \overset{\cdot}{=}\Delta\sum_{i=-\infty}^{\infty}\left(\int_{-\infty}^{+\infty}s(t+iT_{s}){\bf s}(t)^{*}\text{d}t\right)\exp\left(ji\omega\right)\\
 & =\Delta\int_{-\infty}^{\infty}\left(\int_{-\infty}^{\infty}s(t+\tau)s(t)^{*}\text{d}t\right)\\
 & \quad\quad\quad\quad\cdot\left(\sum_{i=-\infty}^{\infty}\delta\left(\tau-iT_{s}\right)\right)\exp\left(j\frac{\omega}{T_{s}}\tau\right)\text{d}\tau.
\end{align*}
 We observe that
\begin{align*}
 & \int_{-\infty}^{+\infty}\left(\int_{-\infty}^{\infty}s(t+\tau)s(t)^{*}\text{d}t\right)\exp\left(j\frac{\omega}{T_{s}}\tau\right)\text{d}\tau\\
= & \left(\int_{-\infty}^{\infty}s(t+\tau)\exp\left(j\frac{\omega}{T_{s}}\left(t+\tau\right)\right)\text{d}\tau\right)\\
 & \quad\quad\left(\int_{-\infty}^{+\infty}s(t)\exp\left(j\frac{\omega}{T_{s}}t\right)\text{d}t\right)^{*}\\
= & \left|S\left(-j\frac{\omega}{T_{s}}\right)\right|^{2}.
\end{align*}
 Since $F_{c}^{k}(\omega)$ corresponds to the Fourier transform of
the signals obtained by uniformly sampling $\int_{-\infty}^{\infty}s(t+\tau)s(t)^{*}\text{d}t$,
we can immediately see that
\begin{equation}
\lim_{\Delta\rightarrow0}F_{c}^{k}(\omega)=\frac{\Delta}{T_{s}}\sum_{i=-\infty}^{\infty}\left|S\left(-j\left(\frac{\omega}{T_{s}}-\frac{i2\pi}{T_{s}}\right)\right)\right|^{2}.
\end{equation}
 If for all $\omega\in\left[-\pi,\pi\right]$, we have
\begin{equation}
\sum_{i=-\infty}^{\infty}\left|S\left(-j\left(\frac{\omega}{T_{s}}-\frac{i2\pi}{T_{s}}\right)\right)\right|^{2}\geq\epsilon_{s}>0
\end{equation}
 for some constant $\epsilon_{s}$, then $\sigma_{\min}\left({\bf C}^{n}\right)=\inf_{\omega}F_{c}^{k}(\omega)\geq\frac{\Delta\epsilon_{s}}{T_{s}}$,
which leads to $\left\Vert \left({\bf C}^{n}\right)^{-1}\right\Vert _{2}\leq\frac{T_{s}}{\Delta\epsilon_{s}}$.

Let $\Xi^{n}={\bf C}^{n}-{\bf S}^{n}{\bf S}^{n*}$. Since ${\bf S}^{n}{\bf S}^{n*}\sim{\bf C}^{n}$,
we can have $\lim_{n\rightarrow\infty}\frac{1}{\sqrt{n}}\left\Vert {\bf \Xi}^{n}\right\Vert _{F}=0$,
which implies that
\begin{align*}
\lim_{n\rightarrow\infty}\frac{1}{\sqrt{n}}\left\Vert {\bf \Xi}^{n}\left({\bf C}^{n}\right)^{-1}\right\Vert _{\text{F}} & \leq\lim_{n\rightarrow\infty}\frac{1}{\sqrt{n}}\left\Vert {\bf \Xi}^{n}\right\Vert _{\text{F}}\left\Vert \left({\bf C}^{n}\right)^{-1}\right\Vert _{2}\\
 & \leq\lim_{n\rightarrow\infty}\frac{T_{s}}{\Delta\epsilon_{s}}\frac{1}{\sqrt{n}}\left\Vert {\bf \Xi}^{n}\right\Vert _{\text{F}}=0.
\end{align*}
 The Taylor expansion of $\left({\bf S}^{n}{\bf S}^{n*}\right)^{-1}$
yields
\begin{align*}
 & \left({\bf S}^{n}{\bf S}^{n*}\right)^{-1}=\left({\bf C}^{n}-{\bf \Xi}^{n}\right)^{-1}\\
= & \left({\bf C}^{n}\right)^{-1}\left({\bf I}+{\bf \Xi}^{n}\left({\bf C}^{n}\right)^{-1}+\left({\bf \Xi}^{n}\left({\bf C}^{n}\right)^{-1}\right)^{2}+\cdots\right).
\end{align*}
 Hence, we can bound
\begin{align*}
 & \lim_{n\rightarrow\infty}\frac{1}{\sqrt{n}}\left\Vert \left({\bf S}^{n}{\bf S}^{n*}\right)^{-1}-\left({\bf C}^{n}\right)^{-1}\right\Vert _{\text{F}}\\
\leq & \lim_{n\rightarrow\infty}\left\Vert \left({\bf C}^{n}\right)^{-1}\right\Vert _{2}\left(\sum_{i=1}^{\infty}\left(\frac{1}{\sqrt{n}}\left\Vert {\bf \Xi}^{n}\left({\bf C}^{n}\right)^{-1}\right\Vert _{\text{F}}\right)^{i}\right)=0.
\end{align*}

\subsection{Proof of Lemma \ref{lem-asymptoticSpectralPropertyGeneralSampling}
\label{sec:Proof-of-Lem-Asymptotic-Spectral-General-Sampling}}

Since $\left({\bf C}^{n}\right)^{-\frac{1}{2}}$ and $\left({\bf \hat{S}}^{n}\right)^{-\frac{1}{2}}$
are both Hermitian and positive semidefinite, we have $\left({\bf C}^{n}\right)^{-\frac{1}{2}}\sim\left(\hat{{\bf S}}^{n}\right)^{-\frac{1}{2}}$.
The asymptotic equivalence allows us to relate $\frac{1}{n}\sum_{i=1}^{n}g(\lambda_{i})$
to the function associated with the spectrum of the circulant matrix
${\bf C}^{n}$ instead of $\hat{{\bf S}}^{n}$. One nice property
is that $\left({\bf C}^{n}\right)^{-\frac{1}{2}}={\bf U}_{c}{\bf \Lambda}_{c}^{-\frac{1}{2}}{\bf U}_{c}^{*}$
is still a circulant matrix. Combining the above results with Lemma
\ref{lemmaAsymptoticEquivalenceEigenvalues} yields
\[
\left({\bf S}^{n}{\bf S}^{n}\right)^{-\frac{1}{2}}\tilde{{\bf H}}^{n}\tilde{{\bf H}}^{n*}\left({\bf S}^{n}{\bf S}^{n*}\right)^{-\frac{1}{2}}\sim\left({\bf C}^{n}\right)^{-\frac{1}{2}}\hat{{\bf H}}^{n}\left({\bf C}^{n}\right)^{-\frac{1}{2}}
\]
Note that $\left({\bf C}^{n}\right)^{-\frac{1}{2}}\hat{{\bf H}}^{n}\left({\bf C}^{n}\right)^{-\frac{1}{2}}$
is simply multiplication of 3 Toeplitz matrices. This allows us to
untangle $F_{c}\left(\omega\right)$ and $F_{\hat{h}}(\omega)$, hence
separating $H(f)$ and $S(f)$. 

Specifically, denote by $F_{c_{0.5}}\left(\omega\right)$, $F_{c}\left(\omega\right)$,
$F_{\hat{h}}\left(\omega\right)$, ${\bf F}_{\tilde{h}}\left(\omega\right)$
the Fourier series related to $\left({\bf C}^{n}\right)^{\frac{1}{2}}$
, ${\bf C}^{n}$ , $\hat{{\bf H}}^{n}$ and $\tilde{{\bf H}}^{n}$,
respectively. We note that $F_{c_{0.5}}\left(\omega\right)$, $F_{c}\left(\omega\right)$
and $F_{\hat{h}}\left(\omega\right)$ are all scalars since their
related matrices are Toeplitz, while ${\bf F}_{\tilde{h}}\left(\omega\right)$
is a $1\times k$ vector since $\tilde{{\bf H}}$ is block Toeplitz.
Then for any continuous function $g(x)$, applying \cite[Theorem 12]{Gray06}
yields
\begin{align*}
 & \lim_{n\rightarrow\infty}\frac{1}{n}\sum_{i=1}^{n}g\left\{ \lambda_{i}\left(\left({\bf S}^{n}{\bf S}^{n}\right)^{-\frac{1}{2}}\tilde{{\bf H}}^{n}\tilde{{\bf H}}^{n*}\left({\bf S}^{n}{\bf S}^{n*}\right)^{-\frac{1}{2}}\right)\right\} \\
= & \lim_{n\rightarrow\infty}\frac{1}{n}\sum_{i=1}^{n}g\left\{ \lambda_{i}\left(\left({\bf C}^{n}\right)^{-\frac{1}{2}}\hat{{\bf H}}^{n}\left({\bf C}^{n}\right)^{-\frac{1}{2}}\right)\right\} \\
= & \frac{1}{2\pi}{\displaystyle \int}_{-\pi}^{\pi}g\left(F_{c_{0.5}}^{-1}\left(\omega\right)F_{\hat{h}}\left(\omega\right)F_{c_{0.5}}^{-1}\left(\omega\right)\right)\text{d}\omega\\
= & \lim_{n\rightarrow\infty}\frac{1}{n}\sum_{i=1}^{n}g\left\{ \lambda_{i}\left(\left({\bf C}^{n}\right)^{-1}\hat{{\bf H}}^{n}\right)\right\} \\
= & \frac{1}{2\pi}{\displaystyle \int}_{-\pi}^{\pi}g\left(\frac{F_{\hat{h}}\left(\omega\right)}{F_{c}\left(\omega\right)}\right)\text{d}\omega.
\end{align*}

Now we only need to show that both $F_{\hat{h}}\left(\omega\right)$
and $F_{c}\left(\omega\right)$ have simple close-form expressions.
We observe that $\hat{{\bf H}}_{n}$ is asymptotically equivalent
to $\tilde{{\bf H}}^{n}\tilde{{\bf H}}^{n*}$, and the eigenvalues
of $\tilde{{\bf H}}^{n}\tilde{{\bf H}}^{n*}$ are exactly the square
of the corresponding singular values of $\tilde{{\bf H}}^{n}$. Hence,
we know from \cite{Tilli98} that for any continuous function $g(x)$:
\begin{align*}
\lim_{n\rightarrow\infty}\frac{1}{n}\sum_{i=1}^{n}g\left\{ \lambda_{i}\left(\hat{{\bf H}}^{n}\right)\right\}  & =\lim_{n\rightarrow\infty}\frac{1}{n}\sum_{i=1}^{n}g\left\{ \sigma_{i}^{2}\left(\tilde{{\bf H}}^{n}\right)\right\} \\
 & =\frac{1}{2\pi}{\displaystyle \int}_{-\pi}^{\pi}g\left(\sigma^{2}\left({\bf F}_{\tilde{h}}(\omega)\right)\right)\text{d}\omega
\end{align*}
 where ${\bf F}_{\tilde{h}}(\omega)$ can be expressed as ${\bf F}_{\tilde{h}}(\omega)=\left[F_{\tilde{h},0}(\omega),\cdots,F_{\tilde{h,}k-1}(\omega)\right]$.
Here, for any $0\leq i<k$:
\begin{align*}
F_{\tilde{h},i}(\omega) & :=\Delta\sum_{u=-\infty}^{+\infty}\tilde{h}_{u,i}\exp\left(ju\omega\right)\\
 & =\Delta\sum_{u\in\mathbb{Z}}\tilde{h}(uT_{s}-i\Delta)\exp\left(ju\omega\right).
\end{align*}
 The above analysis implies that $F_{\hat{h}}\left(\omega\right)=\sigma^{2}\left(F_{\tilde{h}}(\omega)\right)$.

Through algebraic manipulation, we have that
\[
F_{\tilde{h},i}(\omega)=\frac{\Delta}{T_{s}}\sum_{l\in\mathbb{Z}}H\left(-f+lf_{s}\right)\exp\left(-j2\pi\left(f-lf_{s}\right)i\Delta\right),
\]
which yields
\begin{align*}
 & F_{\hat{h}}\left(f\right)=\sigma^{2}\left({\bf F}_{\tilde{h}}(2\pi f)\right)=\sum_{i=0}^{k-1}\left|F_{\tilde{h},i}(2\pi f)\right|^{2}\\
= & \frac{\Delta^{2}}{T_{s}^{2}}\sum_{i=0}^{k-1}\left|\sum_{l=-\infty}^{+\infty}\tilde{H}\left(-f+lf_{s}\right)\exp\left(-j2\pi\left(f-lf_{s}\right)i\Delta\right)\right|^{2}\\
= & \frac{\Delta^{2}}{T_{s}^{2}}\sum_{i=0}^{k-1}\sum_{l_{1},l_{2}}\tilde{H}\left(-f+l_{1}f_{s}\right)\tilde{H}^{*}\left(-f+l_{2}f_{s}\right)\cdot\\
 & \quad\quad\quad\exp\left(-j2\pi\left(l_{2}-l_{1}\right)f_{s}i\Delta\right)\\
= & \frac{\Delta^{2}}{T_{s}^{2}}\sum_{l_{1},l_{2}}\tilde{H}\left(-f+l_{1}f_{s}\right)\tilde{H}^{*}\left(-f+l_{2}f_{s}\right)\\
 & \quad\quad\quad\left[\sum_{i=0}^{k-1}\exp\left(-j2\pi\left(l_{2}-l_{1}\right)\frac{i}{k}\right)\right]\\
= & \frac{\Delta}{T_{s}}\sum_{l}\left|H\left(-f+lf_{s}\right)S\left(-f+lf_{s}\right)\right|^{2}.
\end{align*}
 Similarly, we have
\begin{equation}
F_{c}\left(f\right)=\frac{\Delta}{T_{s}}\sum_{l\in\mathbb{Z}}\left|S\left(-f+lf_{s}\right)\right|^{2}.
\end{equation}
 Combining the above results yields
\begin{align*}
 & \lim_{n\rightarrow\infty}\frac{1}{n}\sum_{i=1}^{n}g\left\{ \lambda_{i}\left(\left({\bf S}^{n}{\bf S}^{n}\right)^{-\frac{1}{2}}\tilde{{\bf H}}^{n}\tilde{{\bf H}}^{n*}\left({\bf S}^{n}{\bf S}^{n*}\right)^{-\frac{1}{2}}\right)\right\} \\
= & T_{s}{\displaystyle \int}_{-f_{s}/2}^{f_{s}/2}g\left(\frac{\sum_{l=-\infty}^{+\infty}\left|H\left(-f+lf_{s}\right)S\left(-f+lf_{s}\right)\right|^{2}}{\sum_{l=-\infty}^{+\infty}\left|S\left(-f+lf_{s}\right)\right|^{2}}\right)\text{d}f
\end{align*}
This completes the proof.

\subsection{Proof of Lemma \ref{lem: FourierSymbolModulationBank} \label{sec:Proof-of-Lemma-Fourier-Symbol-Modulation}}

Denote by ${\bf K}_{\alpha}^{\eta}$ the Fourier symbol associated
with the block Toeplitz matrix ${\bf G}_{\alpha}^{\eta}$. We know
that the Fourier transform of $g_{i}^{\eta}\left(t,\tau\right)$ with
respect to $\tau$ is given by
\begin{align*}
 & {\displaystyle \int}_{\tau}g_{i}^{\eta}\left(t,\tau\right)\exp\left(-j2\pi f\tau\right)\mathrm{d}\tau\\
= & {\displaystyle \int}_{\tau_{2}}{\displaystyle \int}_{\tau_{1}}s_{i}\left(t-\tau_{1}\right)q_{i}\left(\tau_{1}\right)p_{i}\left(\tau_{1}-\tau_{2}\right)\exp\left(-j2\pi f\tau_{2}\right)\mathrm{d}\tau_{1}\mathrm{d}\tau_{2}\\
= & {\displaystyle \int}_{\tau_{2}}p_{i}\left(\tau_{1}-\tau_{2}\right)\exp\left(j2\pi f\left(\tau_{1}-\tau_{2}\right)\right)\mathrm{d}\tau_{2}\\
 & \quad\quad{\displaystyle \int}_{\tau_{1}}s_{i}\left(t-\tau_{1}\right)q_{i}\left(\tau_{1}\right)\exp\left(-j2\pi f\tau_{1}\right)\mathrm{d}\tau_{1}\\
= & P_{i}\left(-f\right)\cdot\left[S_{i}\left(-f\right)\exp\left(-j2\pi tf\right)\cdot\sum_{u}c_{i}^{u}\delta\left(f-uf_{q}\right)\right]\\
= & P_{i}\left(-f\right)\cdot\left[\sum_{u}c_{i}^{u}S_{i}\left(-f+uf_{q}\right)\exp\left(-j2\pi t\left(f-uf_{q}\right)\right)\right].
\end{align*}
Introduce the notation $S^{e}(f)\overset{\Delta}{=}S(f)\exp\left(j2\pi l\tilde{T}_{s}f\right)$.
For any $\left(l,m\right)$ such that $1\leq l\leq a$ and $1\leq m\leq ak$,
the $\left(l,m\right)$ entry of the Fourier symbol ${\bf K}_{\alpha}^{\eta}$
can be related to the sampling sequence of $g_{\alpha}^{\eta}\left(l\tilde{T}_{s},\tau\right)$
at a rate $\frac{\tilde{f}_{s}}{a}$ with a phase shift $m\Delta$,
and hence it can be calculated as follows

\begin{align*}
\left({\bf K}_{\alpha}^{\eta}\right)_{l,m}= & \sum_{v}P_{i}\left(-f+v\frac{f_{q}}{b}\right)\exp\left(j2\pi\left(f-v\frac{f_{q}}{b}\right)m\Delta\right)\\
 & \quad\quad\cdot\left[\sum_{u}c_{i}^{u}S^{e}\left(-f+uf_{q}+v\frac{f_{q}}{b}\right)\right].
\end{align*}
Using the fact that $\sum_{m=0}^{ak-1}\exp\left(j2\pi\left(\left(v_{2}-v_{1}\right)\frac{f_{q}}{b}\right)m\Delta\right)=ak\delta\left[v_{2}-v_{1}\right]$,
we get through algebraic manipulation that 
\begin{align*}
\left({\bf K}_{\alpha}^{\eta}{\bf K}_{\beta}^{\eta*}\right)_{l,d}= & ak\sum_{v}P_{\alpha}\left(-f+v\frac{f_{q}}{b}\right)\cdot P_{\beta}^{*}\left(-f+v\frac{f_{q}}{b}\right)\\
 & \quad\quad\quad\left[\sum_{u_{1}}c_{\alpha}^{u_{1}}S_{\alpha}^{e}\left(-f+u_{1}f_{q}+v\frac{f_{q}}{b}\right)\right]\cdot\\
 & \quad\quad\quad\left[\sum_{u_{2}}c_{\beta}^{u_{2}}S_{\beta}^{e}\left(-f+u_{2}f_{q}+v\frac{f_{q}}{b}\right)\right]^{*}.
\end{align*}
Define another matrix ${\bf F}_{\alpha}^{\eta}$ such that 
\begin{align*}
\left({\bf F}_{\alpha}^{\eta}\right)_{l,v} & =\sum_{u}c_{\alpha}^{u}S_{\alpha}\left(-f+uf_{q}+v\frac{f_{q}}{b}\right)\cdot\\
 & \quad\quad\quad\exp\left(-j2\pi l\tilde{T}_{s}\left(f-uf_{q}-v\frac{f_{q}}{b}\right)\right).
\end{align*}
It can be easily seen that
\[
{\bf K}_{\alpha}^{\eta}{\bf K}_{\beta}^{\eta*}=ak{\bf F}_{\alpha}^{\eta}{\bf F}_{\alpha}^{p}{\bf F}_{\alpha}^{p*}{\bf F}_{\beta}^{\eta*}.
\]
Replacing $P_{\alpha}$ by $P_{\alpha}H$ immediately gives us the
Fourier symbol for ${\bf G}_{\alpha}^{h}{\bf G}_{\beta}^{h*}$.

\bibliographystyle{IEEEtran} \bibliographystyle{IEEEtran}

\bibliographystyle{IEEEtran} \bibliographystyle{IEEEtran}
\bibliography{bibfileCap}

\begin{IEEEbiographynophoto}{Yuxin Chen} (S'09) received the B.S. in Microelectronics with High Distinction from Tsinghua University in 2008, and the M.S. in Electrical and Computer Engineering from the University of Texas at Austin in 2010. He is currently a Ph.D. candidate in the Department of Electrical Engineering and a Master student in the Department of Statistics at Stanford University. His research interests include information theory, compressed sensing, network science and high-dimensional statistics. \end{IEEEbiographynophoto} \begin{IEEEbiographynophoto}{Yonina C. Eldar} Yonina C. Eldar (S'98-M'02-SM'07-F'12) received the B.Sc. degree in physics and  the B.Sc. degree in electrical engineering both from Tel-Aviv University (TAU), Tel-Aviv,  Israel, in 1995 and 1996, respectively, and the Ph.D. degree in electrical engineering and computer science from the  Massachusetts Institute of Technology (MIT), Cambridge, in 2002.

 From January 2002 to July 2002, she was a Postdoctoral Fellow at the  Digital Signal Processing Group at MIT. She is currently a Professor in the Department of Electrical  Engineering at the Technion-Israel Institute of Technology, Haifa and holds the  The Edwards Chair in Engineering. She is  also a Research Affiliate with the Research Laboratory of Electronics at MIT  and a Visiting Professor at Stanford University, Stanford, CA. Her research interests are in the broad areas of  statistical signal processing, sampling theory and compressed sensing,  optimization methods, and their applications to biology and optics.

Dr. Eldar was in the program for outstanding students at TAU from 1992 to 1996. In 1998, she held the Rosenblith Fellowship for study in electrical engineering at MIT, and in 2000, she held an IBM Research Fellowship. From  2002 to 2005, she was a Horev Fellow of the Leaders in Science and  Technology program at the Technion and an Alon Fellow. In 2004, she was  awarded the Wolf Foundation Krill Prize for Excellence in Scientific  Research, in 2005 the Andre and Bella Meyer Lectureship, in 2007 the Henry  Taub Prize for Excellence in Research, in 2008 the Hershel Rich Innovation  Award, the Award for Women with Distinguished Contributions, the Muriel \& David Jacknow Award for Excellence in Teaching, and the Technion Outstanding  Lecture Award, in 2009 the Technion's Award for Excellence in Teaching, in 2010 the Michael  Bruno Memorial Award from the Rothschild Foundation, and in 2011 the Weizmann Prize for Exact Sciences.   In 2012 she was elected to the Young Israel Academy of Science and to the Israel Committee for Higher Education, and elected an IEEE Fellow. She received several best paper awards together with her research students and colleagues.  She is a Signal Processing Society Distinguished Lecturer, a member of the IEEE Bio Imaging Signal Processing technical committee, an Associate Editor for the SIAM Journal on Imaging Sciences, and Editor in Chief of Foundations and Trends in Signal Processing. In the past, she was a member of the IEEE Signal Processing Theory and Methods technical committee, and served as an associate editor for the IEEE Transactions On Signal Processing, the EURASIP Journal of Signal Processing, and the SIAM Journal on Matrix Analysis and Applications.
\end{IEEEbiographynophoto} 

\begin{IEEEbiographynophoto}{Andrea J. Goldsmith}
is the Stephen Harris professor in the School of Engineering and a professor of Electrical Engineering at Stanford University. She was previously on the faculty of Electrical Engineering at Caltech. She co-founded Accelera, Inc., which develops software-defined wireless network technology, and Quantenna Communications Inc., which develops high-performance WiFi chipsets. She has previously held industry positions at Maxim Technologies, Memorylink Corporation, and AT\&T Bell Laboratories. Dr. Goldsmith is a Fellow of the IEEE and of Stanford, and she has received several awards for her work, including the IEEE Communications Society and Information Theory Society joint paper award, the National Academy of Engineering Gilbreth Lecture Award, the IEEE Wireless Communications Technical Committee Recognition Award, the Alfred P. Sloan Fellowship, and the Silicon Valley\/San Jose Business Journal's Women of Influence Award. She is author of the book ``Wireless Communications'' and co-author of the books ``MIMO Wireless Communications'' and ``Principles of Cognitive Radio,'' all published by Cambridge University Press. She received the B.S., M.S. and Ph.D. degrees in Electrical Engineering from U.C. Berkeley.

Dr. Goldsmith is currently on the Steering Committee for the IEEE Transactions on Wireless Communications, and has previously served as editor for the IEEE Transactions on Information Theory, the Journal on Foundations and Trends in Communications and Information Theory and in Networks, the IEEE Transactions on Communications, and the IEEE Wireless Communications Magazine. Dr. Goldsmith participates actively in committees and conference organization for the IEEE Information Theory and Communications Societies and has served on the Board of Governors for both societies. She has been a Distinguished Lecturer for both societies, served as the President of the IEEE Information Theory Society in 2009, founded and chaired the student committee of the IEEE Information Theory society, and currently chairs the Emerging Technology Committee and is a member of the Strategic Planning Committee in the IEEE Communications Society. At Stanford she received the inaugural University Postdoc Mentoring Award, served as Chair of its Faculty Senate, and currently serves on its Faculty Senate and on its Budget Group. 
\end{IEEEbiographynophoto} 
\end{document}